\documentclass[twocolappendix,numberedappendix]{emulateapj}

\usepackage{times}
\usepackage{graphicx}
\usepackage{epstopdf}
\usepackage{subfigure}
\usepackage{color}
\usepackage{verbatim}
\usepackage{bm}
\usepackage{hyperref}

\hypersetup{colorlinks = true, linkcolor = red, anchorcolor = red, citecolor = blue, filecolor = red, urlcolor = red, backref = page}

\begin{document}

\title{Stellar and AGN feedback in isolated early-type galaxies: the role in regulating star formation and ISM properties}

\author
{Ya-Ping Li$^{1}$, Feng Yuan$^{1,2}$, Houjun Mo$^{3,4}$, Doosoo Yoon$^{1}$, Zhaoming Gan$^{1}$, Luis C. Ho$^{5,6}$, Bo Wang$^{7}$, Jeremiah P. Ostriker$^8$, Luca Ciotti$^9$}
\affil{$^{1}$Key Laboratory for Research in Galaxies and Cosmology, Shanghai Astronomical Observatory, Chinese Academy of Sciences, 80
Nandan Road, Shanghai 200030, China; liyp@shao.ac.cn, fyuan@shao.ac.cn, yoon@shao.ac.cn, zmgan@shao.ac.cn}
\affil{$^{2}$School of Astronomy and Space Sciences, University of Chinese Academy of Sciences, No. 19A Yuquan Road, Beijing 100049, China}
\affil{$^{3}$Department of Astronomy, University of Massachusetts, Amherst MA 01003-9305, USA; hjmo@astro.umass.edu}
\affil{$^{4}$Tsinghua Center of Astrophysics and Department of Physics, Tsinghua University, Beijing 100084, China}
\affil{$^{5}$Kavli Institute for Astronomy and Astrophysics, Peking University, Beijing 100871, China; lho.pku@gmail.com}
\affil{$^{6}$Department of Astronomy, School of Physics, Peking University, Beijing 100871, China}
\affil{$^{7}$Key Laboratory for the Structure and Evolution of Celestial Objects, Yunnan Observatories, CAS, Kunming 650216, China; wangbo@ynao.ac.cn}
\affil{$^{8}$Department of Astronomy, Columbia University, 550 W, 120th Street, New York, NY10027, USA; jpo@astro.columbia.edu}
\affil{$^{9}$Department of Physics and Astronomy, University of Bologna, via Piero Gobetti 93/2, 40129 Bologna, Italy; luca.ciotti@unibo.it}

\begin{abstract}
{Understanding how galaxies maintain the inefficiency of star formation with physically self-consistent models is a central problem for galaxy evolution. Although numerous theoretical models have been proposed in recent decades, the debate still exists. By means of high-resolution two-dimensional hydrodynamical simulations, we study the three feedback effects (the stellar wind heating, SNe feedback, and AGN feedback) in suppressing star formation activities on the evolution of early-type galaxies with different stellar masses. AGN feedback models are updated based on \citet{Yuan2018}. The gas sources comes exclusively from the mass losses of dying low-mass stars for most of our models. We find that SNe feedback can keep star formation at a significantly low level for low mass elliptical galaxies for a cosmological evolution time. For the high mass galaxies, AGN feedback can efficiently offset the radiative cooling and thus regulate the star formation activities. Such a suppression of star formation is extremely efficient in the inner region of the galaxies. AGB heating cannot account for this suppression for low and high mass galaxies. The X-ray temperature $T_{\rm X}$ and luminosity $L_{\rm X}$ of hot plasma can be in agreement with the observed data with the inclusion of effective feedback processes.  These results thus suggest that we can use $T_{\rm X}$ and $L_{\rm X}$ to probe the role of different feedback processes. The inclusion of additional gas sources can make the mass scale between SNe and AGN feedback dominating in suppressing star formation decrease to an observationally inferred value of a few $10^{10}~M_{\odot}$.}
\end{abstract}

\section{Introduction}

Early-type galaxies (ETGs) in our local Universe are representative quiescent systems with little ongoing star formation activities, in which almost all of their stars are formed $\sim10$ Gyr ago \citep[e.g.,][]{Kauffmann2003}. The existence of these quiescent galaxies raises two major questions to galaxy formation and evolution models (see \citealt{Naab2017} for a review). The first one is what causes the cessation of the star formation in the first place (e.g., \citealt{Mo2005}, and references therein), and the second one is how to maintain the observed low level star formation rates (SFRs) over the cosmological evolution time. The second one could be essential because the ETGs are embedded in hot X-ray emitting gases \citep{Fabbiano1989,OSullivan2001}. These hot diffuse gases continuously replenished by stellar mass losses will eventually form new stars if an effective heating source is absent to offset the radiative cooling in the entire galaxy \citep[][]{Fabian1984,Croton2006}.
A major challenge for galaxy formation and evolution is thus the reproduction of the overall inefficiency of star formation with physically self-consistent models.

From the observational point of view, the energy deposited in the interstellar medium (ISM) by both stellar feedback via stellar winds and supernovae (SNe) \citep[e.g.,][]{Heckman2000,Pettini2000} and active galactic nucleus (AGN) feedback via the powerful wind and radiation (e.g., \citealt{Lynds1967,Schawinski2007,Feruglio2010,Cicone2012,Tombesi2013,Teng2014,Baron2017}, see also \citealt{McNamara2007,Fabian2012,King2015} for reviews on different aspects) can suppress star formation\footnote{But see also \citet{Bieri2016} for a compilation of the observational evidence of AGN positive feedback on star formation activities.} on galactic scales. Observational evidence of AGN feedback effect in the host galaxy can also be found from the linear correlation between the black hole accretion rate and the star formation activity of galaxies hosting AGNs (both for star-forming and quiescent galaxies, e.g., \citealt{Harris2016,Netzer2016,Yang2017}), although some debates still exist \citep[e.g.,][]{Harrison2012,Rosario2012,Azadi2015,Barger2015,Stanley2015,Suh2017}. The reason for this discrepancy is probably because of the sample selection and the different timescales between the black hole accretion and star formation activities \citep{Hickox2014,McAlpine2017}. These observations suggest that the energetic output from stellar wind, SNe and/or AGNs are indeed important in regulating the star formation activities.

Various theoretical works to solve this long outstanding problem by invoking some forms of feedback process or a combination of them exist in the literatures.
In this way, many semi-analytic works have been conducted in recent decades. It is believed that stellar feedback is an important mechanism to suppress star formation activities in low-mass galaxies and to match the low-mass end of the galaxy luminosity function \citep[e.g.,][]{Somerville1999,Benson2003,Bower2006,Bower2008,Bower2012,Pan2017}, while AGN feedback can have significant impact in more massive systems \citep[e.g.,][for a review]{Silk2012}.
Other scenarios due to external or environmental effect, such as tidal interactions \citep[][]{Moore1996} and ram pressure stripping \citep[e.g.,][]{Gunn1972,Dressler1983,Gavazzi1995,Boselli2006}, are also proposed.

Recently, \citet{Conroy2015} proposed another stellar feedback scenario to prevent star formation in elliptical galaxies based on an analytical analysis. They suggest that the thermalization of the winds from dying low-mass stars (asymptotic giant branch (AGB) stars, red giants, and planetary nebula phases, hereafter we simply term it AGB heating) with pre-existing hot ISM itself can be responsible for heating the gas and hence plays an important role in preventing star formation in quiescent galaxies.
However, the analytical model neglects the mass increment to the ISM from evolved dying low-mass stars and it is hard to capture the complex interaction of the stellar wind with the ambient gas. Detailed numerical simulations are, therefore, necessary to address this issue.

Apart from these semi-analytic works, there exist abundant theoretical works based on large scale cosmological simulations, which can incorporate many physical processes occurring in real galaxies.
Qualitative agreements with some semi-analytic models have been reached.
Specifically, stellar feedback due to stellar winds, photoionization from young stars and the thermal energy from SN incorporated in cosmological simulations is proven to be the chief piece of physics required to limit the efficiency of star formation and reduce star formation in less massive galaxies ($M_{\star}\lesssim 10^{11}~M_{\odot}$; e.g., \citealt{Stinson2006,Scannapieco2008,Schaye2010,Stinson2013,Hopkins2014,Wang2015,Angles2017,Habouzit2017,Prieto2017}).
However, energy feedback from type II SN feedback alone is found to be difficult to produce realistic quiescent galaxies \citep[e.g.,][]{Martizzi2014}.
For high galaxy masses, the energy sources could be provided by radiation, wind and radio jet from the central AGN \citep[e.g.,][]{Croton2006,Lu2007,Choi2015,Dave2016,Weinberger2017,Taylor2017}\footnote{There are also some theoretical works which suggest AGN-induced star formation activities \citep[e.g.,][]{Ciotti2007,Liu2013,Silk2013,Zubovas2013,Bieri2016} or no impact of AGN feedback in the star formation \citep{Roos2015}.}.

Compared to analytical works, cosmological simulations are better to  capture the environmental effect during the cosmological evolution of galaxies. However, the scales on which different feedback processes operate are usually much smaller than the typical resolution of cosmological simulations, although this situation is improving in recent years. For example, the TNG50 run of the latest Illustris TNG simulations can reach a high spatial resolution of 74 pc (for the Illustris TNG simulations method, see, e.g., \citealt{Springel2018})\footnote{http://www.tng-project.org/}. Recently, \citet{Curtis2015} developed a new refinement scheme to increase the spatial and mass resolution around the accreting black hole, and it has been applied in the moving mesh-code \texttt{AREPO} to investigate the black hole growth and different AGN feedback processes. Nevertheless, if the focus of a simulation is a single galaxy with an idealized, isolated setup, then a much higher resolution can be reached, allowing one to pin down how feedback arises on relatively smaller scales to influence galaxy properties. Many simulation works have been carried out in this direction, some of them focused on the effect of feedback by AGN on nuclear activities and host galaxy properties \citep[e.g.,][]{Binney1995,Ciotti1997,Ciotti2007,Novak2011,Choi2012,Gan2014,Ciotti2017,Eisenreich2017,Biernacki2018,Yuan2018}, while others on the role of SN feedback \citep[e.g.,][]{Ciotti1991,Nunez2017,Smith2018}.

As different feedback processes could be responsible for suppressing the star formation activities in different galaxies, it is crucial to isolate different feedback processes to understand their individual roles.
To this aim, we perform two-dimensional (2D) hydrodynamical simulations for an idealized, isolated galaxy with different feedback models to quantify the roles of different feedback processes. We mainly consider three feedback processes, which are AGB heating, SNe feedback, and AGN feedback. The stellar mass of the elliptical galaxies we explore covers a range of $6.9\times10^{9}-9.8\times10^{11}~M_{\sun}$. We then explore how these three different feedback processes can regulate star formation activities in elliptical galaxies differently.

X-ray observations show that the gas temperatures $T_{\rm X}$ of low angular momentum galaxies are consistent just with the thermalization of the stellar kinetic energy estimated from their stellar velocity dispersion $\sigma_{\rm e}$ within the effective radius $r_{\rm eff}$ \citep{Boroson2011,Pellegrini2011,Sarzi2013,Goulding2016}. The physical reason may be related to feedback processes. The X-ray luminosity $L_{\rm X}$ of the hot gas, produced by the bremsstrahlung and metal-line radiation, is the main observable by which galactic haloes are detected.
Both $T_{\rm X}$ and $L_{\rm X}$ are effected by different feedback processes, making them potentially good tools to probe the effect of feedback apart from star formation activities \citep[e.g.,][]{Ciotti1991,Tang2009,Tang2010,Ostriker2010,Pellegrini2011,Choi2012,Pellegrini2012b,Gaspari2014,Negri2014,Choi2015,Eisenreich2017,Pellegrini2018}. Therefore, we will further investigate the effect of different feedback processes on $T_{\rm X}$ and $L_{\rm X}$, and build a connection with the star formation suppression. This is another goal of this paper.

As a series of works, we here mainly focus on a comparison of these three different feedback models in regulating star formation and related ISM properties for different galaxies. For a detailed discussion of AGN feedback, we refer the interested readers to \citet{Yuan2018}. Compared to previous works, the main feature of this work is that they have incorporated the most updated AGN physics into simulations, namely the correct descriptions of radiation and wind from the AGN for any given accretion rates. This is obviously crucial to correctly evaluate the effect of AGN feedback.  \citet{Yoon2018} extended \citet{Yuan2018} to the case of high angular momentum galaxies.

The paper is structured as follows. In Section \ref{sec:model}, we introduce the physics included in our model. The numerical results for different feedback models in different galaxies are presented in Section \ref{sec:results}. Discussions and summary of our work are given in Section \ref{sec:summary}.

\section{Model}\label{sec:model}

In this section, we introduce the feedback model with a highlight on stellar physics in our models. Most of the input physics is the same as those in \citeauthor{Novak2011} (\citeyear{Novak2011}; and see also \citealt{Gan2014}) except the AGN physics, which is taken from \citet{Yuan2018} (see the Appendix for a description).

\subsection{Galaxy Model}
Following \citet{Ciotti2007}, we choose the galaxy parameters to be consistent with the edge-on view of the fundamental plane and the Faber-Jackson relation \citep{Faber1976}.
The structural and dynamical properties of the galaxy models adopted here can be found in \citet{Ciotti2009}. Specifically, the \citet{Jaffe1983} stellar models embedded in a dark matter (DM) halo plus a central suppermassive black hole (SMBH) of mass $M_{\rm BH}$ are used so that the total mass density profile is described by a singular isothermal sphere. The stellar density profile is

\begin{equation} \label{eq:jaffe-dstar}
\rho_{\star}  = {M_{\star} r_{\star}\over 4\pi r^2( r_{\star}+r)^2},
\end{equation}
where $ M_{\star} $ and $r_{\star}$ are the total stellar mass and the scale-length of the galaxy, respectively, and the effective radius is $r_{\rm eff}=0.7447r_{\star}$. The total mass density profile is given by
\begin{equation} \label{eq:dtot}
\rho_{\rm T}  =\frac{v_{\rm c}^2}{4\pi Gr^2},
\end{equation}
where $v_{\rm c}$ is the constant circular velocity, and is related to the central projected velocity dispersion as $\sigma_{\circ}=v_{\rm c}/\sqrt{2}=\sqrt{GM_{\star}/2r_{\star}}$. The gravitational potential associated with this mass density profile is then described as

\begin{equation} \label{eq:pot}
\phi_{\rm T}  ={v_{\rm c}^2}\ln (r/r_{\star}).
\end{equation}

An important ingredient is the energetics of the gas flows, namely the thermalization of the stellar wind which depends on its radial profile of the stellar velocity dispersion. For the isotropic model we consider here, the energy density associated with the stellar wind is given by
\begin{eqnarray}
\nonumber \rho_{\star}\sigma_{\star}^2 &=&\rho_{\star}\sigma^2_{\star\circ}+ {GM_{\star}M_{\rm BH}\over 4\pi r_{\star}^4} \\
            &\times& \left[{1-2s+6s^2+12s^3\over 3s^3(1+s)} -4\ln\left(1+{1\over s}\right)\right],
\label{eq:jeanstot}
\end{eqnarray}
where $s\equiv {r\over r_{\star}}$, and
\begin{eqnarray}
\sigma_{\star\circ}^2 &=& \sigma_{\circ}^2 (1+s)^{2}s^2
            \left[\frac{1-3s-6s^2}{s^2(1+s)}+6\ln\frac{1+s}{s} \right],
\label{eq:jeans}
\end{eqnarray}
where $\sigma_{\star\circ}$ is the isotropic one-dimensional stellar velocity dispersion without the contribution of the central SMBH.

For the secular evolution cases we consider, the gas density over the galaxy is initially very low\footnote{The initial gas density is set to be $n_{0}=10^{-10}~{\rm cm^{-3}}$ (ref. to Equation~\ref{eq:gas}), which is negligible compared to the final gas density, e.g., shown in Figure~\ref{fig:ism_c12}.} so that the gas in the simulations comes almost exclusively from stellar evolution, e.g., mass losses from evolved stars as described in Section~\ref{sec:sse}. For several models, we also explore the effect of the external gas supply by including a gaseous component throughout the entire galaxy in the initial setup as shown in Section~\ref{sec:results:igas}.

\subsection{Radiative Cooling and Heating}
Radiative cooling process are computed by using the formulae in \citet{Sazonov2005}. When AGN feedback is ignored in our simulations, it corresponds to the case of the ionization parameter $\xi=0$ \citep{Sazonov2005,Ciotti2012}. In particular,
bremsstrahlung losses, line and continuum cooling, are taken into account for solar metallicity.
The net gas energy change rate per unit volume for $T \gtrsim 10^4$~K
is given by:
\begin{equation}
 C \equiv n^2 (S_{\rm brem} + S_{\rm recomb}),
\label{eq:cooling}
\end{equation}
where $n$ is the Hydrogen number density. All quantities are expressed in cgs units. The bremsstrahlung loss is given by
\begin{equation}
S_{\rm brem} = 3.8\times 10^{-27}\sqrt{T}.
\label{eq:brem}
\end{equation}
The sum of line and recombination continuum cooling is
\begin{equation}
S_{\rm recomb} = 10^{-23}\frac{Z}{Z_{\sun}}a(T),
\label{eq:recomb}
\end{equation}
where the solar metallicity is used in this work, and
\begin{equation}
a(T)=\frac{18}{  e^{25 (\log T -4.35)^2}}
    +\frac{80}{  e^{5.5(\log T -5.2)^2}}
    +\frac{17}{  e^{3.6(\log T -6.5)^2}}.
\end{equation}

For the case of AGN feedback included, we use the cooling and heating function as presented in \citet{Sazonov2005} except an updated Compton temperature (see also \citealt{Ciotti2012,Yuan2018}, and a brief description in the Appendix.). Specifically, the Compton heating/cooling and photoionization heating are incorporated when AGN feedback is included.

\subsection{Stellar Secular Evolution and SN Ia heating}\label{sec:sse}
In ETGs, the gas is lost by evolved stars mainly during the red giant, AGB, and planetary nebula phases at a rate of $\dot{M}_{\star}$. These ejecta, with an initial velocity of the parent star, then interact with the mass lost from other stars or with the hot ISM and mix with it. Thus, stellar winds are heated to X-ray temperatures by thermalization
of the kinetic energy of collisions between stellar ejecta. According to single burst stellar population synthesis models \citep{Maraston2005}, the evolution of $\dot{M}_{\star}$
for solar metal abundance, after an age of $\gtrsim 2$ Gyr, can be approximated as \citep{Pellegrini2012}:
\begin{equation}
\dot{M}_{\star} (t) = 10^{-12}\, A\,\times M_{\star} \,\, t_{12}^{-1.3}
\quad\quad ({M_{\sun}~{\rm yr}^{-1}}),
\label{eq:mdots_agb}
\end{equation}
where $M_{\star}$ is the galactic stellar mass in solar masses at an age of 12 Gyr\footnote{Here, we do not consider the time evolution of $M_{\star}$, in terms of both the spatial distribution and the total stellar mass, since the total stellar mass loss integrated after 2 Gyr is only $\sim6\%$ of the initial value for a Kroupa IMF.}, $t_{12}$ is the age in units of 12 Gyr, and $A=2.0$ or 3.3 for a
Salpeter or Kroupa IMF (the latter is adopted here). The energy input per unit time into the ISM due to the thermalization (simply termed the AGB heating rate) is

\begin{equation}\label{eq:lsn_agb}
  L_{\star}(t)=\frac{3}{2}\int\dot{\rho}_{\star}(t)\,\sigma_{\star}^2\textrm{d}V,
\end{equation}
where $\dot{\rho}_{\star}(t)$ is the stellar mass loss rate density. The stellar mass loss and energy injection rates, which are proportional to the stellar mass density $\rho_{\star}$ in each grid, are added into the computational domain.\footnote{We do not consider the contribution of newly formed stars to AGB heating because of the minor effect.}

The luminosity defined above can be converted into an equivalent temperature as \citep{Pellegrini2011}
\begin{equation}\label{eq:tsigma}
T_{\sigma}=\frac{\mu m_{\rm p}}{k_{\rm B}M_{\star}}\int\rho_{\star}\sigma_{\star}^2\textrm{d}V,
\end{equation}
where $k_{\rm B}$ is the Boltzmann constant, $m_{\rm p}$ is the proton mass, and $\mu$ is the mean molecular weight.

SN Ia explosions also provide mass and energy to the ISM, where the mass loss due to SN Ia is $\dot{M}_{\rm SN}
(t)=1.4M_{\sun}\, R_{\rm SN} (t)$. Here $R_{\rm SN} (t)$ (in yr$^{-1}$) is the time
evolution of the explosion rate, and each SN Ia ejects
$1.4M_{\sun}$. Following the latest observational and theoretical results \citep{Totani2008,Maoz2010,Maoz2012,Wang2012,Graur2013,Liu2018},
we parameterize the time evolution of the SN Ia rate as
\begin{equation}
R_{\rm SN} (t)=1.6\times10^{-13}\frac{L_{\rm B}}{L_{\rm B\sun}}\left(\frac{t}{13.7~\rm Gyr}\right)^{\rm -s} \,\, ({\rm yr}^{-1}) ,
\label{eq:rsn}
\end{equation}
where $L_{\rm B}$ is the present epoch B-band galaxy luminosity, and $s=1.1$ characterizes the secular evolution \citep{Pellegrini2012}. This is roughly consistent with the SN Ia rate adopted in the previous works \citep{Novak2011,Gan2014}. Then the corresponding mass loss rate is
\begin{equation}
\dot{M}_{\rm SN}(t)=2.2\times10^{-13}\frac{L_{\rm B}}{L_{\rm B\sun}}\left(\frac{t}{13.7~\rm Gyr}\right)^{\rm -s}~(M_{\sun}~{\rm yr^{-1}}),
\label{eq:mdots_snia}
\end{equation}
which leads to $\dot{M}_{\rm SN}(13.7~{\rm Gyr})$ almost 100 times smaller than the ``quiescent" stellar mass loss $\dot{M}_{\star}(13.7~{\rm Gyr})$ in Equation~(\ref{eq:mdots_agb}).
Assuming each type Ia SN event releases an energy of $E_{\rm SN}=10^{51}~{\rm erg}$ \citep{Iwamoto1999,Thielemann2004}, and a fraction of $\eta_{\rm SN}=0.85$ is thermalized in the surrounding ISM\footnote{Based on numerical simulations, a fraction of $\sim90\%$ of the injected energy from SN can be thermalized into the tenuous hot ISM of ETGs \citep{Cho2008,Agertz2013}.}, the energy input rate over the whole galaxy is
\begin{equation}
L_{\rm SN} (t)=5.0\times10^{30}\eta_{\rm SN}\frac{L_{\rm B}}{L_{\rm B\sun}}\left(\frac{t}{13.7~\rm Gyr}\right)^{\rm -s} \,\, ({\rm erg~s}^{-1}) .
\label{eq:lsn_ia}
\end{equation}

\subsection{Star Formation and SN II Heating}
The star formation rate per unit volume at each radius $r$ is estimated from the equation
\begin{equation}
  \dot{\rho}_\mathrm{SF} = \frac{\eta_{\rm SF} \rho }{\tau_{\rm SF}} \, ,
  \label{eq:rhodot_sf}
\end{equation}
where $\rho$ is the local gas density, $\eta_{\rm SF}$ is the star formation efficiency, which is in the range of $0.02-0.4$ (e.g., \citealt{Elmegreen1997,Weinberg2002,Cen2006,Stinson2006,Shi2011}, here a value of 0.1 is adopted),
and
\begin{equation}
  \tau_{\rm SF} = \max(\tau_{\rm cool}, \tau_{\rm dyn}),\quad
  \label{eq:tau_sf}
\end{equation}
where
\begin{equation}
  \tau_{\rm cool} = {E\over C}, \quad
  \mathbf{\tau_{\rm dyn} = \min(\tau_{\rm Jeans}, \tau_{\rm rot})},\quad
  \label{eq:tau_sf1}
\end{equation}
and
\begin{equation}
  \tau_{\rm Jeans} = \sqrt{\frac{3\pi}{32 G \rho}}, \quad   \tau_{\rm rot} = \frac{2\pi r}{v_{\rm c}(r)}.
  \label{eq:tau_sf2}
\end{equation}
Here $E$ and $C$ are the gas internal energy and the effective cooling per unit volume, $G$ is the Newtonian gravitational constant, $r$ is the distance from the galaxy center, $\tau_{\rm rot} $ is an estimate of the radial epicyclic period, and $v_{\rm c}(r)$ is the galaxy circular velocity in the equatorial plane. Then the star formation rate (SFR) is simply expressed as $\mathrm{SFR}=\int\dot{\rho}_\mathrm{SF}dV$. We do not consider the migration of newly formed stars, which means that the stars stay at the place where they initially form. It should be noted that we do not impose temperature and/or Jeans' mass limiters for star formation algorithm, which may lead to an overestimate of SFR in the high temperature and low density regions (e.g., the outer region of our simulated galaxies). This will be studied in the future in more detail.

Star formation removes mass, momentum and energy from each grid, but also injects new mass and energy by type II SN with a delay time of $2\times10^{7}~{\rm yr}$. Assuming the newly formed star follows a Salpeter IMF, the mass return from SN II progenitors is $20\%$ of the newly formed stellar mass. By assuming each SN II typically injects an energy of $E_{\rm SN}$ \citep{Woosley1995}, the SN II energy injection efficiency (the injected energy divided by the rest mass energy of the ejector from type II SN) is $3.9\times10^{-6}\eta_{\rm SN}=3.3\times10^{-6}$.

\subsection{Hydrodynamics}

The evolution of the galaxy is governed by the following time--dependent Eulerian equations of hydrodynamics (see \citeauthor{Ciotti2012} \citeyear{Ciotti2012} for a full description) when AGN feedback is not incorporated:

\begin{equation} \label{eq:massconsvr}
   \frac{\partial \rho}{\partial t} + \nabla\cdot(\rho{\bf v})
        = \alpha\rho_{\star} + \dot{\rho}_{\rm II} - \dot{\rho}_{\star}^{+},
\end{equation}
\begin{equation} \label{eq:momconsvr}
   \frac{\partial {\bf m}}{\partial t} + \nabla\cdot({\bf m v})
        = - \nabla p_{\rm gas} + \rho {\bf g} -\dot{\bf m}^{+}_{\star},
\end{equation}
\begin{equation} \label{eq:engconsvr}
   \frac{\partial E}{\partial t} + \nabla\cdot(E{\bf v})
        =  -p_{\rm gas} \nabla \cdot {\bf v} - C  + \dot{E}_{\rm S} +\dot{E}_{\rm I}+\dot{E}_{\rm II} -\dot{E}^{+}_{\star},
\end{equation}
\noindent where $\rho$, ${\bf m}$ and $E$ are the gas mass, momentum and internal energy per unit volume, respectively, and ${\bf v}$ is the velocity, $p_{\rm gas}=(\gamma-1) E$ is the gas pressure, and we adopt an adiabatic index $\gamma=5/3$. Here $C$ is the net rate of radiative cooling, ${\bf g}$ is  the gravitational field of the galaxy (i.e., stars, dark matter, plus the central SMBH). For simplicity, we do not take into account effects of the self-gravity of the ISM or the gravitational effects of the mass redistribution due to the stellar mass losses and star formation\footnote{The gravitational effect due to such stellar mass losses and star formation is minor because the stellar mass variation resulting from them is $\lesssim10\%$.}.
Here we just recall some main points in the previous works (\citealt{Ciotti2012}; see also \citealt{Gan2014,Yuan2018}).
In the energy equation above, we include AGB stellar wind thermalization $\dot{E}_{\rm S}$, and supernovae heating $\dot{E}_{\rm I}$ and $\dot{E}_{\rm II}$.
This just corresponds to the SNe feedback cases (with a suffix .agb+sn for the model names in Table~\ref{tab:parameters}). While in the AGB heating case (with a suffix .agb for the model names), we only consider late-type stellar winds in the mass equation and their thermalization in energy equation, i.e., no $\dot{E}_{\rm I}$ and $\dot{E}_{\rm II}$ in the energy equation, no $\dot{\rho}_{\rm II}$ term in the mass equation. $\alpha=\alpha_{\star}+\alpha_{\rm SN}=\dot{M}_{\star}/M_{\star}+\dot{M}_{\rm SN}/M_{\star}$ is calculated for every simulated grid for SNe feedback models, and only the first term is included for AGB heating models. $\dot{\rho}_{\star}^{+}$, $\dot{\bf m}^{+}_{\star}$, and $\dot{E}^{+}_{\star}$ are the mass, momentum, and energy sink terms associated with star formation, respectively.

When AGN feedback is considered in this work (with a suffix .agb+sn+agn for the model names), we follow the most updated AGN physics as described in \citet{Yuan2018}. Specifically, the initial black hole mass is determined according to the updated $M_{\rm BH}-M_{\star}$ relation, which is given by (\citealt{Kormendy2013})
\begin{equation}\label{eq:mbh}
  M_{\rm BH}=4.9\times10^{8}M_{\odot}\left(\frac{M_{\star}}{10^{11}M_{\odot}}\right)^{1.17}.
\end{equation}
Note that $M_{\rm bulge}=M_{\star}$ for elliptical galaxies. For a typical stellar mass $M_{\star}=3.0\times10^{11}M_{\odot}$, the black hole mass $M_{\rm BH}=1.8\times10^{9}M_{\odot}$.  The black hole accretion rate $\dot{M}_{\rm Bondi}$ is calculated from the inflow rate at the innermost grid of the simulation domain. Depending on whether $\dot{M}_{\rm Bondi}$ is larger than $2\%\dot{M}_{\rm Edd}$ ($\dot{M}_{\rm Edd}$ is the Eddington accretion rate) or not, the accretion flow will stay in the
cold or hot mode. The corresponding wind and radiation outputs in each mode will then be determined.
The radiative heating/cooling processes and the radiation force associated with AGN emission are incorporated accordingly. See the Appendix for a further description of the AGN feedback model.

We perform 2D hydrodynamic simulations using \emph{ZEUS}-MP/2 \citep{Hayes2006} code in spherical coordinates ($r$, $\theta$, $\phi$) with an axisymmetric configuration. Following \citet{Novak2011} and \citet{Gan2014}, the mesh in the $\theta$ direction is divided uniformly into 30 cells, while a logarithmic mesh in the radial direction with 120 bins is used in the range of 2.5 pc--250 kpc. The simulated galaxies are setup at an age of 2 Gyrs, and evolved in isolation for 12 Gyrs.

We use the standard ``outflow boundary condition" in the \emph{ZEUS} code for the inner/outer radial boundary \citep{Stone1992}. This allows the gas to flow out and flow in from the boundary depending on the state of the gas. For the $\theta$ direction, a reflecting boundary condition at each pole is assumed.

\begin{table*}[htb]
  \begin{center}
  \caption{\bf Galaxy Parameters and Some Simulation Results for Different Feedback Models} \label{tab:parameters}
  \resizebox{0.95\textwidth}{!}{
  \begin{tabular}{lcccccccr}
    \hline\hline
    Model & $\sigma_{\circ}$ & $r_{\rm eff}$ & $M_{\star}$  & $M_{\star}/L_{\rm B}$ & $n_{0}$ & $M_{\rm gas}^{\dag}$ & $M_{\rm wind}^{\dag}$ & Notes \\
    & $(\rm km~s^{-1})$ & $(\rm kpc)$ & $(M_{\sun})$ & $(M_{\sun}/L_{B\sun})$ & $({\rm cm^{-3}})$ & $(M_{\sun})$ & $(M_{\sun})$ & \\
     (1) & (2) & (3) & (4)  & (5) & (6) & (7) & (8) & (9) \\

    \hline
    E100.agb  & $100$  & $1.1$   & $6.9\times10^{9}$ & $3.1$  & $-$ & $1.3\times10^7$ & $1.0\times10^9$ & only AGB heating${^a}$   \\
     \\
     E100.agb+sn  & $100$  & $1.1$   & $6.9\times10^{9}$ & $3.1$  & $-$ & $1.8\times10^5$ & $1.7\times10^8$ & stellar feedback${^b}$    \\
     \\
     E190.agb  & $190$  & $3.6$   & $8.0\times10^{10}$ & $4.8$  & $-$ & $2.0\times10^8$ & $1.4\times10^9$ & only AGB heating    \\
     \\
     E190.agb+sn  & $190$  & $3.6$   & $8.0\times10^{10}$ & $4.8$  & $-$ & $1.0\times10^7$ & $1.8\times10^8$ & stellar feedback    \\
     \\
     E220.agb  & $220$  & $4.9$   & $1.5\times10^{11}$ & $5.3$  & $-$ & $4.3\times10^8$ & $1.8\times10^9$ & only AGB heating    \\ %
     \\
     E220.agb+sn  & $220$  & $4.9$   & $1.5\times10^{11}$ & $5.3$  & $-$ & $2.9\times10^7$ & $2.1\times10^8$ & stellar feedback    \\
     \\
     E260.agb & $260$  & $6.9$   & $3.0\times10^{11}$ & $5.8$  & $-$ & $9.8\times10^8$ & $1.1\times10^9$ & only AGB heating    \\
     \\
     E260.agb+sn & $260$  & $6.9$   & $3.0\times10^{11}$ & $5.8$  & $-$ & $3.5\times10^9$ & $6.9\times10^9$ & stellar feedback     \\ %
     \\
     E260.agb+sn+agn  & $260$  & $6.9$   & $3.0\times10^{11}$ & $5.8$  &  $-$ & $4.4\times10^8$ & $9.4\times10^9$ & AGN feedback ${^c}$  \\
     \\
     E340.agb  & $340$  & $13.5$   & $9.8\times10^{11}$ & $7.2$  & $-$ & $6.9\times10^9$ & $1.9\times10^9$ & only AGB heating    \\
     \\
     E340.agb+sn  & $340$  & $13.5$   & $9.8\times10^{11}$ & $7.2$  & $-$ & $1.6\times10^{10}$ & $7.7\times10^9$ & stellar feedback    \\
     \\
     E340.agb+sn+agn  & $340$  & $13.5$   & $9.8\times10^{11}$ & $7.2$  &  $-$ & $2.1\times10^{10}$ & $6.0\times10^9$ & AGN feedback    \\
     ...  & ...  & ...   & ... & ...  &  ... & ... & ... & ...    \\
     \hline \\
     E190.agb+low gas  & $190$  & $3.6$   & $8.0\times10^{10}$ & $4.8$  & $0.01$ & $4.4\times10^8$ & $2.7\times10^9$ & only AGB heating \\
     &&&&&&&&+ initial low $n_{0}$   \\
     E190.agb+sn+low gas  & $190$  & $3.6$   & $8.0\times10^{10}$ & $4.8$  & $0.01$ & $1.0\times10^7$ & $1.8\times10^8$ & stellar feedback \\
     &&&&&&&&+ initial low $n_{0}$    \\
     E190.agb+high gas  & $190$  & $3.6$   & $8.0\times10^{10}$ & $4.8$  & $0.16$ & $4.3\times10^8$ & $1.3\times10^9$ & only AGB heating \\
     &&&&&&&&+ initial high $n_{0}$   \\
     E190.agb+sn+high gas  & $190$  & $3.6$   & $8.0\times10^{10}$ & $4.8$  & $0.16$ & $1.3\times10^9$ & $7.0\times10^9$ & stellar feedback \\
     &&&&&&&&+ initial high $n_{0}$    \\
     E220.agb+low gas  & $220$  & $4.9$   & $1.5\times10^{11}$ & $5.3$  & $0.01$ & $1.1\times10^9$ & $3.3\times10^9$ & only AGB heating \\
     &&&&&&&&+ initial low $n_{0}$    \\
     E220.agb+sn+low gas  & $220$  & $4.9$   & $1.5\times10^{11}$ & $5.3$  & $0.01$ &  $6.3\times10^8$ & $9.1\times10^9$ & stellar feedback \\
     &&&&&&&&+ initial low $n_{0}$    \\
     E220.agb+high gas  & $220$  & $4.9$   & $1.5\times10^{11}$ & $5.3$  & $0.10$ & $5.6\times10^8$ & $1.3\times10^9$ & only AGB heating \\
     &&&&&&&&+ initial high $n_{0}$    \\
     E220.agb+sn+high gas  & $220$  & $4.9$   & $1.5\times10^{11}$ & $5.3$  & $0.10$ &  $2.1\times10^9$ & $6.5\times10^9$ & stellar feedback \\
     &&&&&&&&+ initial high $n_{0}$    \\
     E260.agb+sn+gas & $260$  & $6.9$   & $3.0\times10^{11}$ & $5.8$  & $0.05$ & $3.9\times10^9$ & $6.3\times10^9$ & stellar feedback \\
     &&&&&&&&+ initial $n_{0}$    \\
     E260.agb+sn+agn+gas  & $260$  & $6.9$   & $3.0\times10^{11}$ & $5.8$  &  $0.05$ & $9.4\times10^8$ & $5.1\times10^9$ & AGN feedback \\
     &&&&&&&& initial $n_{0}$ \\
     E340.agb+sn+gas  & $340$  & $13.5$   & $9.8\times10^{11}$ & $7.2$  & $0.04$ & $2.0\times10^{10}$ & $6.6\times10^9$ & stellar feedback    \\
     &&&&&&&&+ initial $n_{0}$    \\
     E340.agb+sn+agn+gas  & $340$  & $13.5$   & $9.8\times10^{11}$ & $7.2$  & $0.04$ & $1.7\times10^{10}$ & $6.0\times10^9$ & AGN feedback \\
     &&&&&&&&+ initial $n_{0}$    \\
    \hline\hline
  \end{tabular}
  }
 \end{center}
 \tablecomments{$^{\dag}$ $M_{\rm gas}$ is the total ISM mass remaining within $10~r_{\rm eff}$ of the galaxies in the end of the simulations, and $M_{\rm wind}$ is the ISM mass driven beyond $10~r_{\rm eff}$ in the galaxies.
 $(a)$. only winds from evolved stars and their thermalization included (model name with a suffix .agb).
 $(b)$. stellar feedback includes AGB heating, SN Ia and SN II feedback (model name with a suffix .agb+sn).
 $(c)$. including AGN feedback in addition to AGB heating and SNe feedback (model name with a suffix .agb+sn+agn).
 }
\end{table*}

\section{Results}\label{sec:results}
The relevant galaxy parameters and descriptions about different feedback models are shown in Table~\ref{tab:parameters}.

\subsection{Energy Balance and Star Formation}

We will study the energy balance and star formation activities for both low and high mass galaxy models in the following two subsections and then extend to the study for other different galaxy models.

\subsubsection{The Low Mass Galaxy Model (E220)}

We first consider a galaxy with $M_{\star}=1.5\times10^{11}~M_{\odot}$. In this case, we only study two feedback models, namely AGB heating model (only AGB heating) and SNe feedback model (include AGB heating, type Ia SNe and type II SNe), which correspond to model names of E220.agb and E220.agb+sn, respectively.

\begin{figure*}[htb]
\centering
\includegraphics[height=0.5\textwidth,width=0.4\textwidth]{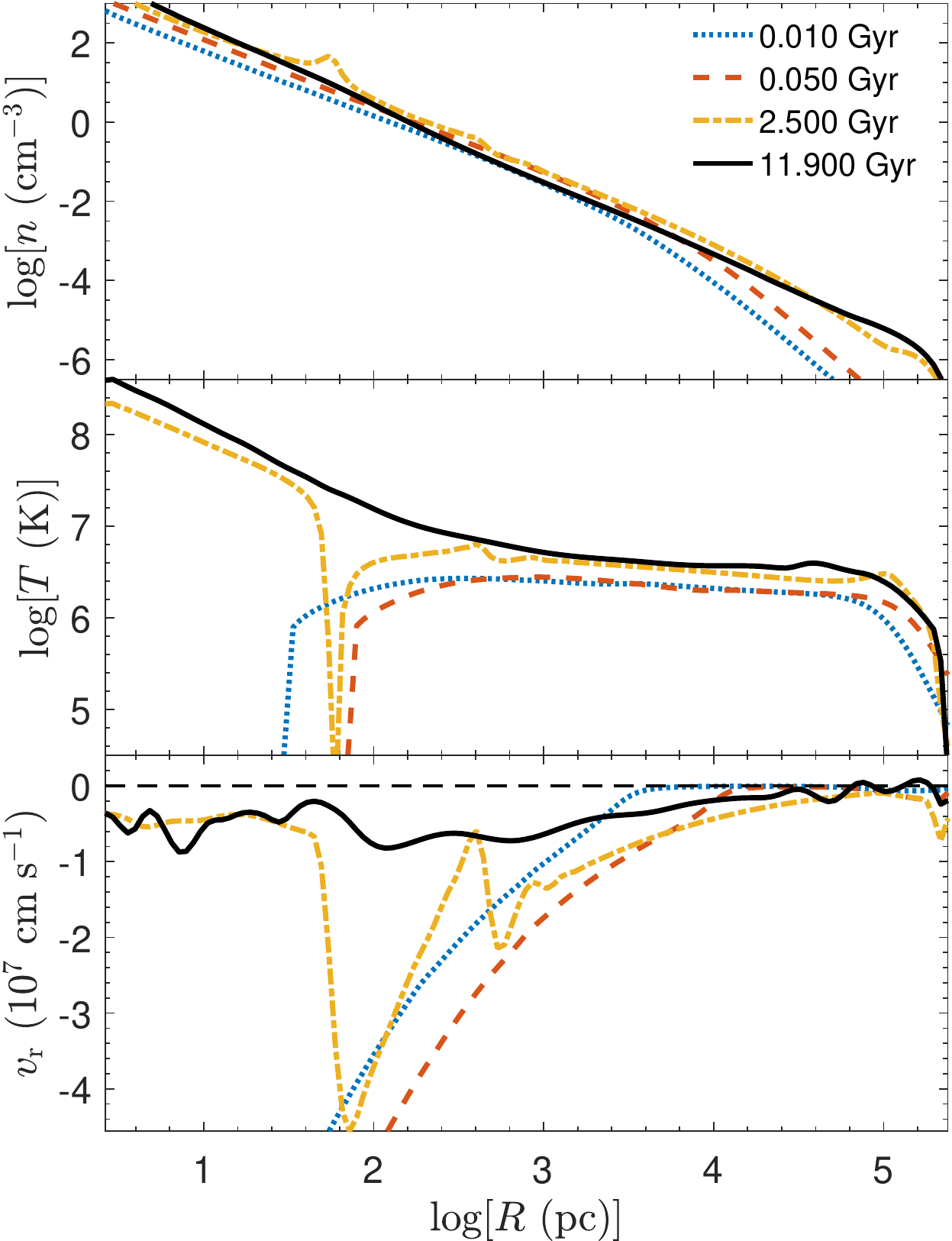}
\hskip 0.5truecm
\includegraphics[height=0.5\textwidth,width=0.4\textwidth]{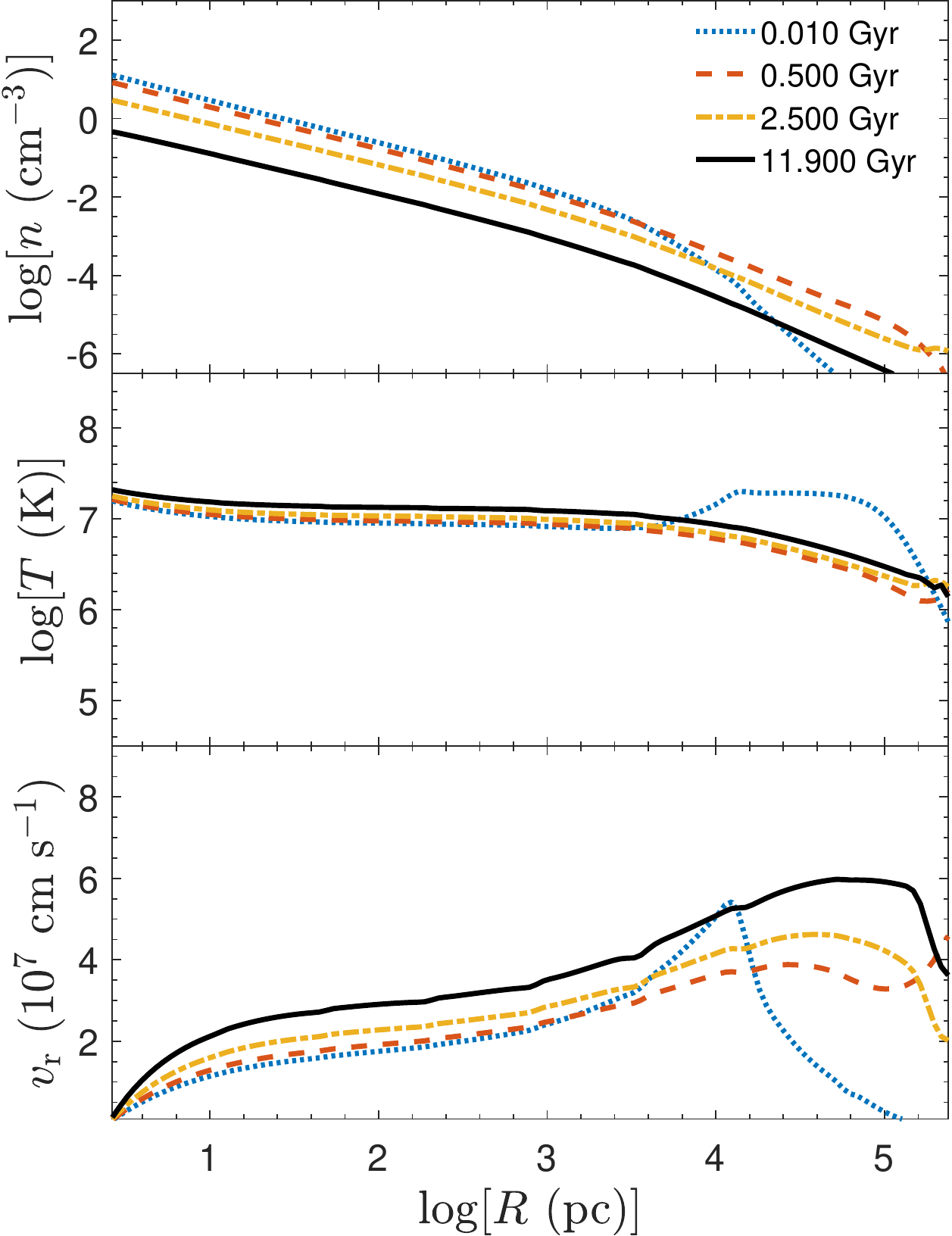}
\caption{Radial profiles of density, temperature, and velocity of the gas around the simulated galaxy E220 at four representative times. The left (right) panel corresponds to the model E220.agb (E220.agb+sn). The gas density is averaged over $\theta$ direction, while the temperature and velocity values at the equatorial plane are adopted. The four time periods selected can cover from almost close to the initial stage (0.01 Gyr) to the end of the simulation (11.9 Gyr). }\label{fig:ism_c12}
\end{figure*}

We first study how the hot gaseous halo is built up under the initial conditions and the input physics for the AGB heating model. We show the radial profiles of the gas density, temperature, and radial velocity for the model E220.agb at four representative times in the left panel of Figure~\ref{fig:ism_c12}. The gas is apparently inflowing during almost the whole evolution period. The density in the inner region of the galaxy continuously becomes larger because of the accumulation of the inflowing gas. The gas temperature is initially very low and the inflowing velocity is very high due to the gravitational acceleration. After an interval of about 3 Gyrs, the gas temperature jump around 100 pc disappears and the temperature in the inner region can reach a very high value, while the inflow velocity in the inner region decreases to $\sim100~{\rm km~s^{-1}}$. This arises from the heating provided by the compression work of the inflowing gas. In this stage, it is mainly the thermal pressure gradient that balances the gravity force, which is dominated by the rapidly growing SMBH in the inner region because of a large inflowing gas. We should note that AGB heating cannot be responsible for such a high gas temperature, as we will see below that AGB heating cannot effectively balance cooling in the entire galaxy.

\begin{figure}[htb]
\centering
\includegraphics[height=0.6\textwidth,width=0.33\textwidth]{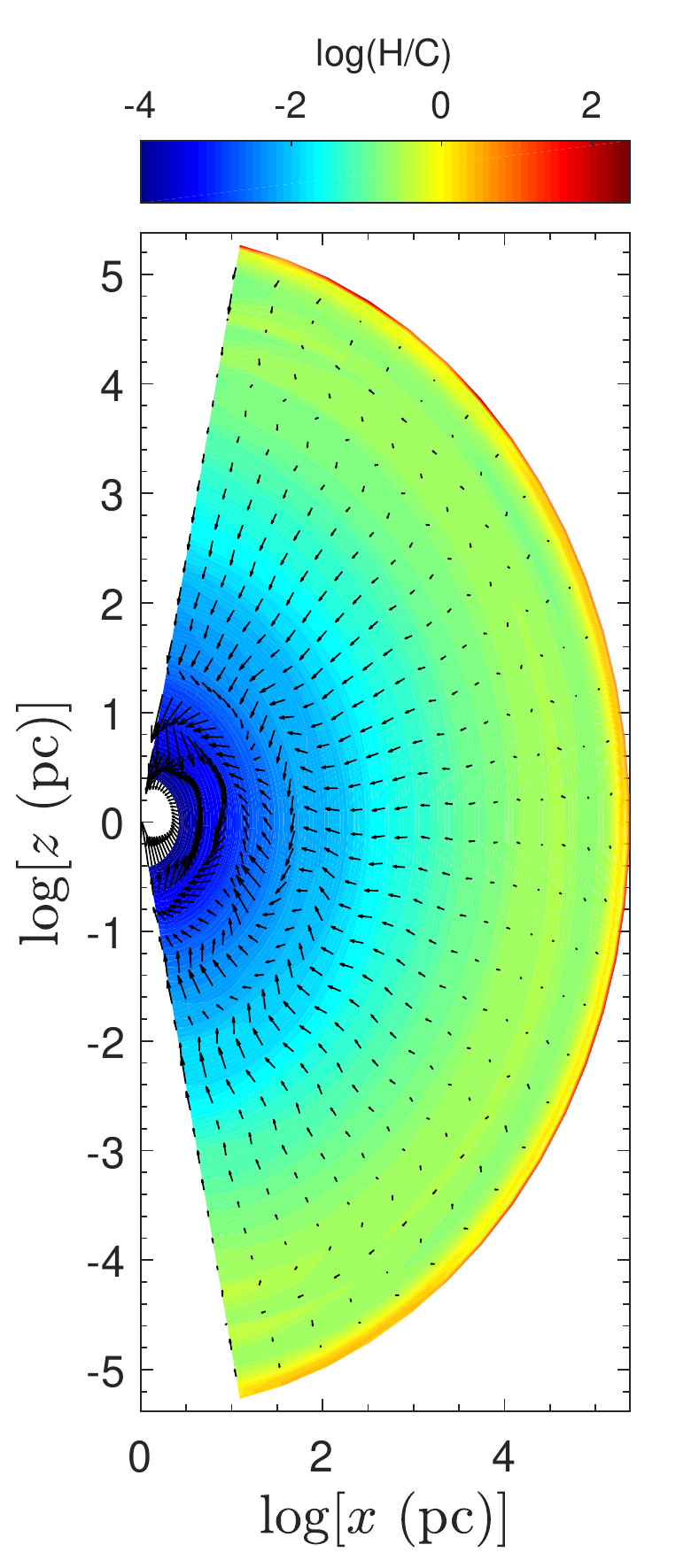}
\caption{Heating over cooling rate ratio for model E220.agb at $t= 11.9$ Gyr, which is close to the end of the simulation. The heating source is contributed by the thermalization of the stellar wind. The cooling sources include the bremsstrahlung loss, line and recombination continuum cooling processes.  Arrows are normalized to the velocity field at the $\phi=0$ plane.}\label{fig:eb_c12_agb_snapshot}
\end{figure}

We then show the results of the heating and cooling rates for the AGB heating model in Figure~\ref{fig:eb_c12_agb_snapshot}.
Since heating and cooling rates fluctuate in different spatial regions, the local heating (cooling) rate is more meaningful than the integrated global term in determining whether the AGB heating is important or not for balancing cooling losses.
The heating sources ($H$) include the thermalization from winds of low-mass dying stars, while the cooling terms ($C$) are contributed by bremsstrahlung losses, line and recombination continuum cooling. Note that the heating contributed by the compression work and wind shocks is not included in the calculation of $H$, although these terms can influence the gas temperature and density as we have discussed above.
We calculate the ratio $H/C$ of every simulated grid in the whole galaxy.
One snapshot of the heating over cooling ratio is shown in Figure~\ref{fig:eb_c12_agb_snapshot}, the time of which approaches the end of the simulation. The arrows show the velocity field at $\phi=0$ plane. The red color corresponds to a heating-dominated region while blue refers to a cooling region. As we can see, most regions of the simulated galaxy are cooling dominated in our selected period. From the velocity field overlaid in the plots, there exists remarkable inflow towards the galactic center due to the strong cooling (see also the left panel of Figure~\ref{fig:ism_c12}). In the absence of an effective heating source, the gravitation potential can accelerate the gas to a large velocity, which is basically due to the singular isothermal potential we adopt. Therefore, a large temperature in the inner region shown above is mainly related to the unbalanced cooling behaviour.

\begin{figure*}[htb]
\centering
\includegraphics[height=0.5\textwidth,width=0.45\textwidth]{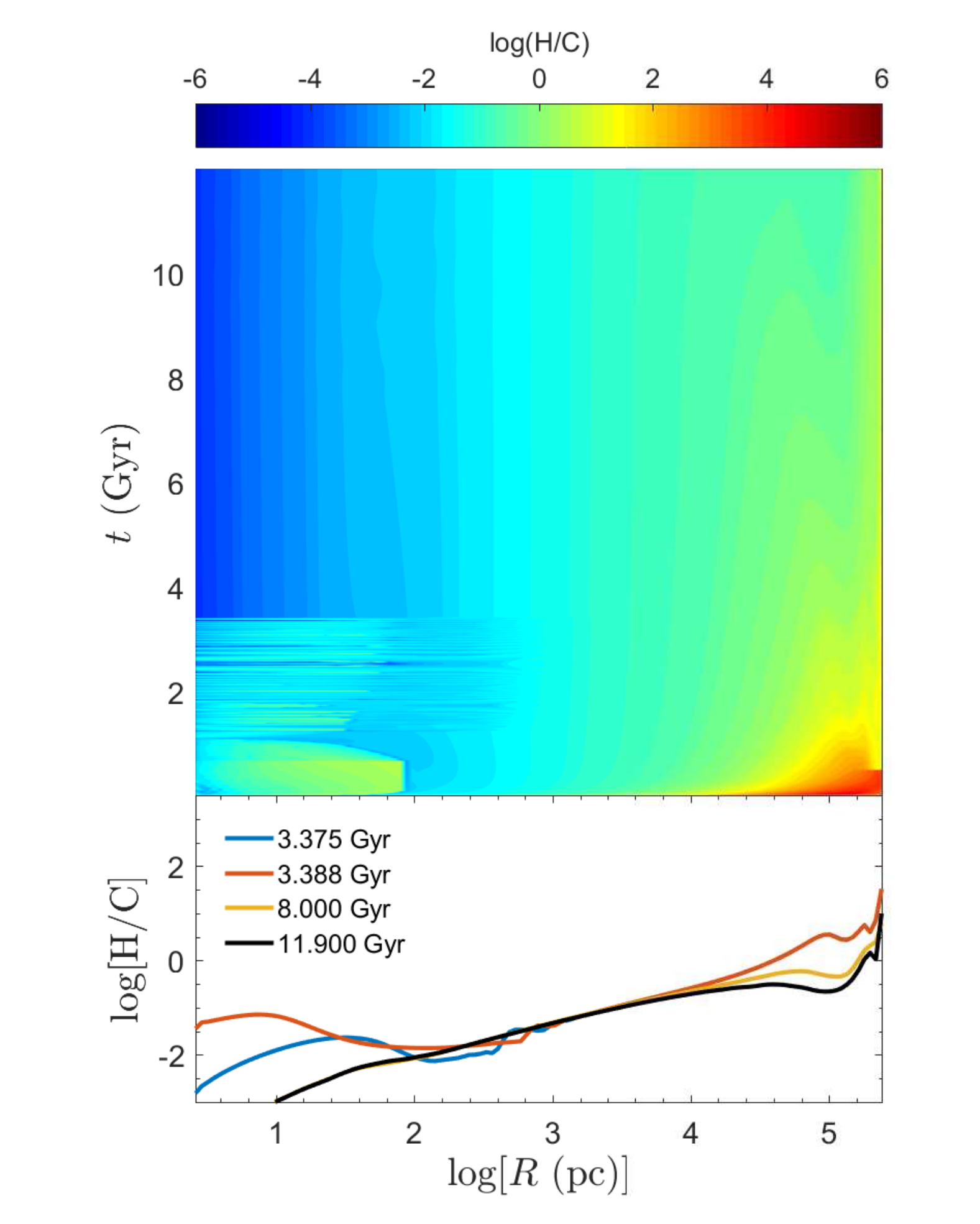}
\includegraphics[height=0.5\textwidth,width=0.45\textwidth]{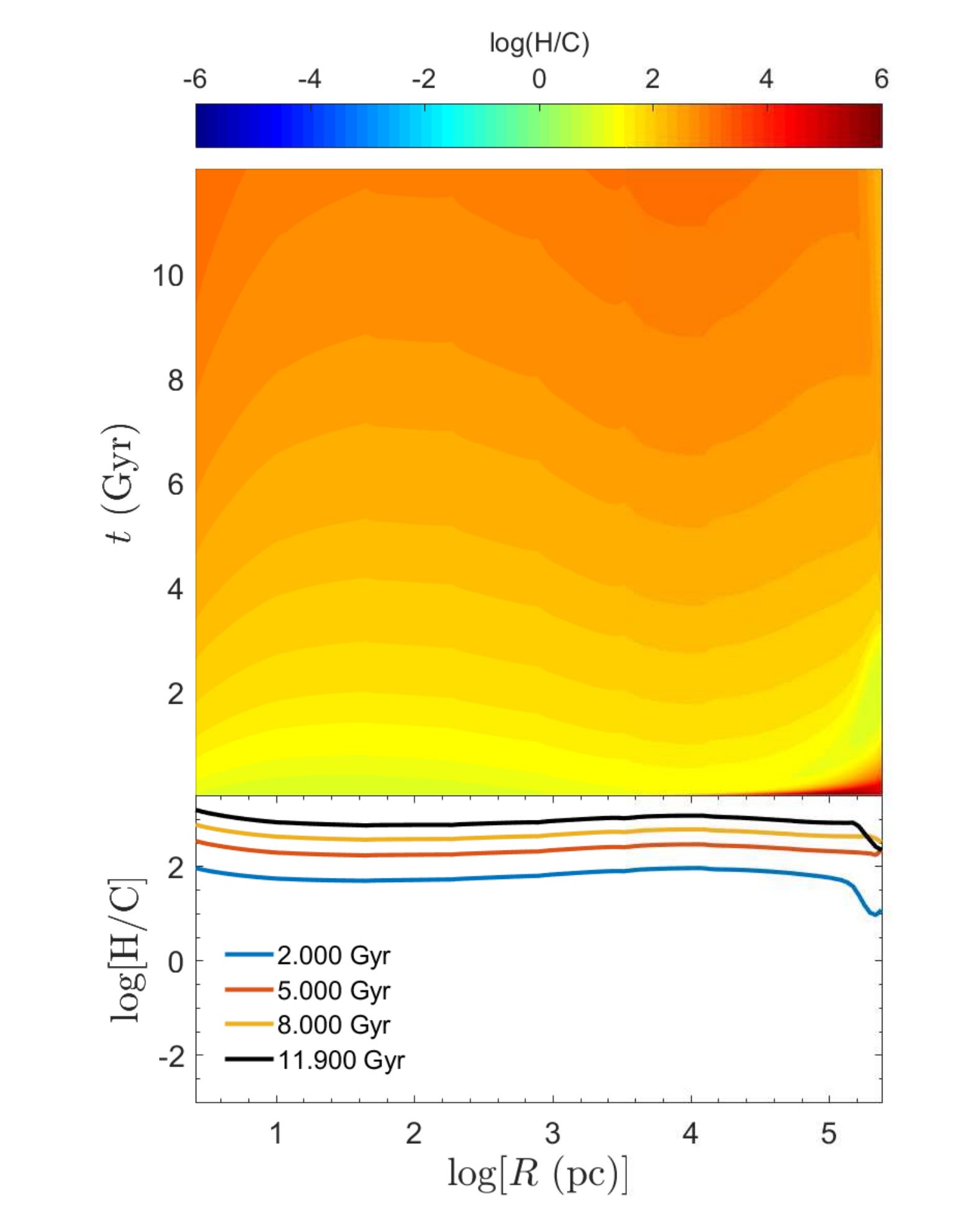}
\includegraphics[height=0.3\textwidth,width=0.36\textwidth]{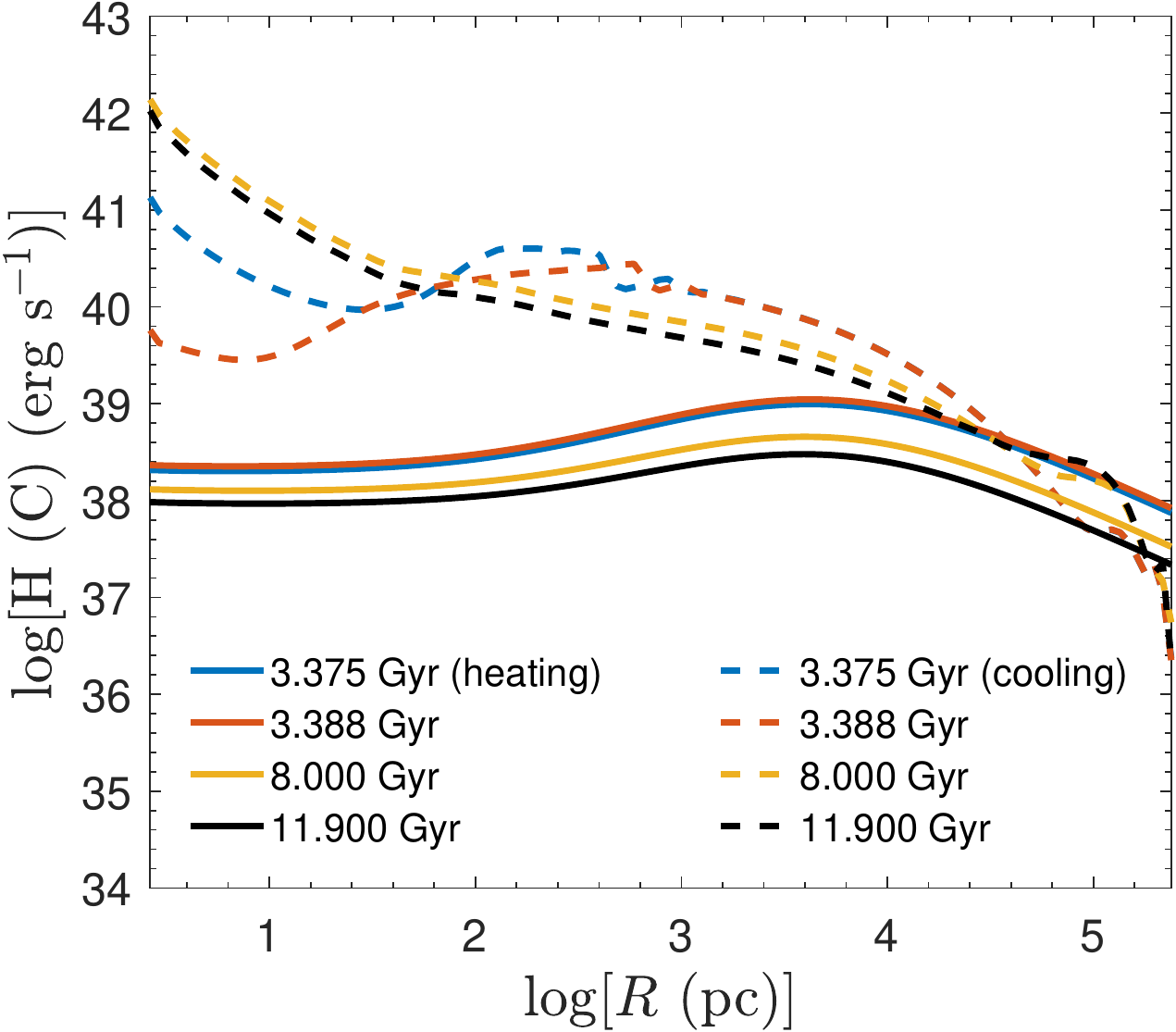}
\hskip 1.6truecm
\includegraphics[height=0.3\textwidth,width=0.36\textwidth]{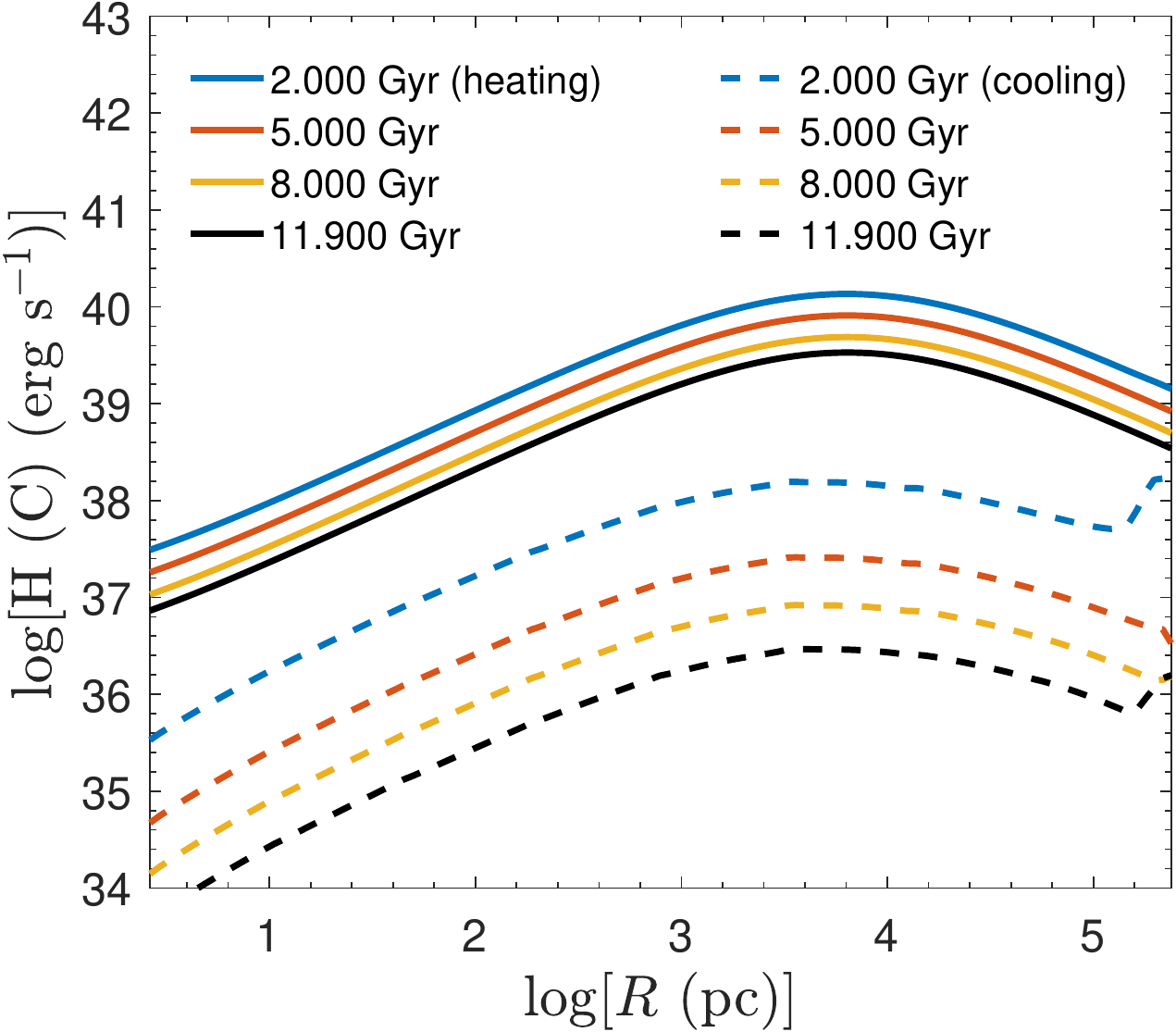}
\caption{Radial profiles of heating over cooling rate ratios and their time evolutions for models E220.agb (left panel), E220.agb+sn (right panel). The middle panels specifically show the ratio at four epoches in the simulations. In the bottom panels, we also show the radial profiles of the absolute values of the heating (solid lines) and cooling (dashed lines) rates for two models. Different colors correspond to different times.}\label{fig:eb_c12_r_t}
\end{figure*}

After integrating the heating and cooling rates over the polar angel $\theta$,  we show the time evolution of the spatial map of the $H/C$ ratio in the left panel of Figure~\ref{fig:eb_c12_r_t}. When we calculate the heating and cooling rates in each radial ring, this is obtained by $\dot{E}_{\rm r}(r)=\int2\pi\,r^2\sin\theta\delta r\dot{E}d\theta$, where $\dot{E}$ could be heating ($H$) and cooling ($C$) rates for each term.
The same method is applied to the calculation of the radial distribution of the specific SFR ($\mathrm{sSFR}$).

The left panel of Figure~\ref{fig:eb_c12_r_t} corresponds to the case with only AGB heating included. The ratio $H/C<1$ is satisfied in almost all radial bins during the whole evolution stage. To specifically demonstrate this, we present the $H/C$ ratio in our selected periods in the middle panel. The absolute values of $H$ and $C$ are shown in the bottom panels.  It is clearly shown that $C$ is larger than $H$ even in the whole region of the galaxy, especially in the inner region of the galaxy, where cooling is remarkably strong.  Therefore, AGB heating alone cannot balance the cooling in the entire galaxy for the galaxy with $M_{\star}=1.5\times10^{11}~M_{\odot}$.

When type Ia and II SNe heating are incorporated, the evolution of the gas properties are shown in the right panel of Figure~\ref{fig:ism_c12}. In comparison to the AGB heating model, the gas is persistently blown out and escapes from the galaxy, as shown from the density and velocity plot. The SNe driven wind pushes the gas out of the galaxy, and also heats the gas effectively as the time evolves.

\begin{figure*}[htb]
\centering
\includegraphics[height=0.6\textwidth,width=0.33\textwidth]{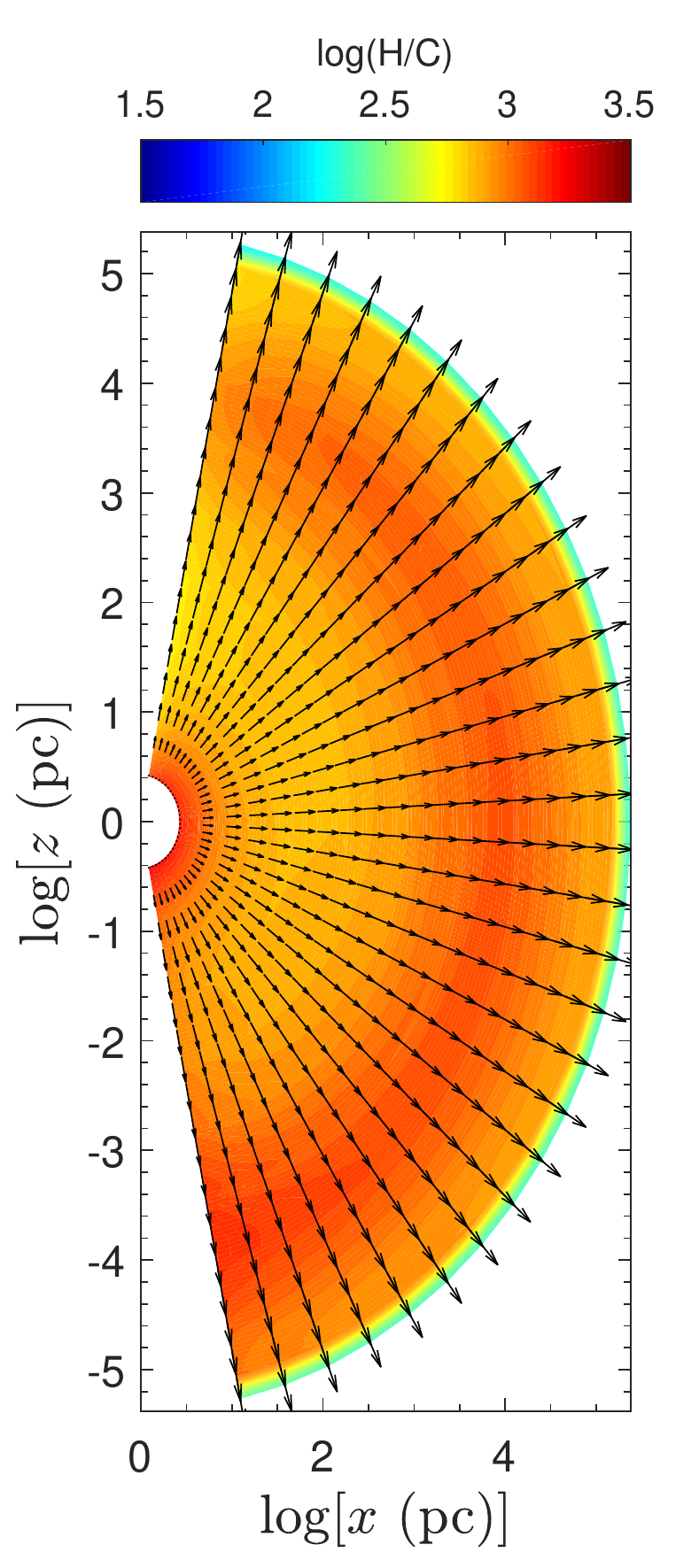}
\includegraphics[height=0.6\textwidth,width=0.33\textwidth]{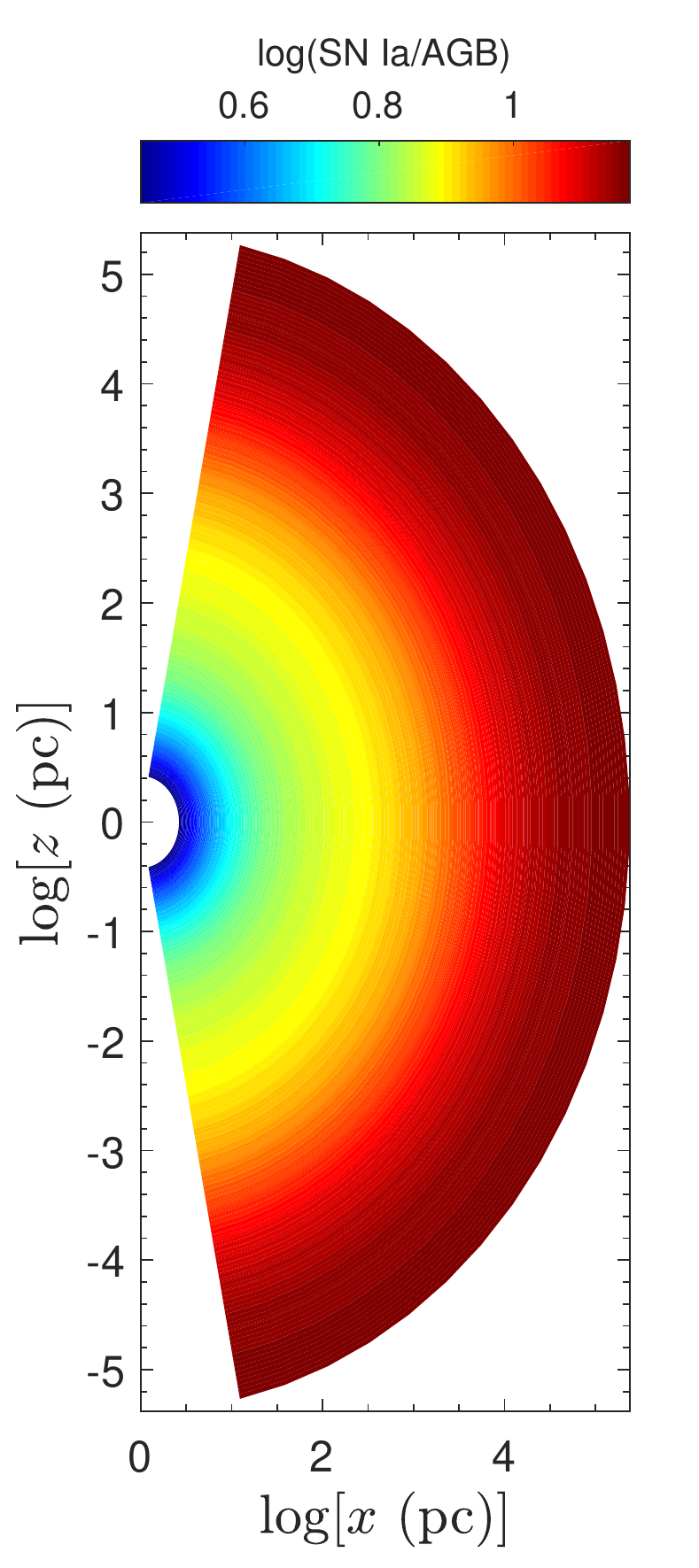}
\caption{Left panel: heating over cooling rate ratios for model E220.agb+sn near the end of the simulation $t=11.9$ Gyr. The heating sources include AGB, SN Ia and SN II heating. Right panel: the ratio of type Ia SN and AGB heating at $t=11.9$ Gyr.  Arrows in the left panel are normalized to the velocity field at the $\phi=0$ plane.}\label{fig:eb_c12_snapshot}
\end{figure*}

For the heating and cooling rate, we choose one snapshot, which approaches the end of the simulation, to show the 2D $H/C$ ratio in the whole galaxy. The results are presented in the left panel of Figure~\ref{fig:eb_c12_snapshot}.
In this case, the heating terms are contributed by AGB, SN Ia and SN II heating, while SN II is found to be unimportant compared to SN Ia because of the low level of star formation activities as we will show below.

As apposed to the model with only AGB heating included, it clearly shows that the total heating rate is much higher than the cooling term in every spatial region. This results in massive outflows as indicated by the velocity field overlaid in the left plot of Figure~\ref{fig:eb_c12_snapshot} (see also the right panel of Figure~\ref{fig:ism_c12}). We further calculate that the ratio of SNe heating (mainly contributed by type Ia SN) and AGB heating to quantify the role of SNe and AGB heating in making the heating effective. The 2D map of SN Ia/AGB heating ratio is shown in the right panel of Figure~\ref{fig:eb_c12_snapshot}. As expected, the heating rate of SN Ia is much larger than that of AGB throughout the whole galactic region.
It thus indicates that the SNe feedback is extremely effective in heating the ISM and driving galactic outflow \citep{Ciotti1991}, which will in turn suppress star formation presented below.

In a similar way, we show the time evolution of the $\theta-$integrated $H/C$ ratio in the right panel of Figure~\ref{fig:eb_c12_r_t}. The heating rate is significantly larger than the cooling term up to about two orders of magnitude at most times and in almost all spatial regions, consistent with the results shown in Figure~\ref{fig:eb_c12_snapshot}. From the bottom panel of Figure~\ref{fig:eb_c12_r_t}, we can see that the cooling rate is significantly suppressed due to the SNe driven wind, which causes the decrease of the gas density, as shown in the right panel of Figure~\ref{fig:ism_c12}.

\begin{figure*}[htb]
\includegraphics[height=0.5\textwidth,width=0.45\textwidth]{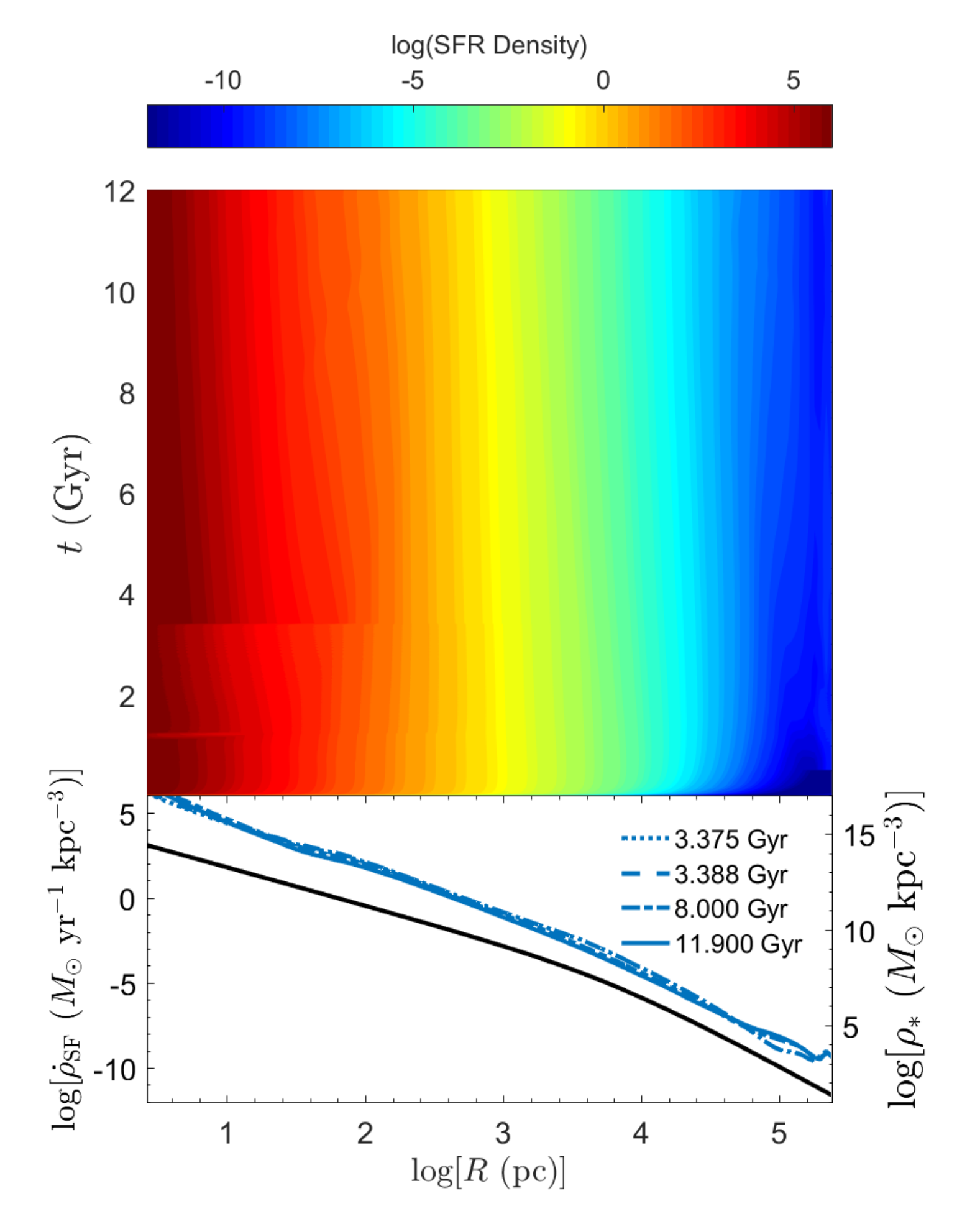}
\includegraphics[height=0.5\textwidth,width=0.45\textwidth]{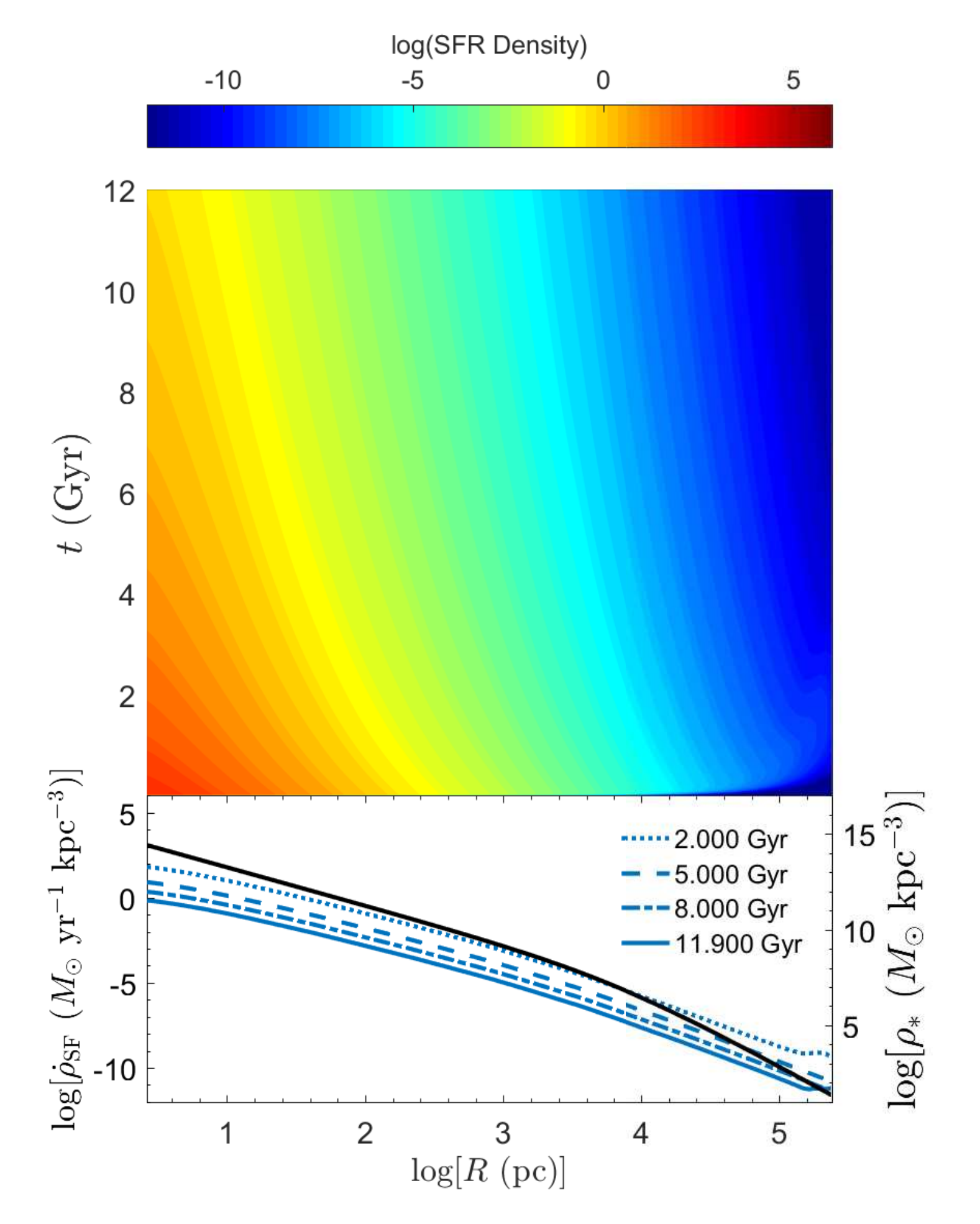}
\caption{Two-dimensional contour of SFR densities at each radial bin of the simulation grid for models E220.agb (left panel), E220.agb+sn (right panel). The bottom small plots show the radial profiles of SFR densities at four selected epoches. As a comparison, the stellar radial mass distributions are also shown as black solid lines with the labels in the right axes. }\label{fig:sfrd_c12_r_t}
\end{figure*}

Since the star formation activities depend on the amount of cold gas material available and the state of the gas, both are heavily influenced by the heating/colling balance state, a straightforward question is how these heating processes will influence the star formation activity.
In Figure~\ref{fig:sfrd_c12_r_t}, we show the 2D map of SFR density to demonstrate the time evolution of the star formation activities. The black solid line in the two panels of Figure~\ref{fig:sfrd_c12_r_t} shows the stellar density distribution in the corresponding radial bin, which is assumed to be constant throughout the simulation, for a comparison. We can see that there is a significant excess of star formation activity in the inner galactic region for the model of E220.agb (the left panel) because of the strong cooling-induced inflow. There is no apparent SFR suppression signature as time evolves (the lower plot in the left panel). When the SNe heating is invoked (the right panel), the SFR density decreases with time because of a gradual removal of  a large amount of heated gas from the galactic potential as we will show in Section~\ref{sec:ism}.

\begin{figure*}[htb]
\centering
\includegraphics[width=0.7\textwidth]{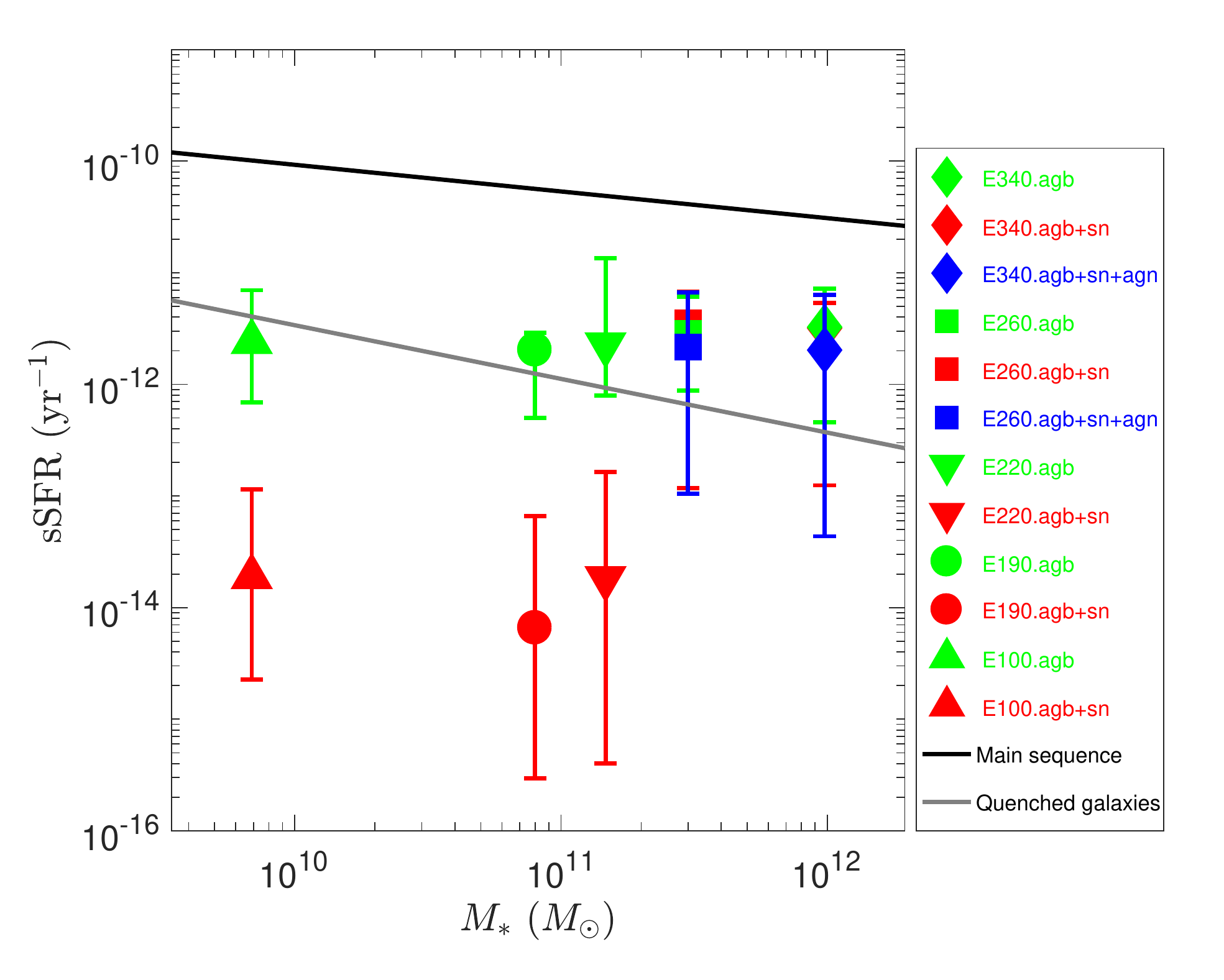}
\caption{sSFR as a function of stellar mass.  For each model, we show the mean values of the time evolution sequences, and the error bars represent their maximum and minimum values to indicate the variabilities. Different shapes of the symbols (i.e., diamonds (E340), squares (E260), downward triangles (E220), circles (E190), and upward triangles (E100)) represent galaxies with different stellar masses, and symbols with different colors correspond to different feedback models. The observed data for the main sequence of star-forming galaxies and quenched galaxies are shown as a comparison \citep{Renzini2015}. Note that the scatters of sSFR for these two populations are  $\sim0.5$ dex.
} \label{fig:ssfr_mstar}
\end{figure*}

As is shown with the downward filled triangles in Figure~\ref{fig:ssfr_mstar}, sSFR (defined as ${\rm sSFR}\equiv {\rm SFR}/M_{\star}$) in the entire galaxy with only AGB heating included (green symbol) is remarkably higher than that with SNe feedback (red symbol) due to the lack of effective heating sources to offset cooling. The error bars for each symbol demonstrate the maximum and minimum sSFR during the whole evolution. This further suggests that AGB heating cannot be effective in preventing star formation. Nevertheless, we find that the stellar feedback by SN Ia is sufficient to balance the radiative losses and hence reduce star formation significantly. The reduced sSFRs are then expected to be comparable to or lower than the observed values for quenched galaxies \citep{Renzini2015}.\footnote{Although the sSFR in the model E100.agb without SNe feedback can stay at a low value comparable to the observed one, it is attributed to the compression work. The inclusion of SNe feedback can make the sSFR much lower.}

We have found that the inclusion of AGN feedback does not change the results (i.e., energy balance and SFR) significantly with respect to model E220.agb+sn. This is because SN Ia feedback can effectively heat the ISM close to the virial temperature. In this case, the accretion rate of the black hole will be very low, thus the accretion cannot provide energetic outputs.

\subsubsection{The High Mass Galaxy Model (E260)}

The ISM mainly comes from the mass losses of evolved stars, and is then related to the stellar mass, therefore, the cooling, which is determined by the ISM density, depends strongly on the stellar mass (Equations~(\ref{eq:cooling},\ref{eq:mdots_agb})). It is obvious that both the AGB heating and SNe heating depend sensitively on the stellar mass (Equations~(\ref{eq:mdots_agb},\ref{eq:lsn_agb},\ref{eq:lsn_ia})). To explore how the results above could change with the stellar mass, we then move to a more massive galaxy with $M_{\star}=3.0\times10^{11}~M_{\odot}$.

\begin{figure*}[htb]
\centering
\includegraphics[height=0.6\textwidth,width=0.33\textwidth]{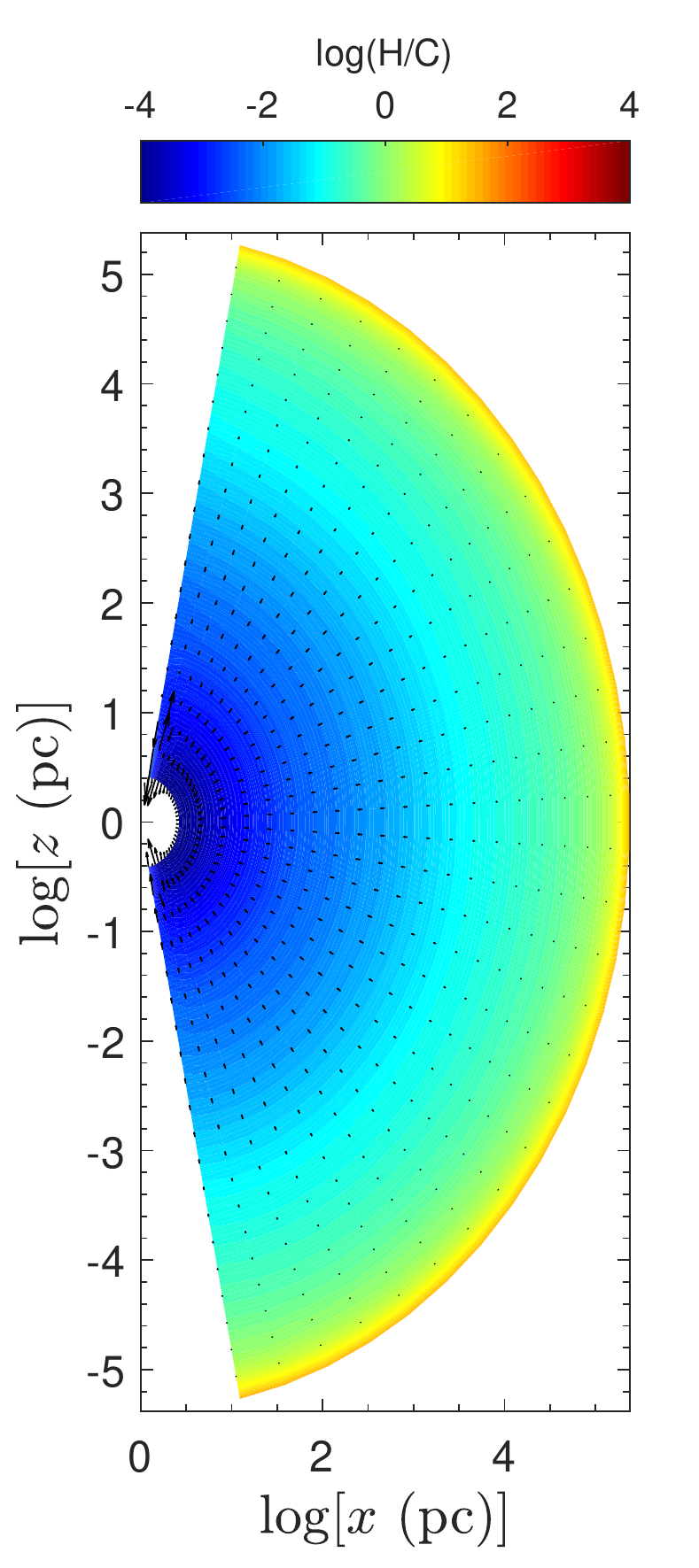}
\includegraphics[height=0.6\textwidth,width=0.33\textwidth]{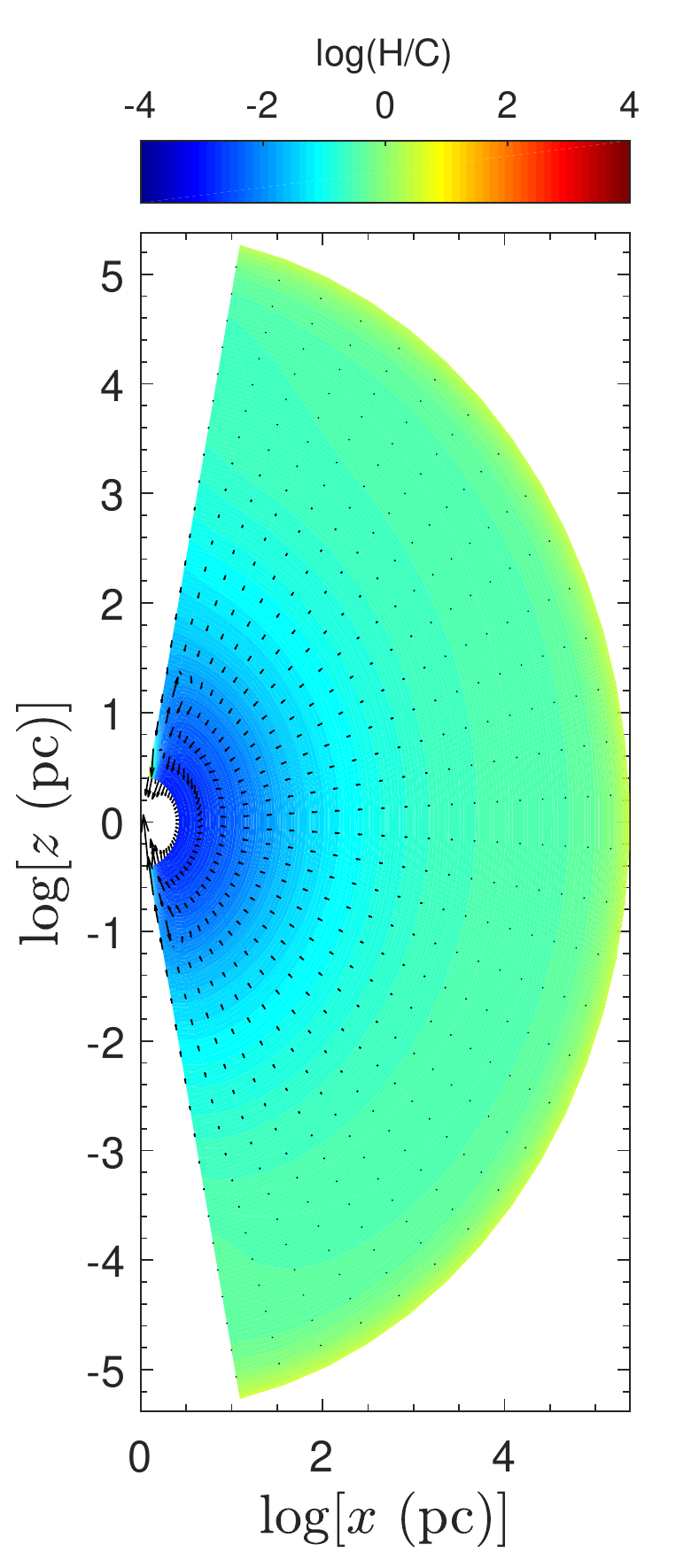}
\includegraphics[height=0.6\textwidth,width=0.33\textwidth]{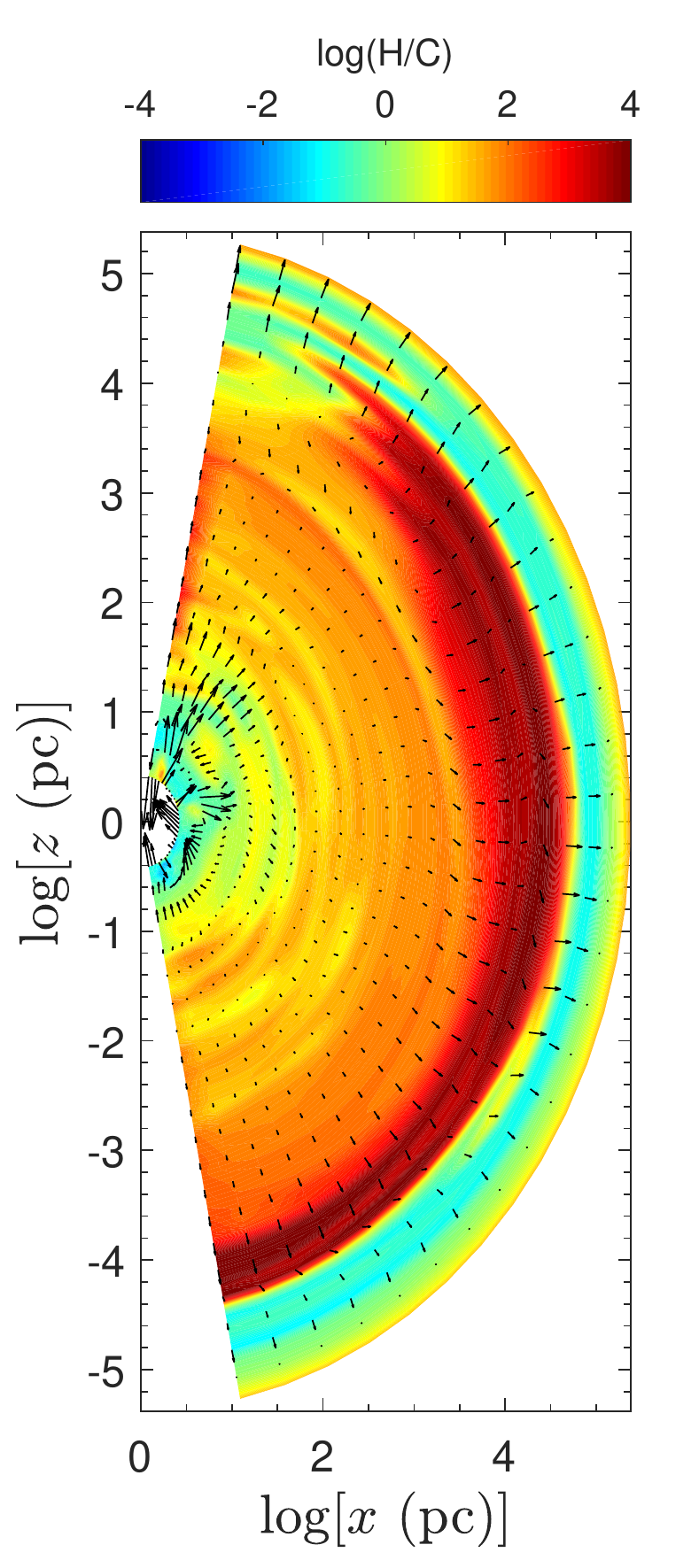}
\caption{Heating over cooling rate ratios for model E260.agb (left panel), E260.agb+sn (middle panel), and E260.agb+sn+agn (right panel) close to the end of the simulation ($t=11.9$ Gyr).  Arrows are normalized to the velocity field at the $\phi=0$ plane.}\label{fig:eb_f3_snapshot}
\end{figure*}

Two models are run at first, one with only AGB heating (E260.agb), another one also incorporating SNe (SNe Ia and II) feedback (E260.agb+sn).
We do not show the analysis for the gas thermal and dynamical properties here to avoid duplication since they are qualitatively similar to the results above.  The 2D snapshots of the heating over cooling ratio ($H/C$) for the two models are shown in the left and middle panels of Figure~\ref{fig:eb_f3_snapshot}. Only one representative snapshot close to the end of the simulation (11.9 Gyr) is shown here since we find that the results are qualitatively similar at other time epoches. As expected, for the model of E260.agb, AGB heating cannot balance the cooling as shown in the left panel of Figure~\ref{fig:eb_f3_snapshot}, which is the same as in the AGB heating model in less massive galaxy discussed above (i.e., E220.agb). For the SNe feedback model of E260.agb+sn, the total heating rate still cannot offset the total cooling rate as shown in the middle panel of Figure~\ref{fig:eb_f3_snapshot}.

\begin{figure*}[htb]
\includegraphics[height=0.43\textwidth,width=0.33\textwidth]{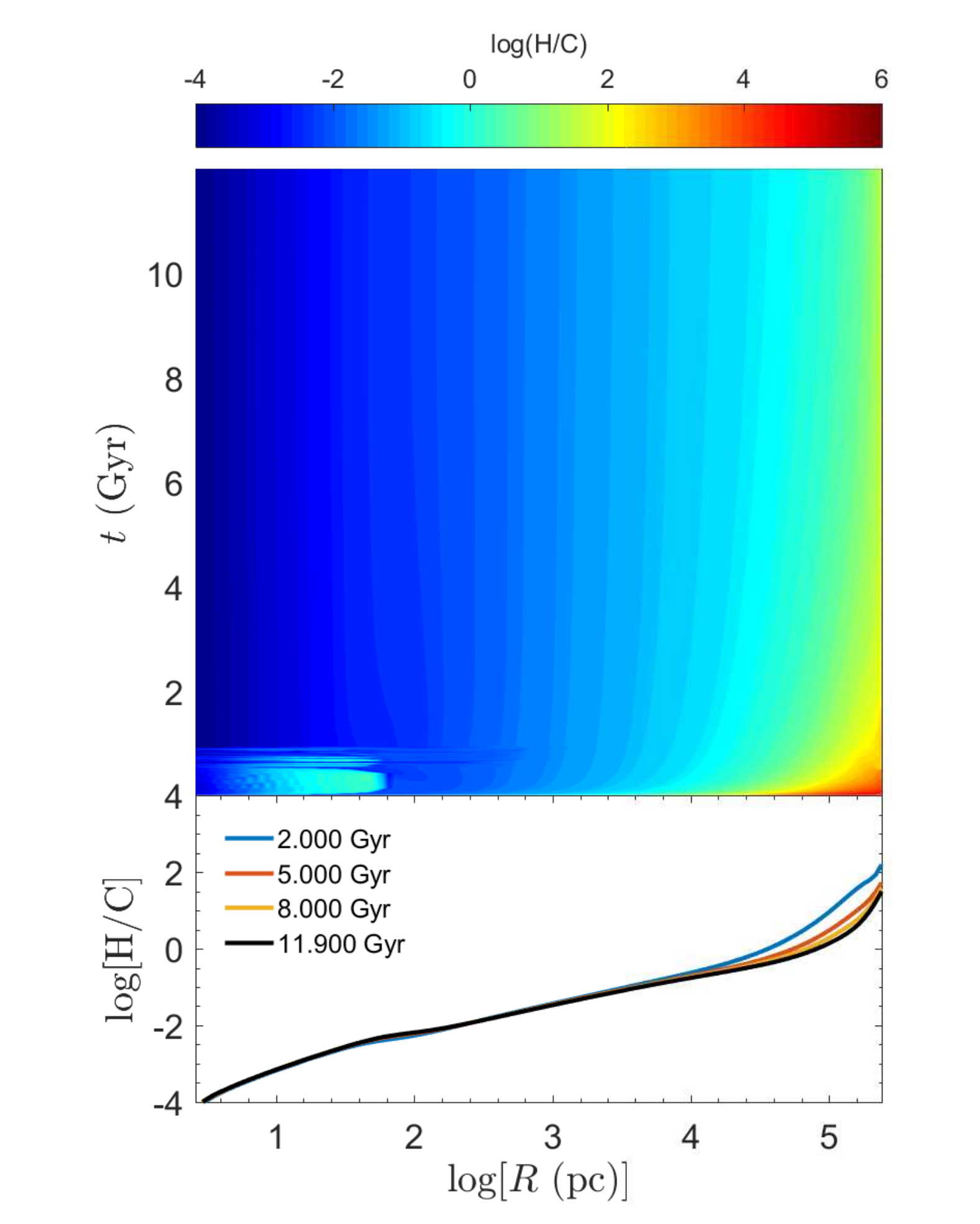}
\includegraphics[height=0.43\textwidth,width=0.33\textwidth]{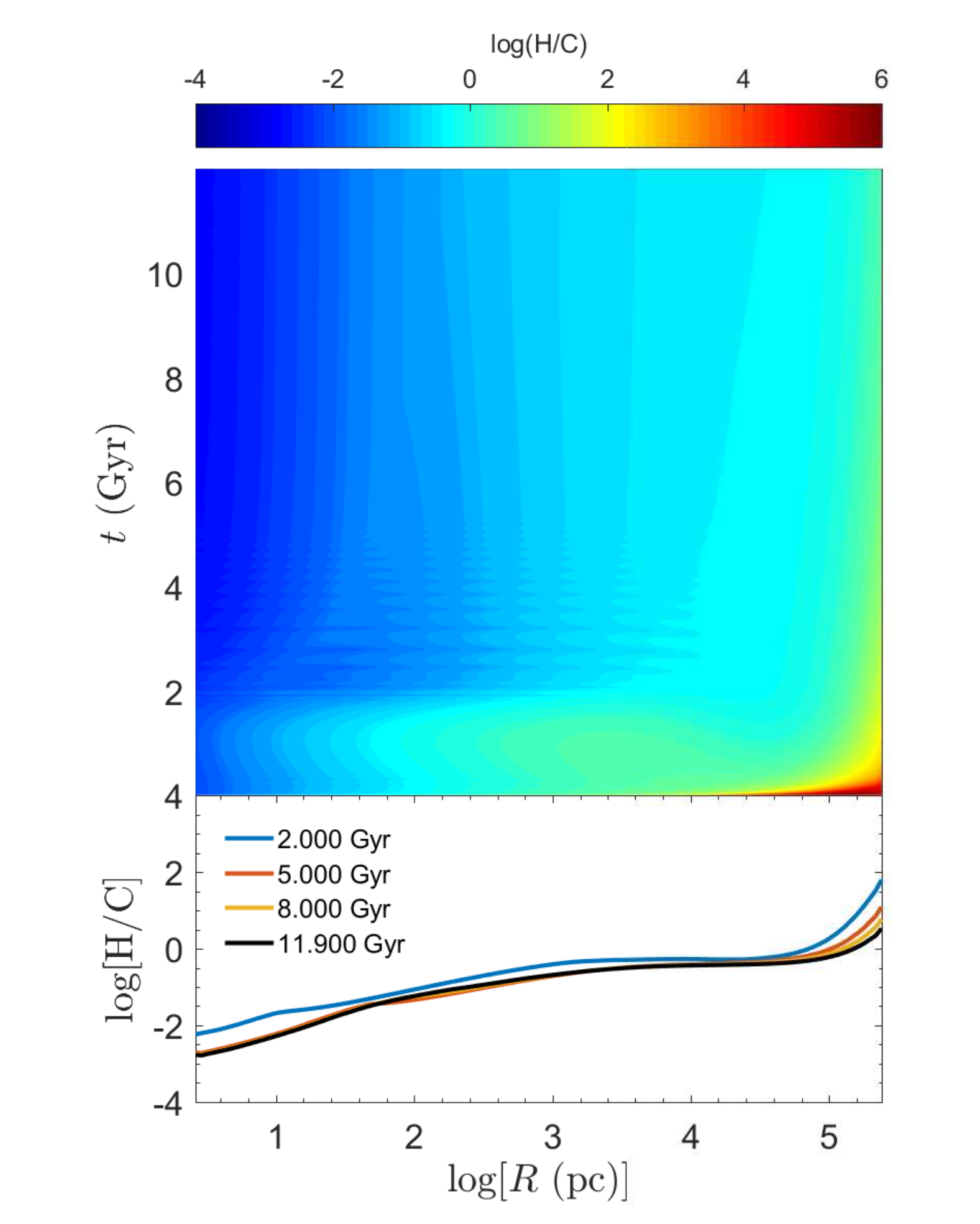}
\includegraphics[height=0.43\textwidth,width=0.33\textwidth]{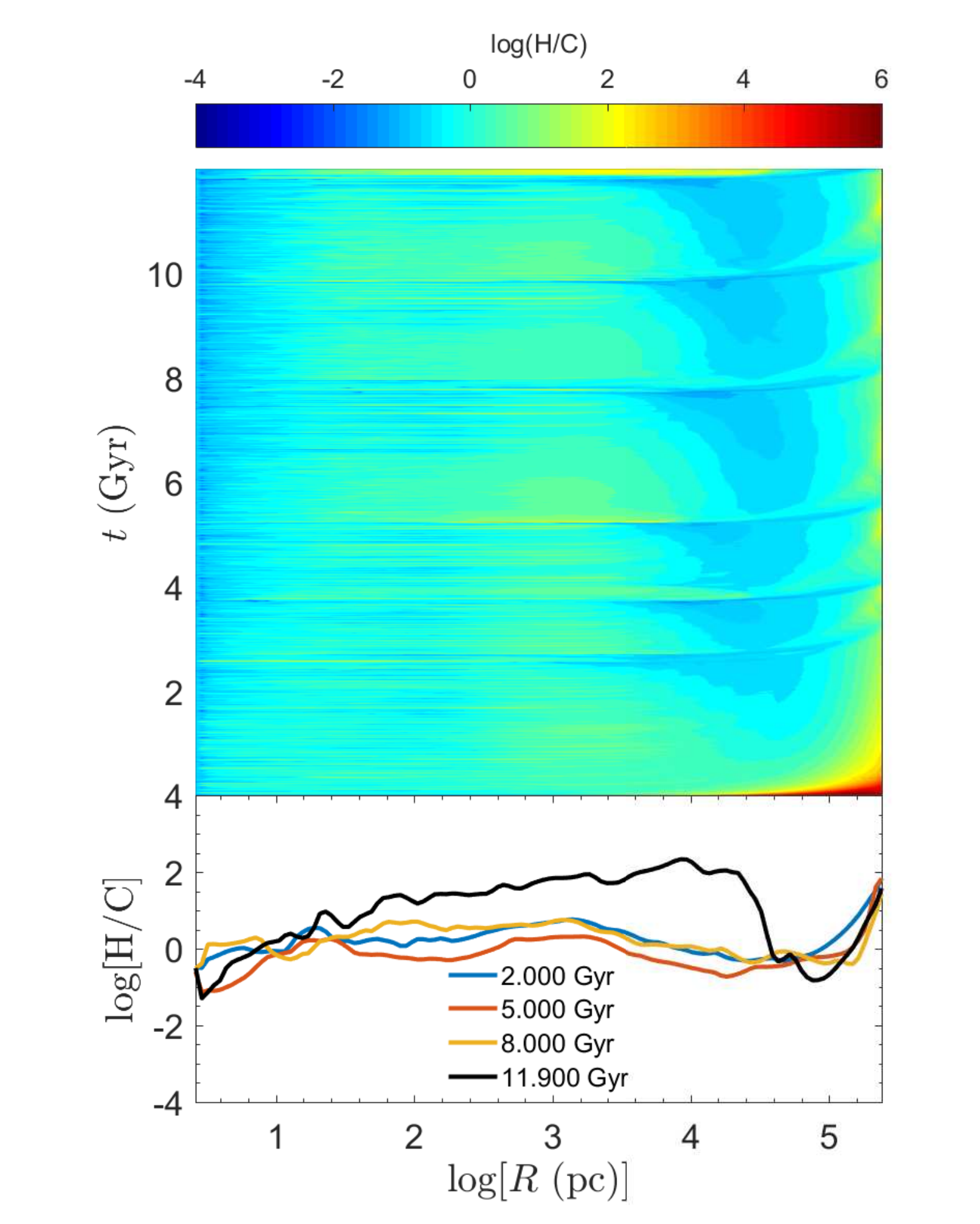}
\vskip 0truecm
\hskip 0.5truecm
\includegraphics[height=0.23\textwidth,width=0.27\textwidth]{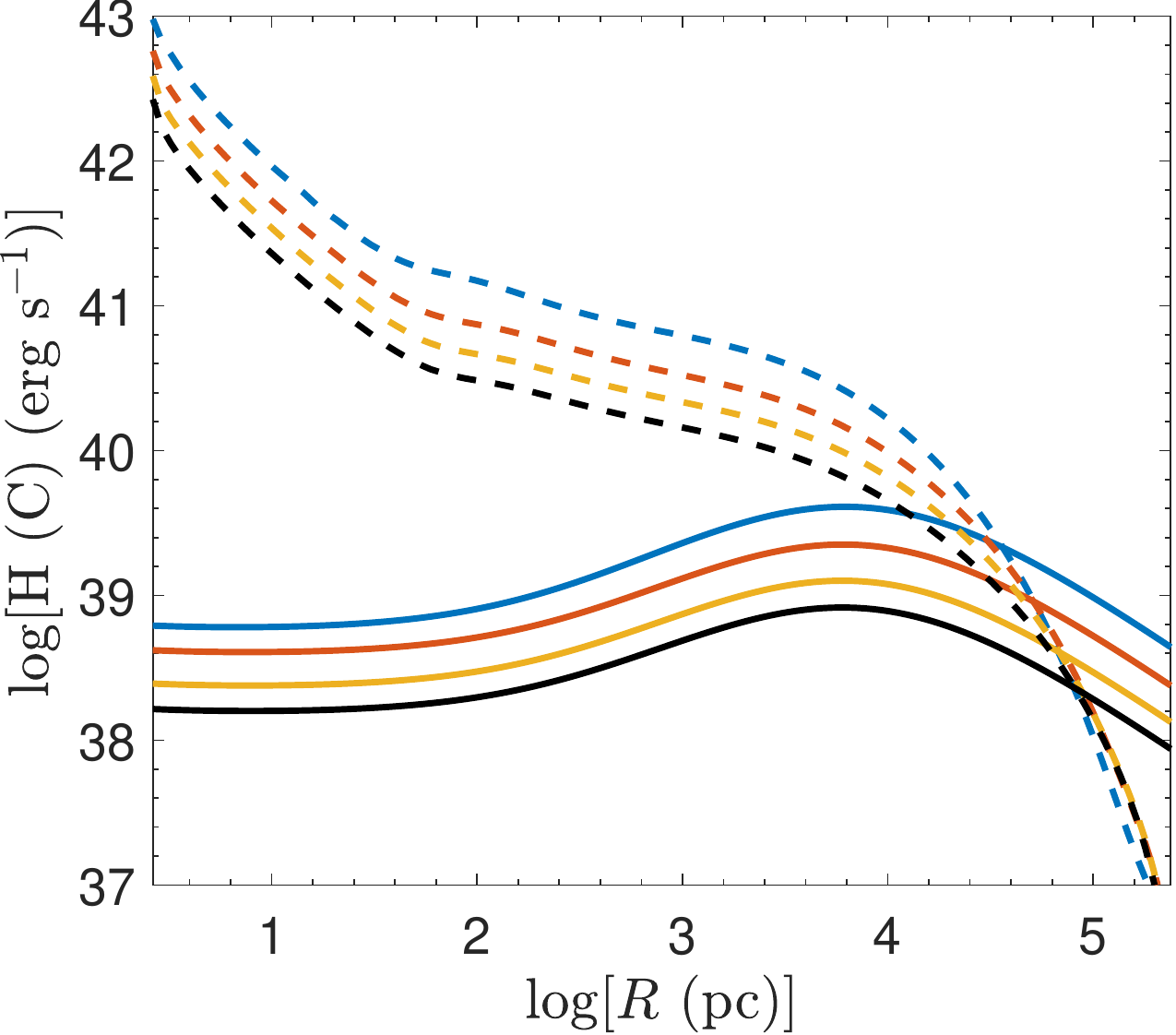}
\hskip 1.2truecm
\includegraphics[height=0.23\textwidth,width=0.27\textwidth]{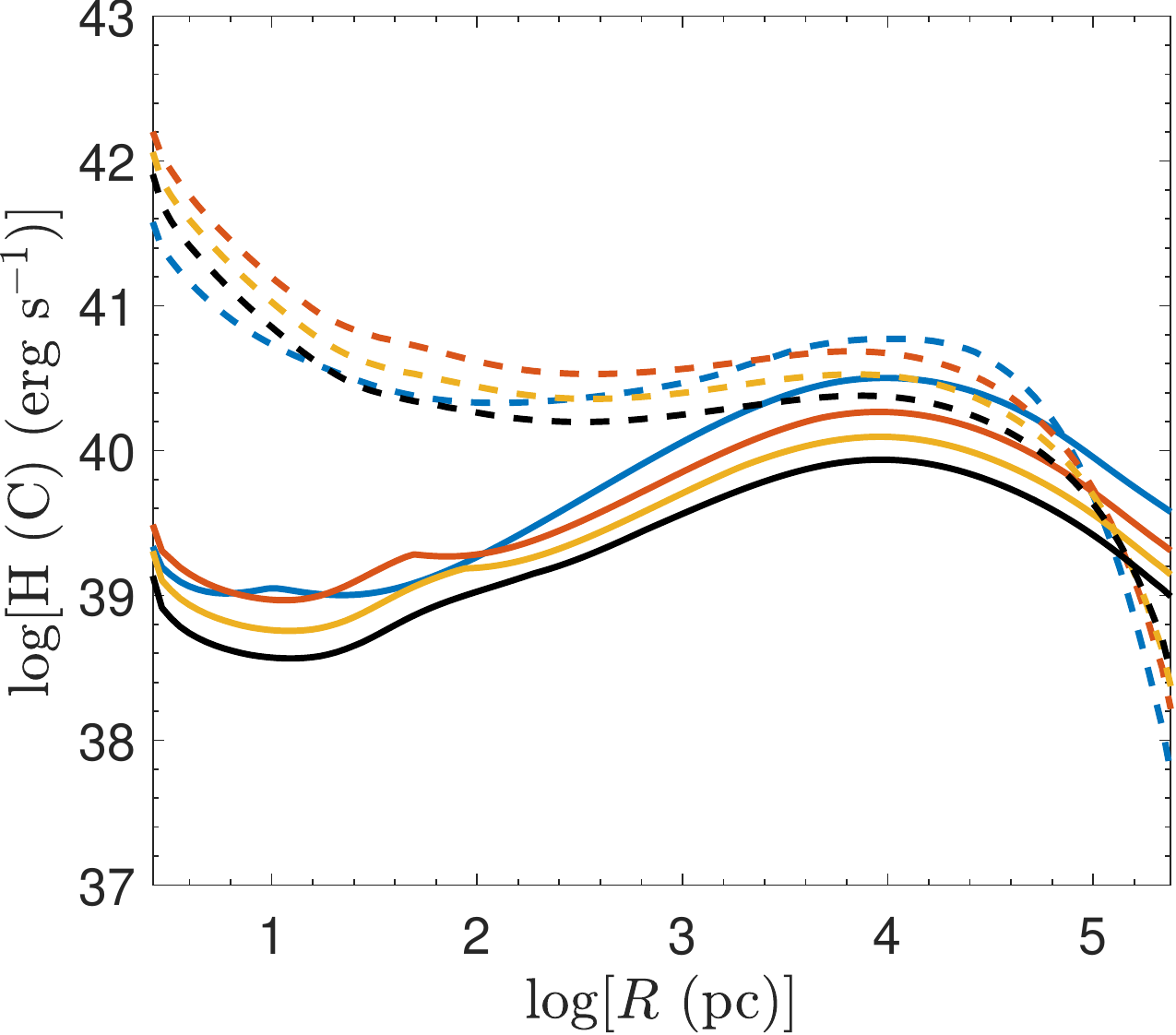}
\hskip 1.0truecm
\includegraphics[height=0.23\textwidth,width=0.27\textwidth]{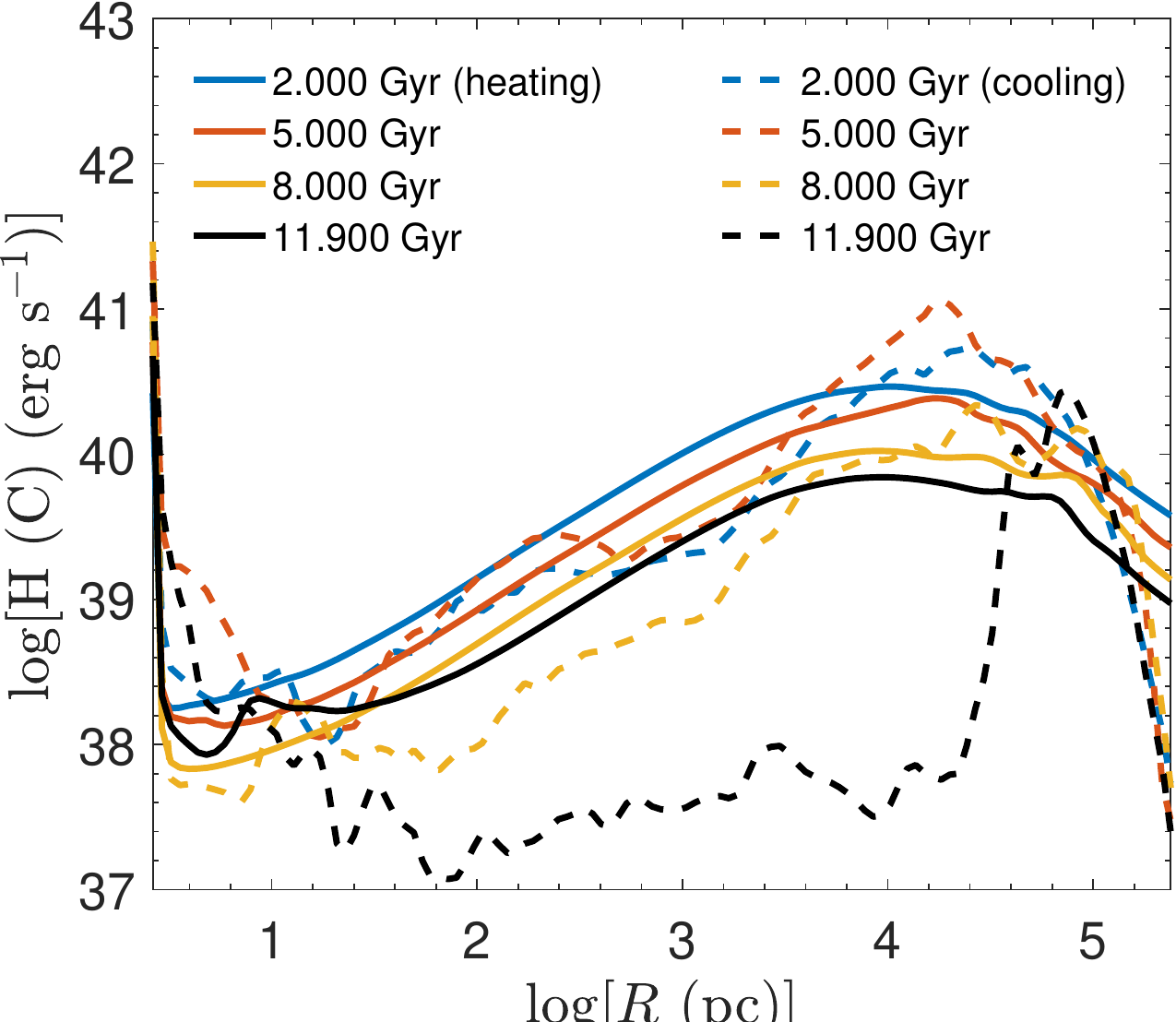}
\caption{Heating over cooling rate ratios as in Figure~\ref{fig:eb_c12_r_t} but for the galaxy model E260. The left, middle, and right panels correspond to models of E260.agb, E260.agb+sn, and E260.agb+sn+agn, respectively.  The heating/cooling sources for AGN feedback model also include Compton heating/cooling. The lines in left and middle panels show the same meanings as those in the the right panel.}\label{fig:eb_f3_r_t}
\end{figure*}

The spatial map of the $\theta-$integrated heating over cooling rate ratio is shown in Figure~\ref{fig:eb_f3_r_t}. The left panel corresponds to the case of AGB heating, while the middle panel for the case including SNe feedback. For both cases, the local heating rate (both AGB and SNe) cannot offset the cooling in most regions at all simulated times, although the inclusion of SNe feedback can relieve the deficiency of heating slightly. The heating rate can marginally balance the cooling rate only in the outskirts of the galaxy ($\gtrsim10$ kpc) in the initial evolution stage.  The regions that suffer from strong cooling even extend to more peripheral areas in the later stage.
These numerical results suggest that AGB heating and even SNe feedback are not effective sources to balance the cooling for a very massive elliptical galaxy, and AGN feedback may be required.

\begin{figure*}[htb]
\includegraphics[width=0.3\textwidth]{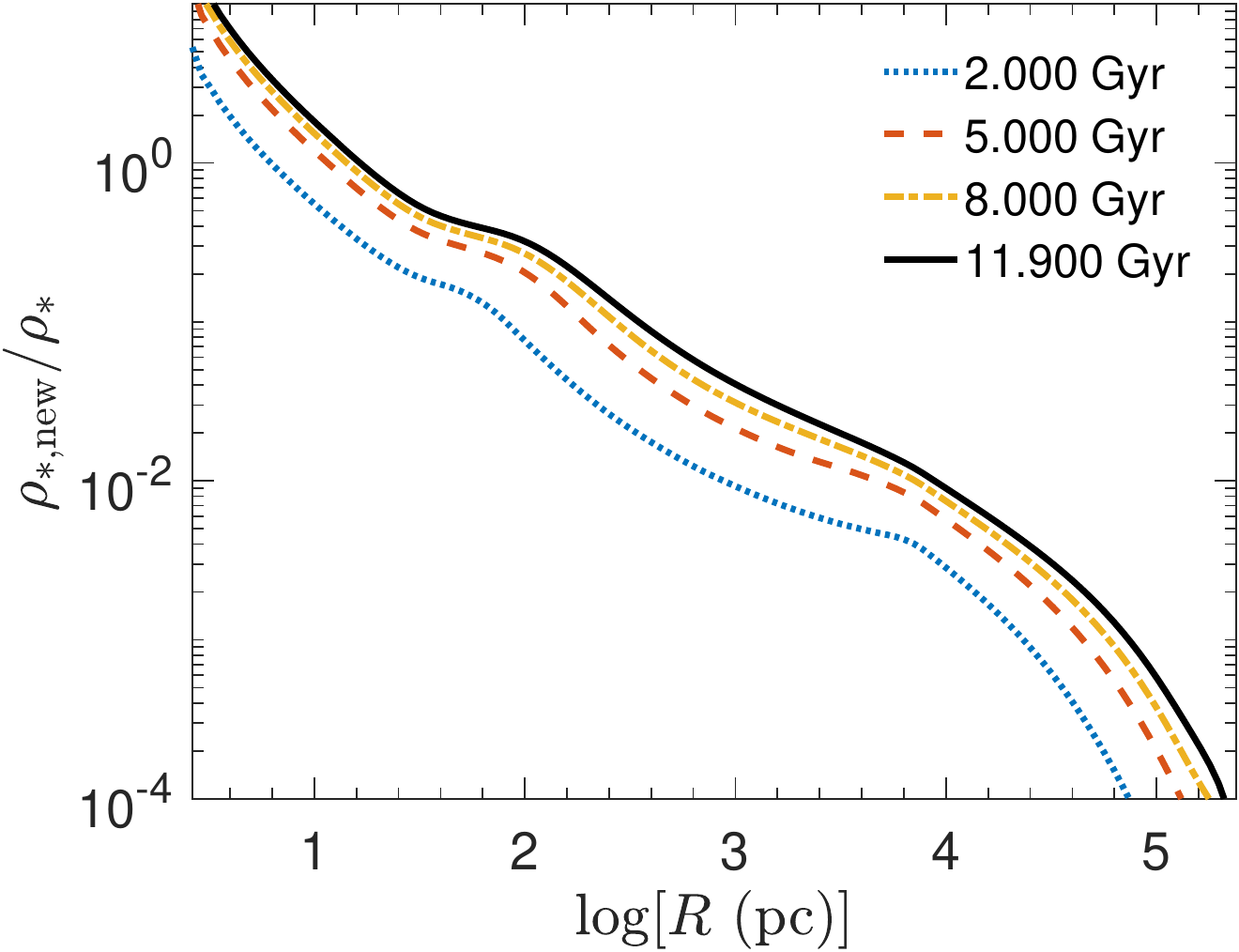}
\includegraphics[width=0.3\textwidth]{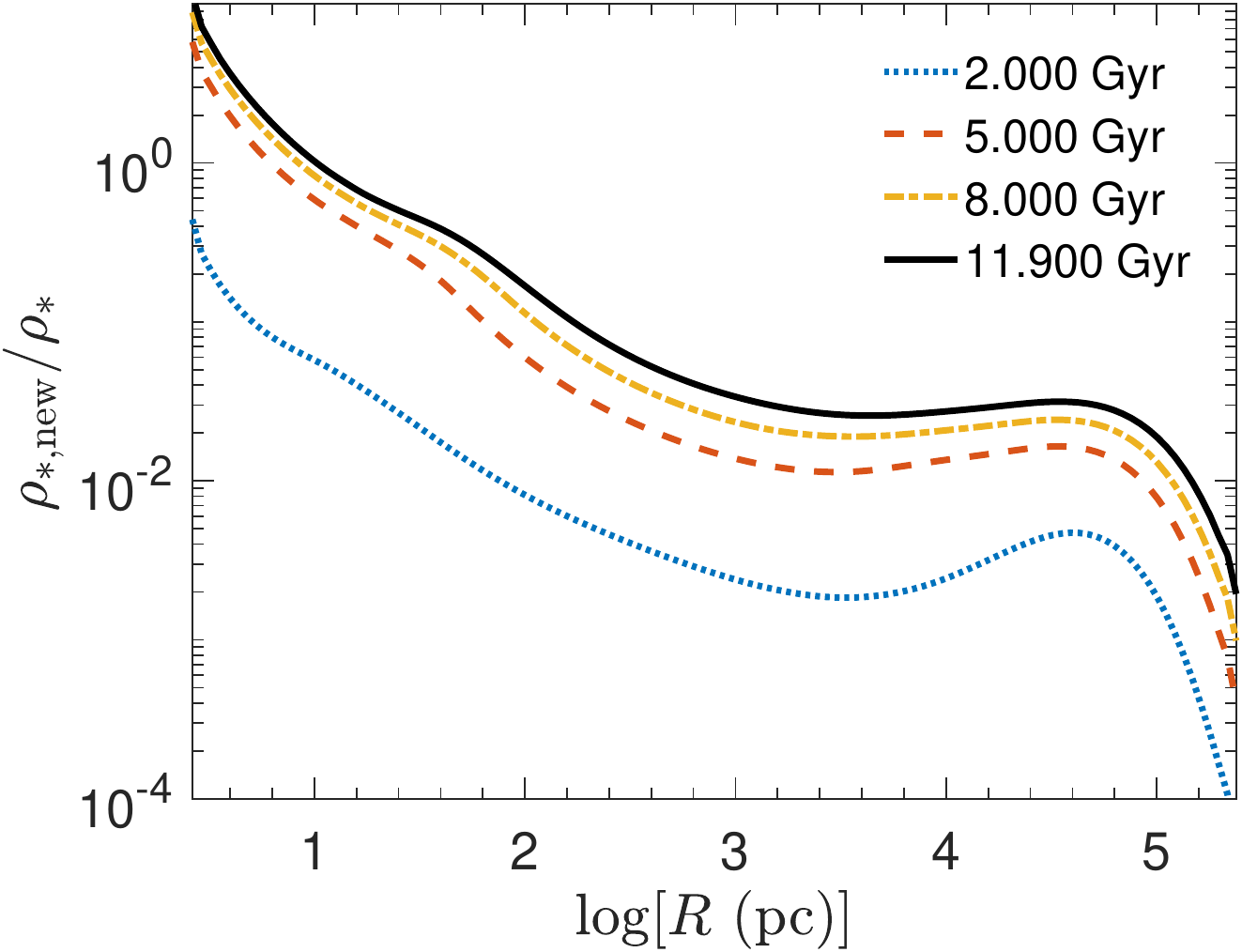}
\includegraphics[width=0.3\textwidth]{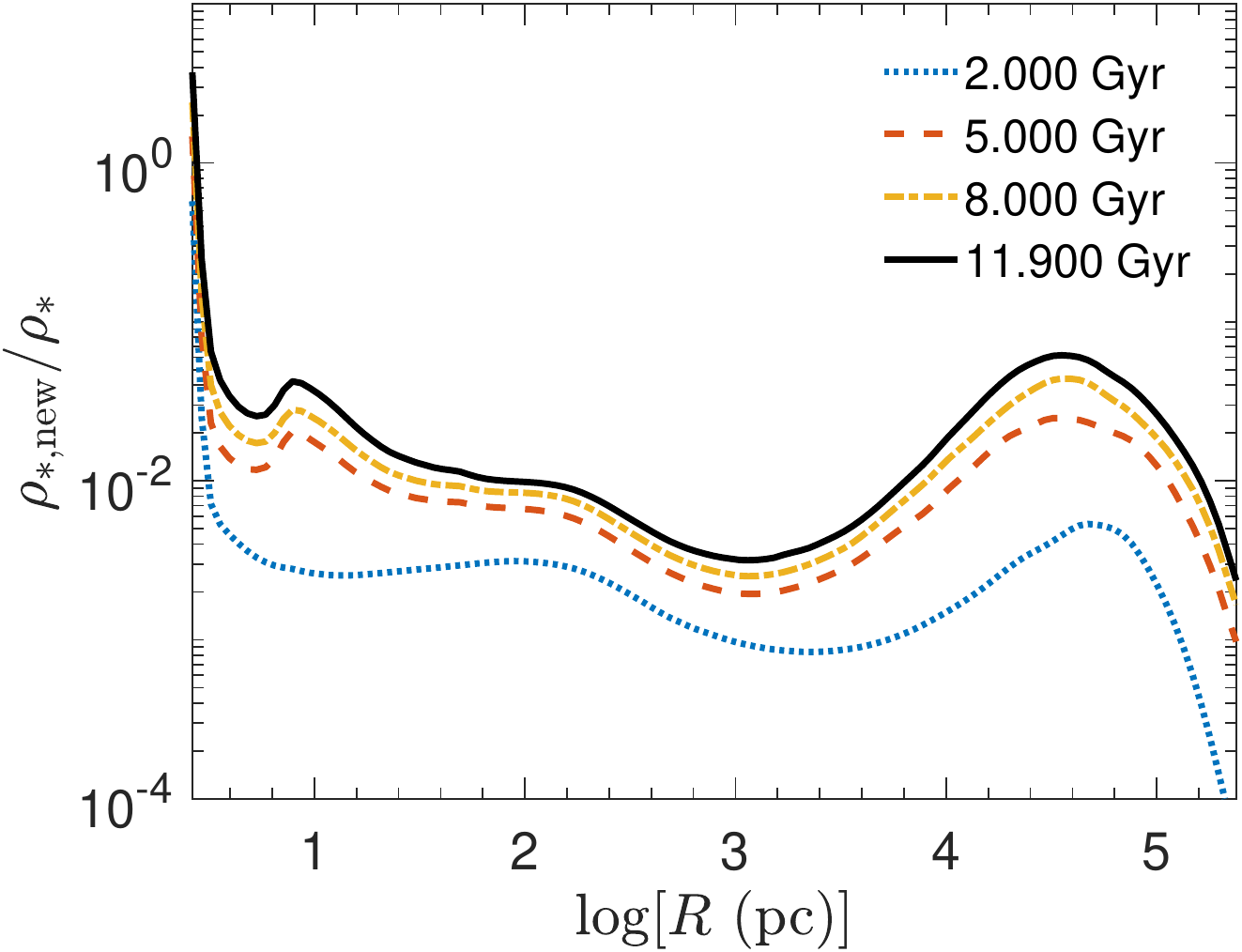}
\caption{Radial profiles of newly formed stellar density $\rho_{\rm \star,new}$ normalized by original stellar population mass density $\rho_{\star}$ for models E260.agb (left panel), E260.agb+sn (middle panel), and E260.agb+sn+agn (right panel) at four selected epoches.  Note that $\rho_{\rm \star,new}$ is the time-integral of SFR density from the beginning of the simulation to the given time, so it monotonously increases with time at all radii.}\label{fig:sfr_f3_r4}
\end{figure*}

Accompanied by the heating and cooling analysis, we then calculate SFR throughout the whole galaxy. We show the $\theta-$averaged SFR density normalized by initial old stellar population mass density $\rho_{\rm \star,new}/\rho_{\star}$ in each radial grid and their time evolution in Figure~\ref{fig:sfr_f3_r4}. The left and middle panels represent the models of E260.agb and E260.agb+sn, respectively. $\rho_{\rm \star,new}$ is calculated by integrating SFR density in each radial grid with time from the beginning of the simulation to the given time. Note that we do not consider the migration of the newly formed stars. Therefore, $\rho_{\rm \star,new}/\rho_{\star}$  monotonously increases with time at any radii of the galaxy.

We can see that $\rho_{\rm \star,new}/\rho_{\star}>1$ in the inner region, which suggests that there exists a large star formation excess in the inner region of the galaxy for both models when we compare $\rho_{\rm \star,new}$ with the stellar mass distribution $\rho_{\star}$.\footnote{The gravitational effect of this newly formed star can be negligible due to the following reasons.  On the one hand, the star formation excess only exists in the very inner region of the galaxies ($<10$ pc) where the galactic potential is dominated by the central black hole.  This still holds when we consider the external gas sources in Section~\ref{sec:results:igas}. On the other hand, the mass of the newly formed stars is compensated by the mass loss of the old stellar population. This makes the total stellar mass in all simulation grids close to the initial value when we do not take into account of the redistribution of the newly formed stars.}
As the galaxy evolves, SFR also declines in the outer region of the galaxy.
This is basically consistent with the scenario that the cold gas material flows into the center region as the local heating rate cannot balance cooling effectively, which then results in the increase of SFR in the inner region and its decrease in the outskirt region.
The stronger deficiency in heating for the AGB heating model (E260.agb) makes $\rho_{\rm \star,new}/\rho_{\star}$ in the inner region larger compared with SNe feedback model (E260.agb+sn).
This is also confirmed by the velocity field as shown in the left and middle panel of Figure~\ref{fig:eb_f3_snapshot}.

From the above analysis, we can see that for the more massive system, both AGB heating and SNe feedback cannot reverse the strong cooling and hence prevent star formation in the entire galaxy effectively. So another feedback process, namely AGN feedback, is invoked for massive elliptical galaxies.

In the right panel of Figure~\ref{fig:eb_f3_snapshot}, we show a snapshot of the heating and cooling rate ratio $H/C$ at 11.9 Gyr for the AGN feedback model (E260.agb+sn+agn). Other terms, such as Compton heating/cooling, and photoionization heating induced by radiation from the accreting supermassive black hole are also included.
Note that the AGN wind power is not explicitly included in the heating rate $H$, since we inject the wind power into the innermost cells of the simulated grid and then calculate the radial transport. The inclusion of the wind power can further offset the cooling rate, and then result in $H/C\gtrsim1$ even in the very central region.\footnote{The cooling is dominated by the bremsstrahlung loss instead of the inverse Compton process, although some simulations have shown that inverse Compton cooling can have an impact on the effectiveness of AGN winds \citep{Bourne2015}, which can be used to reproduce the $M_{\rm BH}-\sigma$ relation \citep{King2003,King2005}. However, the role of inverse Compton cooling is still under debate \citep{Faucher2012,Bourne2013}.} The effect of the wind feedback on the energy balance can be inferred from the thermal state of the ISM (e.g., the X-ray temperature and luminosity).

In comparison with the relatively smooth $H/C$ map in models E260.agb and E260.agb+sn, the $H/C$ ratio for E260.agb+sn+agn shows some remarkably fluctuations in different regions (especially close to the galactic center) as shown in the right panel of Figure~\ref{fig:eb_f3_snapshot}. These are simply arising from the effect of AGN radiation and wind on ISM properties, which play a dominant role in the central region of the galaxy, as discussed in \citet{Yuan2018}.
In addition, the $H/C$ ratio approaches unity even in the galactic central region, which indicates that the heating can now reverse the cooling with the inclusion of AGN feedback. In the right panel of Figure~\ref{fig:eb_f3_r_t}, we show the radial distribution of spatial $H/C$ ratio and its time evolution. We can see that the local heating can balance the cooling in most regions of the galaxy during most of the time. Four time epoches are selected to specifically show the radial profile of $H/C$, which confirms the results in the 2D snapshot $H/C$ map in Figure~\ref{fig:eb_f3_snapshot}. From the heating and cooling rate plots in the bottom panels of Figure~\ref{fig:eb_f3_r_t}, we find that the decrease of the cooling rate is remarkable, when AGN feedback is included. However, the changes of the heating rates are insignificant for different models except in the innermost region. This is because the strong galactic wind driven by AGN activities leads to the decrease of gas density.  As shown from the snapshot above, the heating dominant region should be more prominent if we consider the power of the disk wind from the central AGN in this plot.

In the right panel of Figure~\ref{fig:sfr_f3_r4}, we show the corresponding spatial distribution of newly formed stellar density normalized by old stellar population density ($\rho_{\rm \star,new}/\rho_{\star}$).
The evolution pattern is different from the cases with the AGB heating and AGB+SNe feedback models (right panel vs. left and middle ones in this Figure), i.e., the radial distribution of newly formed stars and its time evolution.
The suppression of SFR in the galactic central region ($\lesssim$ a few kpc) by AGN feedback is remarkable compared with both AGB heating and SN feedback models. Although $\rho_{\rm \star,new}/\rho_{\star}$ is still larger than unity in the innermost region, this value is now much lower and declines more sharply with radius than models without AGN feedback.
In addition, the star formation activities become stronger in the outer region of the galaxy.

This is the signature of AGN feedback in action because the AGN wind and radiation push ISM outward and heat up the gas in the central region, both of which can prevent the star formation close to the central region (see \citet{Yuan2018} for a detailed discussion for this issue).
There are also some localized regions where the SFR is higher than that of model E260.agb+sn at some times (e.g., around 30 kpc from the center at 5 Gyr as shown in the right panel of Figure~\ref{fig:sfr_f3_r4}). The reason is that the compression of the outflowing gas can make dense cold clumps be formed there. The formation of such cold clumps then results in intense star formation activities. This suggests that AGN feedback can sometimes trigger star formation activities in some galactic regions, which is already shown by some observational and theoretical works \citep{Ciotti2007,Liu2013,Silk2013,Zubovas2013,Bieri2016,Zubovas2017,Mukherjee2018}.

\begin{figure}[htb]
\includegraphics[width=0.45\textwidth]{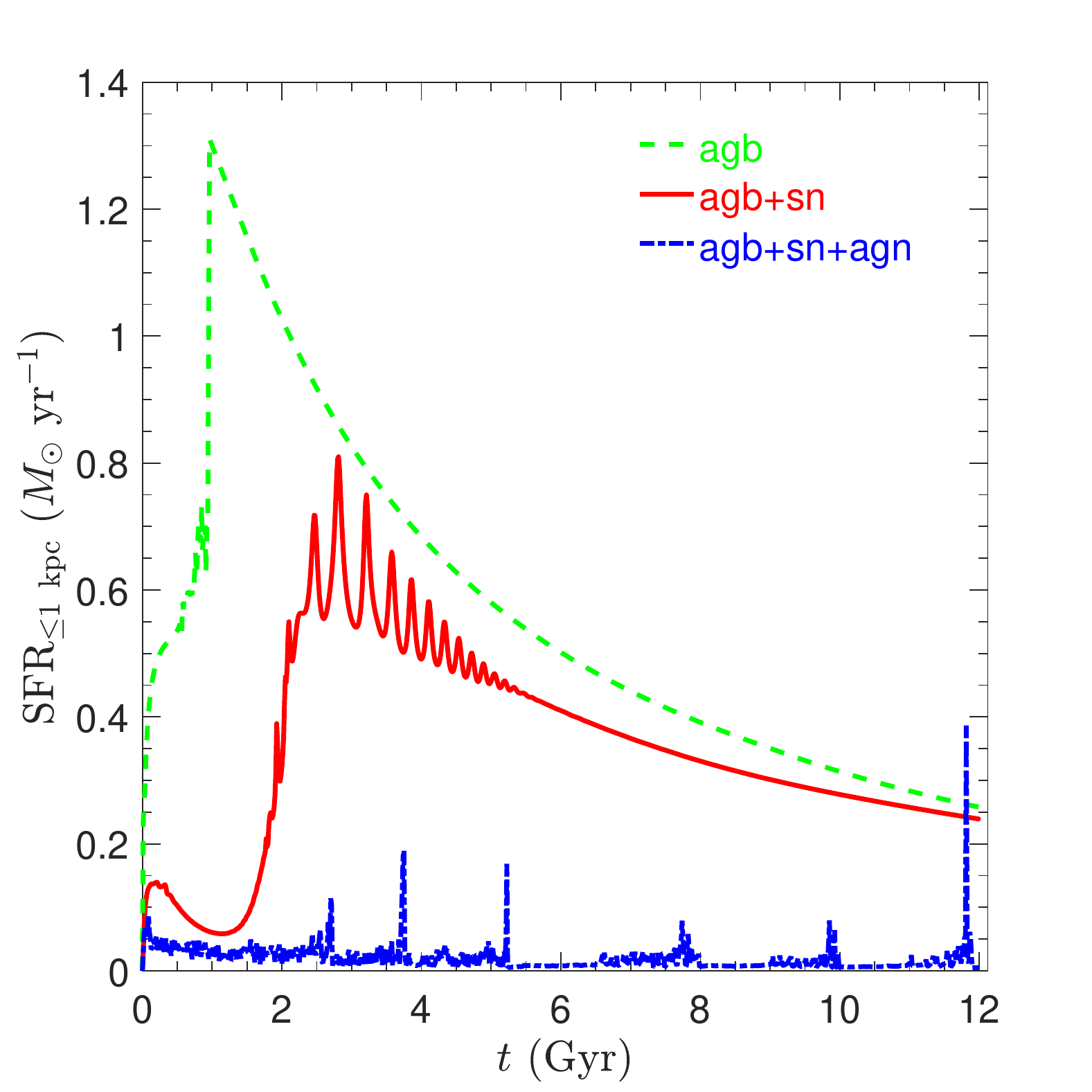}
\caption{Time evolution of SFRs within the inner 1~kpc for three different feedback models for the galaxy E260. The green, red, and blue curves correspond to AGB heating, AGB+SN feedback, and full feedback (AGB+SN+AGN) models, respectively. }\label{fig:sfrinner_f3}
\end{figure}

Based on the analysis of star formation activities in different feedback processes for the galaxy model E260, we can see that AGN feedback plays a significant role in suppressing SFR, especially in the central region of the galaxy. To quantitatively address this issue, here we show the time evolution of SFRs in the inner 1~kpc for three feedback models in Figure~\ref{fig:sfrinner_f3}. This is more close to the spatial scale where SFRs are measured in observations. It clearly demonstrates that the star formation activities in the central region are significantly suppressed and remain at a very low level at most times when AGN feedback is included \citep[e.g.,][]{Choi2015}, even though several star bursts appear during some evolutionary stages.

For the total SFR integrated over the whole galaxy, we show the mean sSFR during the whole evolution in Figure~\ref{fig:ssfr_mstar} as filled squares. Different colors correspond to different feedback models. The mean sSFR of the AGN feedback model (blue symbol) is about a factor of two lower than the cases of stellar feedback models (E260.agb: green; E260.agb+sn: red).  As expected, when AGN feedback is involved, the sSFR oscillates more remarkably, which can be seen from the larger error bar compared to other feedback models. The observed very low rate of star formation in most massive elliptical galaxies \citep{Harwit2015,Renzini2015,Belli2017} essentially rules out the AGB and AGB+SN feedback models.

\subsubsection{Other Galaxy Models}

Several additional galaxy models  are simulated with their  model parameters listed in Table~\ref{tab:parameters}. Here we do not show the detailed energy balance analysis to avoid duplication. The star formation suppression signature can be directly seen from the sSFR in the entire galaxies based on the results above.

The total sSFRs for different galaxies are plotted in Figure~\ref{fig:ssfr_mstar}. For a comparison, we also show the observed data for the main sequence and quenched galaxies from \citet{Renzini2015}. The error bars for different points demonstrate their time variabilities during the whole evolution. It can be seen that the suppression of star formation in less massive galaxy is very significant. More importantly, we find that this suppression is mainly contributed by SN Ia feedback rather than the stellar wind heating when we compare with different stellar feedback models \citep[a similar conclusion is made in 1D simulation by][]{Ciotti1991}.
For more massive galaxies, the stellar feedback alone cannot effectively suppress star formation activities, as is the case of the galaxy model E260. AGN feedback can regulate SFR to a lower value with a few $M_{\odot}~{\rm yr^{-1}}$, which are consistent with observations of star formation activities in massive quiescent galaxies \citep{Harwit2015,Renzini2015,Belli2017}.
One thing we should mention is that in the simulation setup, all the ISM comes exclusively from the mass losses from evolved stars, i.e., there is no gaseous component initially in the whole galaxy.  As the gaseous content is essential for the star formation process, the initial gas mass in the galaxy could be another important factor to effect the results above. In the Section~\ref{sec:results:igas}, we will study how the above results will depend on the initial gas density in different galaxies.

From the results for different galaxies, the mass scale below which stellar feedback becomes important is $M_{\star}\simeq(1\sim3)\times10^{11}~M_{\odot}$, which corresponds to a B-band stellar luminosity for the galaxy of $(2\sim6)\times10^{10}~L_{\odot}$. Above this mass scale, AGN feedback plays a dominant role in balancing the cooling losses and further regulating SFRs. SN Ia feedback, instead of AGB heating, is the main heating source to prevent the SFRs for low mass galaxies. The physical reason of such transition is as follows. The gas source is mainly contributed by the mass loss of evolved stars during the red giant, AGB, and planetary nebula phase. This mass loss is proportional to the stellar mass $M_{\star}$ of the galaxy. The gas density supplied by this mass loss then increases with $M_{\star}$ linearly. Since the cooling rate is related with the square of the gas density, and then related with the square of $M_{\star}$, while the heating rate provided by AGB and SNe (mainly SNe Ia) are linearly proportional to $M_{\star}$ (Equations~\ref{eq:lsn_agb},\ref{eq:lsn_ia}), these heating sources, therefore, cannot balance the cooling when $M_{\star}$  increases to a threshold. Above this mass scale, the heated gas cannot escape from the galaxy potential and so flow into the galactic center to trigger the AGN activity. This scenario is similar to the one found by \citet{Bower2017}.

The results above are qualitatively consistent with the picture for the galaxy evolution model, but with a slightly different transition mass scale \citep[e.g.,][]{Silk2012,Choi2015, Taylor2017}. In the following, we will find that this mass scale can be compatible with the observed value when a small fraction of gas source is added in the entire galaxy initially. However, the results above are contrary to the suggestion by \cite{Conroy2015}, who proposed that AGB heating could play an important role in preventing star formation in quiescent galaxies. We suspect that the main reason for this discrepancy is the gas reservoir of the simulated galaxy. \citet{Conroy2015} only considered the energy input from AGB wind heating, but ignored the addition of the mass losses from AGB stars into the ISM in the galaxy. The inclusion of the AGB mass losses into the galactic gas mass, which is found to be substantial, inevitably introduces strong cooling, which can then alter the heating/cooling balance and the star formation activities as well.

The physical reason of SNe (and AGN) heating being more effective than AGB heating in reversing cooling is that, the specific AGB heating rate (the heating rate divided by the stellar mass, which mainly links to the stellar velocity dispersion $\sigma_{0}$) is roughly bound to the gravitational potential of the galaxy, while SNe (AGN) heating is apparently unaware of such limitation. Therefore, SNe (AGN) heating could be, in principle, energetically important in thermalizing the ISM and acting as a more efficient heating mechanism.  Actually, the specific energy output from SNe (AGN) is indeed much larger than that from AGB for the galaxies we consider.

\subsection{ISM Properties}\label{sec:ism}

The gaseous component in the galaxy is the total material available for star formation and black hole accretion, both of which can in turn affect the ISM densities and temperatures by some feedback processes.
The heating/cooling energy balance and star formation activities should trace the properties of ISM since the former strongly depend on the ISM densities and temperatures. ISM temperatures and luminosities can in turn be used to diagnose the feedback processes in action. So we will discuss how the ISM mass (or its radial distribution), hot gas temperature and luminosity can be used to discriminate different feedback processes in the following.

The X-ray emission in the $0.3-8$ (and $0.5-2$) keV \textit{Chandra} band, and the temperature weighted by the X-ray emission from the hot plasma, are calculated as \citep[e.g.,][]{Negri2014}
\begin{equation}
L_{\rm X}=\int \varepsilon_{\rm X}{\rm d}V,
\label{eq:lx}
\end{equation}
\begin{equation}
T_{\rm X}=\frac{\int T\varepsilon_{\rm X}{\rm d}V}{L_{\rm X}},
\label{eq:tx}
\end{equation}
respectively, where $\varepsilon_{\rm X}$ is the thermal emissivity in the energy band $0.3-8$ (and $0.5-2$) keV of a hot, collisionally ionized plasma, which is obtained from the spectral
fitting package \textsc{XSPEC}\footnote{http://heasarc.nasa.gov/xanadu/xspec/} (spectral model \textsc{APEC}).

\subsubsection{ISM Mass}

In Table~\ref{tab:parameters}, we list the ISM mass in the end of the simulation within $10~r_{\rm eff}$ of the galaxy, beyond which the ISM can be recognized as a galactic wind.

As we have discussed above, SNe feedback in less massive galaxies can effectively drive the ISM to the region far from the galactic center to become the galactic wind.  Therefore, the depletion of ISM becomes significant when effective SNe  feedback is considered. The effect is more important at the lower end of the galaxy stellar mass range because the ISM in a shallower gravitational potential can be more easily expelled. This is confirmed by examining the ISM mass remaining within $10~r_{\rm eff}$ of the galaxies as shown in Table~\ref{tab:parameters}, i.e., the gas masses of SNe feedback models are much lower than those of AGB heating models (e.g., E220.agb+sn vs. E220.agb, E100.agb+sn vs. E100.agb).
This explains why the plasma X-ray luminosity $L_{\rm X}$ and temperature $T_{\rm X}$ in SNe feedback models is even lower compared to those with only AGB heating models for less massive galaxies, as we will discuss below.

For the more massive system, the ISM mass for the case of AGN feedback models is even higher than those with only AGB heating cases. This is a complex mutual effect of cooling-induced inflow and heating-induced outflow. With only AGB heating and SNe feedback, the gas inflowing is significant due to the strong cooling. This reduces the total ISM mass in the whole galaxy because a large fraction of gas mass is accreted onto the central point mass (black hole). While AGN feedback can drive the gas to the outer region of the galaxies \citep{Choi2015}, and self-regulate the ISM inflowing to the central black hole (accretion) at a low level. This can result in the higher gas mass both within and beyond $10~r_{\rm eff}$ of the galaxy (e.g., E340.agb+sn+agn vs. E340.agb and E340.agb+sn) as shown in Table~\ref{tab:parameters}.

\subsubsection{X-ray Temperature}\label{sec:tx}

Some \emph{Chandra} X-ray observations show that the gas temperature $T_{\rm X}$ of ETGs is roughly consistent with $T_{\sigma}$ from the thermalization of the stellar kinetic energy estimated from Equation~(\ref{eq:tsigma}) \citep[e.g.,][]{Matsushita2001,Nagino2009,Boroson2011,Pellegrini2011,Posacki2013,Sarzi2013,Anderson2015,Goulding2016}. This has been used to support AGB heating as a dominant mechanism in suppressing star formation in ETGs \citep[e.g.,][]{Conroy2015}.

\begin{figure*}[htb]
\centering
\includegraphics[width=0.35\textwidth]{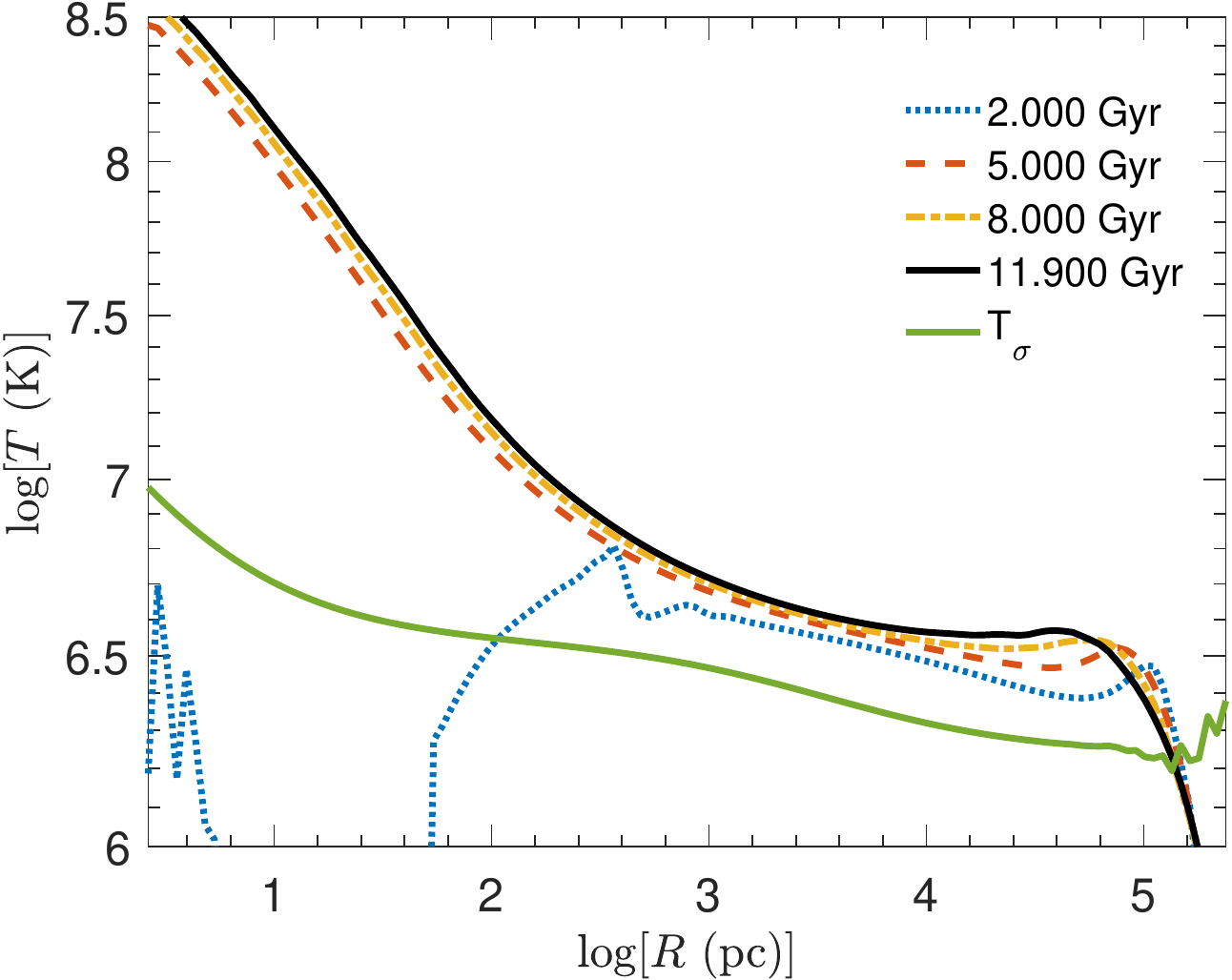}
\hskip 0.5truecm
\includegraphics[width=0.35\textwidth]{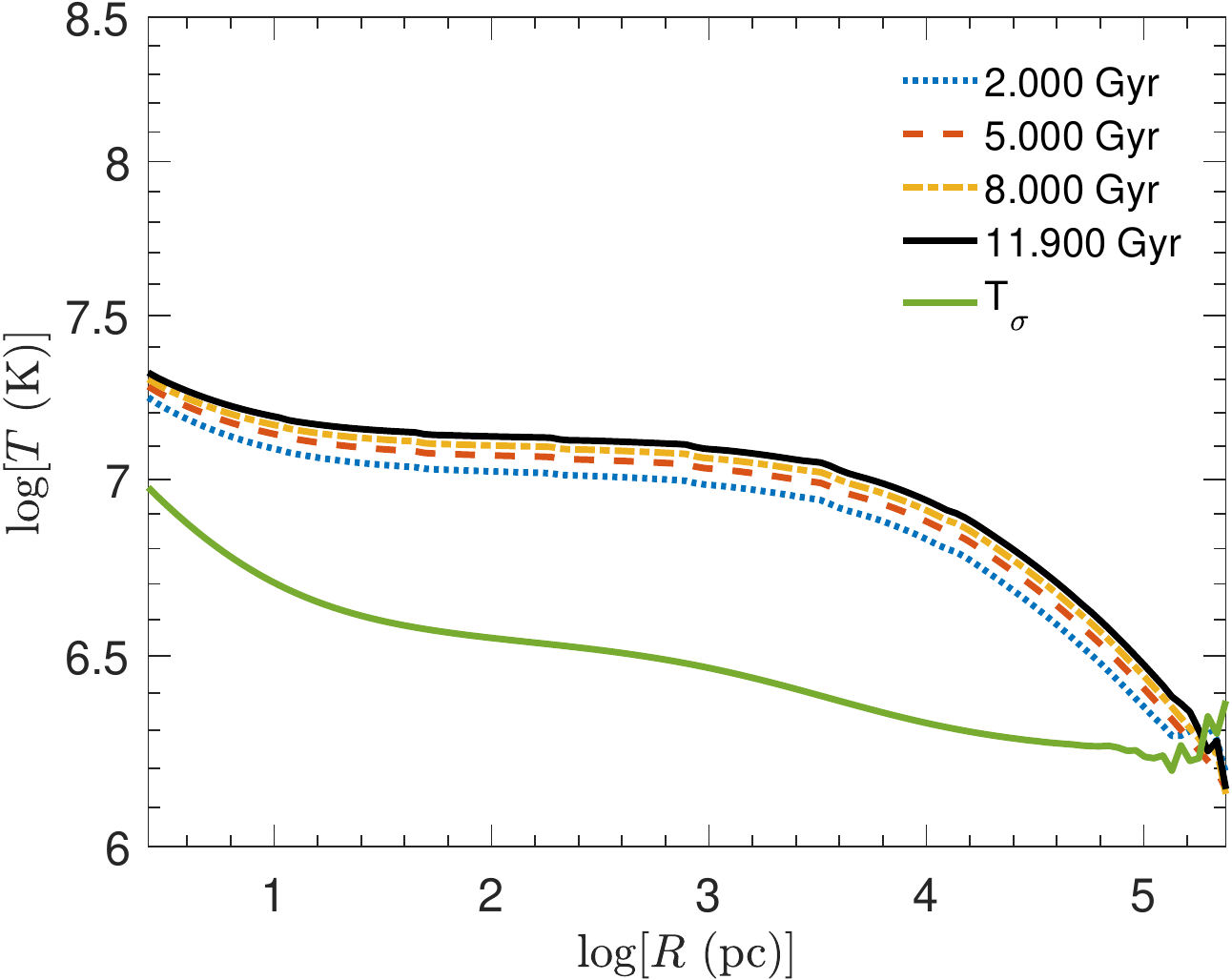}
\caption{Radial profiles of $0.3-8$ keV emission-weighted ISM temperature $T_{\rm X}$ at four simulated epoches for galaxy model E220. The left (right) panel represents the model E220.agb (E220.agb+sn). The green lines show the $T_{\sigma}$ values based on Equation~(\ref{eq:tsigma}). }\label{fig:tx_c12_r4}
\end{figure*}

\begin{figure*}[htb]
\centering
\includegraphics[width=0.35\textwidth]{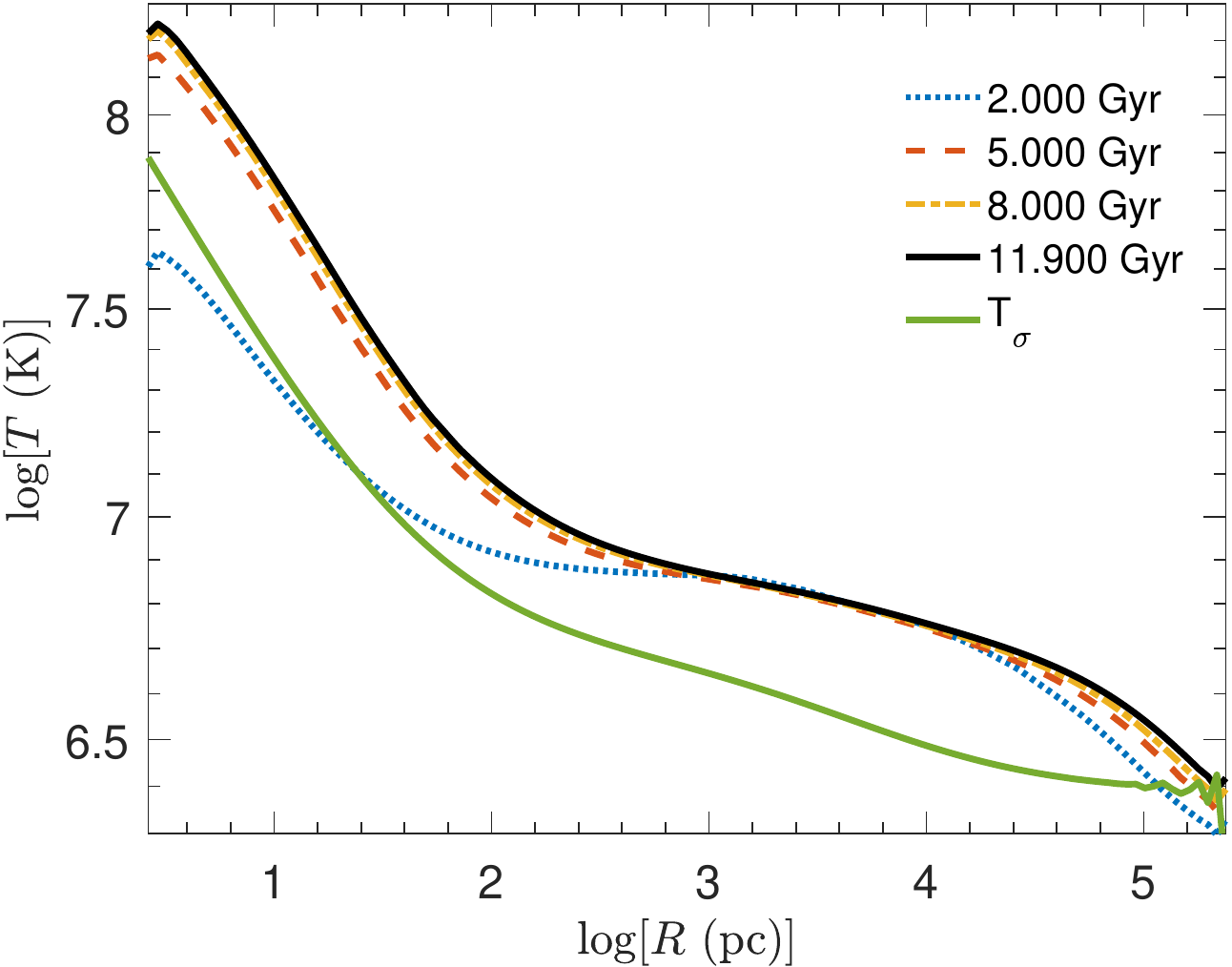}
\hskip 0.5truecm
\includegraphics[width=0.35\textwidth]{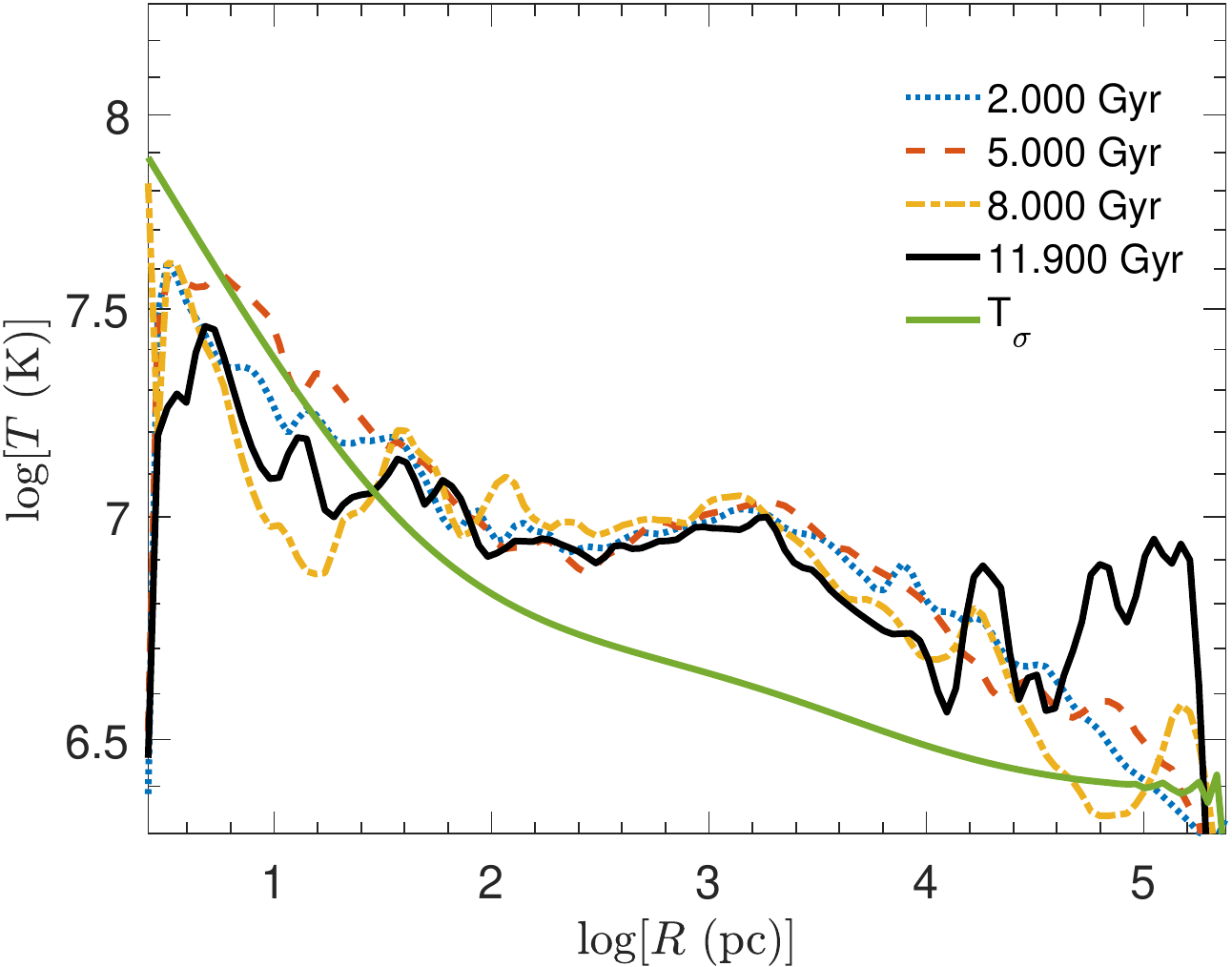}
\caption{Same as Figure~\ref{fig:tx_c12_r4} but for the galaxy E260. The left (right) panel corresponds to the model E260.agb+sn (E260.agb+sn+agn).}\label{fig:tx_f3_r4}
\end{figure*}

Here we calculate the $0.3-8$ keV luminosity-weighted temperature $T_{\rm X}$ using Equation~(\ref{eq:tx}). We show the radial profiles of $T_{\rm X}$ (X-ray luminosity-weighted in each radial bin of the simulated grid) for the galaxy model E220 in Figure~\ref{fig:tx_c12_r4}. The left (right) panel corresponds to model E220.agb (E220.agb+sn). We also select four time epoches to check the time variations of the radial profiles. For each model, the radial profile of $T_{\sigma}$ calculated using Equation~(\ref{eq:tsigma}) at the initial stage is superposed for a comparison\footnote{$T_{\sigma}$ mainly depends on the galactic structure. On the one hand, the evolution of the old stellar population is negligible compared to the initial setup. On the other hand, although the growth of $M_{\rm BH}$ can be significant due to the strong inflowing gas towards the center when only AGB heating is considered, we prefer to use the initial $M_{\rm BH}$ in calculating $T_{\sigma}$ to compare with our simulated results. This is because the initial $M_{\rm BH}$ is already consistent with $M_{\star}-M_{\rm BH}$ relation in the nearby universe and this relation is not expected to evolve too much since a redshift of $z\sim3$ \citep{Kormendy2013,Yang2018}. The very massive $M_{\rm BH}$ in the end of the simulations for the AGB heating models, e.g., $M_{\rm BH,final}=1.2\times10^{10}~M_{\sun}$ for E220.agb model, are highly inconsistent with the $M_{\star}-M_{\rm BH}$ relation. This is additional evidence to rule out the pure AGB heating model in terms of black mass growth, although it is out of the scope of this work (see \citet{Yuan2018} for the discussion of the black hole growth).}.
The discrepancy between $T_{\rm X}$ and $T_{\sigma}$ is remarkable for the model E220.agb. A large $T_{\rm X}$ shown in the late stage of the inner galactic region is similar to the results shown in the left panel of Figure~\ref{fig:ism_c12}.

When the SNe feedback is incorporated, the temperature discrepancy decreases although $T_{\rm X}$ is still slightly larger than $T_{\sigma}$ by about $0.3-0.5$ dex. The reason is as follows. For the AGB heating model, when the heating cannot offset the catastrophic cooling, the gas will flow towards the galactic center, which acts as an extremely effective heating source, i.e., the compression work and shock wave, heat gas in the central region. Such a high temperature due to gravitational heating of the inflowing gas is related to the singular isothermal gravitational potential we adopt here, which has been studied in previous works \citep[e.g.,][]{Guo2014}. While the SN Ia feedback is very efficient in heating the gas up to a few $\times {\rm keV}$, and preventing the gas flowing into the center, which can help to avoid the overheating of the gas by the compression work. This can be also confirmed by the gas mass remaining in the galaxies as discussed above. Such a high plasma temperature $T_{\rm X}$ in the inner region of the galaxy due to gravitational heating is unreasonable. Therefore, SNe feedback is required to avoid the overheating of the gas in the central region of the galaxy and then build up the similarity between $T_{\rm X}$ and $T_{\sigma}$.

For the more massive galaxy E260, we show the radial profiles of $T_{\rm X}$ and $T_{\sigma}$ in Figure~\ref{fig:tx_f3_r4}. The left (right) panel corresponds to model E260.agb+sn (E220.agb+sn+agn). As we have found that AGB heating model is far from being an effective heating mechanism at this mass scale, we do not show its temperature profile here. When AGB heating and SNe feedback are included, there is a remarkable difference between $T_{\rm X}$ and $T_{\sigma}$ in the inner region. This is attributed to the same reason as above (i.e., the compression work by the cooling-induced inflow). Interestingly, $T_{\rm X}$ closely matches $T_{\sigma}$, especially in the galactic center region, when AGN feedback is involved. The evolution of the black hole mass is insignificant due to the effective heating of the inflowing gas \citep{Yuan2018}. This is an indication of the hydrostatic equilibrium built up by the inclusion of AGN feedback. Therefore, it strongly suggests the necessity of AGN feedback in regulating ISM properties in massive elliptical galaxies, where it can balance the cooling even in the inner region.

\begin{figure*}[htb]
\centering
\includegraphics[width=0.95\textwidth]{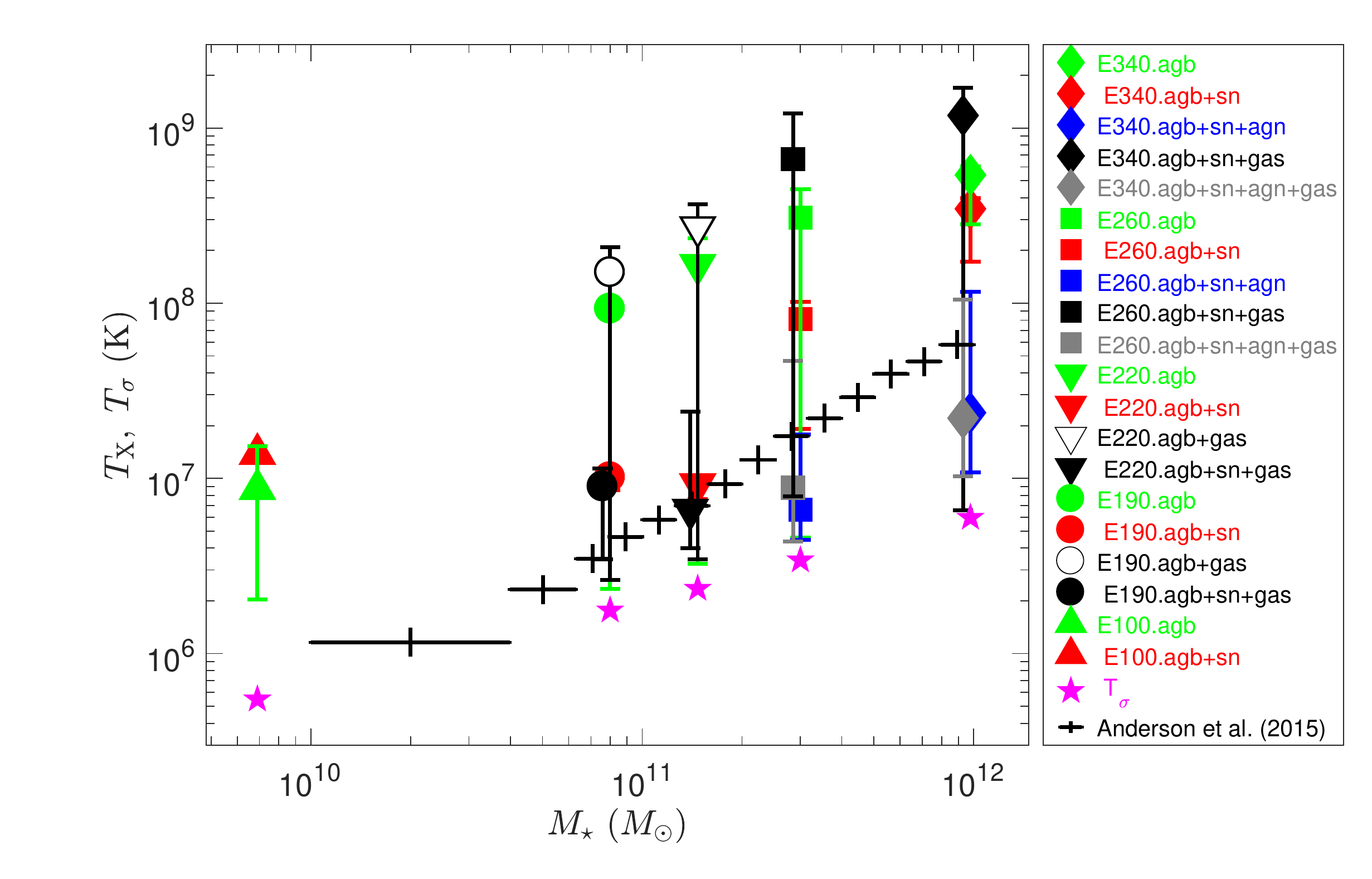}
\caption{$0.3-8$ keV emission-weighted ISM temperature $T_{\rm X}$ as a function of stellar mass $M_{\star}$. For each model, we show the time-averaged value with the error bar corresponding to the maximum and minimum values reached during the evolution sequence. The green symbols represent the AGB heating models, red ones for the stellar feedback (SN+AGB) models, and blue ones for full feedback (AGB+SN+AGN feedback) models. Symbols with different shapes (diamonds, squares, downward triangles, circles, and upward triangles) correspond to galaxy models with different stellar masses. For some galaxy models, we also incorporate a gas density initially throughout the entire galaxy to compare with the secular evolution cases. The filled black markers indicate the stellar feedback models, open black ones for AGB heating models, and filled grey ones for full feedback models. Note that we only show the low gas density cases for galaxy E190 and E220. Several simulated $T_{\rm X}$ points with the same $M_{\star}$ are slightly shifted horizontally for the purpose of presentation. The $T_{\sigma}$ values for different galaxies are shown as purple symbols. The observed $T_{\rm X,obs}$ data (black crosses) are adapted from \citet{Anderson2015}. }\label{fig:tx_mstar}
\end{figure*}

For the integrated $T_{\rm X}$ and $T_{\sigma}$ obtained from Equations~(\ref{eq:lx},\ref{eq:tx}) and Equation~(\ref{eq:tsigma}), their values for different galaxy models with different feedback processes are all shown in Figure~\ref{fig:tx_mstar}. One representative X-ray observational data set $T_{\rm X,obs}$ \citep[][and see also the references therein]{Anderson2015} is chosen to compare with our simulated data. The discrepancy between $T_{\sigma}$ and $T_{\rm X,obs}$ could be due to different stellar kinematics and gas dynamical states \citep{Sarzi2013,Negri2014,Goulding2016}.  For each model, we show the time-averaged value of $T_{\rm X}$ with the error bar corresponding to the maximum and minimum values that can be reached during the evolution sequence. The mean value $T_{\rm X}$, which indicates the temperature of gas where it stays in most of time, is used below to compare with the observational data $T_{\rm X,obs}$.

Our numerical results show that $T_{\rm X}$ is higher for a more massive system when only AGB heating models are considered, but systematically larger with respect to both $T_{\sigma}$ and $T_{\rm X,obs}$.
When additional feedback mechanisms are considered, the tendency of $T_{\rm X}$ in different galaxies is different.
With the inclusion of SNe feedback in low mass galaxies (the red circle and red downward triangle in Figure~\ref{fig:tx_mstar}), the gas temperature is close to the temperature $T_{\sigma}$ defined by the gravitational potential with an exception of galaxy model E100\footnote{Such a large discrepancy could be due to the rather effective SNe feedback to blow out most of the gas material, as shown in Table~\ref{tab:parameters}.}. More importantly, our simulated $T_{\rm X}$ values are in agreement with observed ones as well.
Although there is still a slight discrepancy due to overheating by SNe, the inclusion of additional gas content will potentially resolve this problem, as we will discuss in Section~\ref{sec:results:igas}.

For more massive galaxies with stellar masses larger than $\sim3\times 10^{11}~M_{\odot}$, it simply shows the sequence of $T_{\rm X,agb}> T_{\rm X,agb+sn}\gg T_{\rm X,agb+sn+agn}\approx T_{\sigma}$.
The high $T_{\rm X}$ for the cases with AGB heating and SNe feedback models is expected because of the compression work of the inflowing gas as discussed above.  With the inclusion of AGN feedback, $T_{\rm X}$ values are located in the region between $T_{\sigma}$ and $T_{\rm X,obs}$, shown as the blue symbols in Figure~\ref{fig:tx_mstar}.
It thus suggests that AGN feedback can regulate the plasma temperature to be close to the X-ray observations (\citealt{Pellegrini2012b}, see \citealt{Boroson2011,Sarzi2013,Anderson2015} for observations).

To summarize, we find that when an effective heating source can offset cooling over the galaxy, it can play a dominant role in causing the consistency between $T_{\rm X}$ and $T_{\sigma}$ (and $T_{\rm X,obs}$). Specifically, in low mass galaxies, SNe feedback could help to make $T_{\rm X}$ close to $T_{\sigma}~(T_{\rm X,obs})$, while the heating source in high mass galaxies comes from AGN feedback. Both of them play their dominant role in reversing the cooling in their respective galaxies. This suggests a logical connection between $T_{\rm X}-T_{\rm X,obs}$ similarity and the energy balance state.

\subsubsection{X-ray Luminosity}\label{sec:lx}

\begin{figure*}[htb]
\centering
\includegraphics[width=0.95\textwidth]{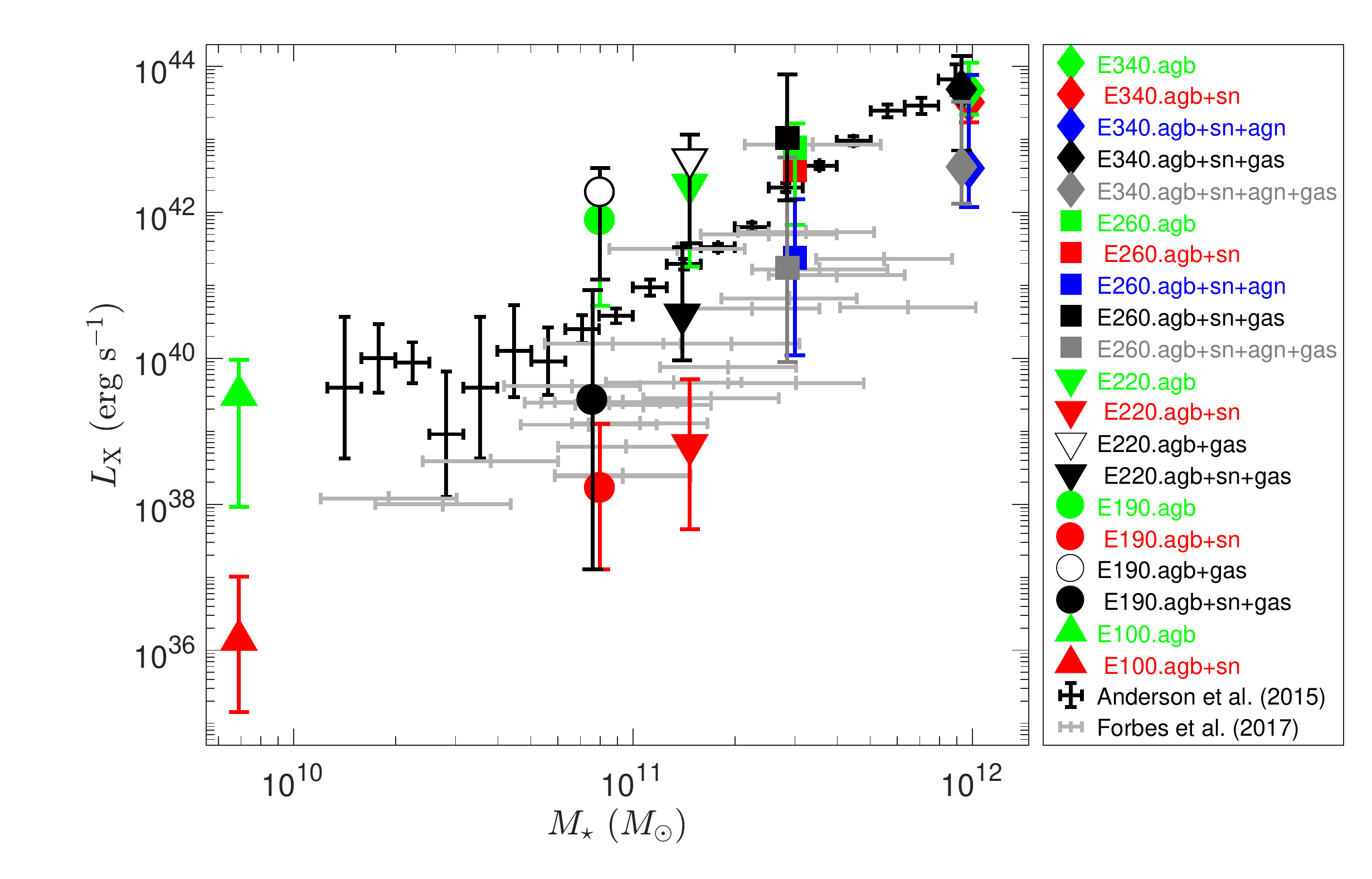}
\caption{ISM X-ray luminosity $L_{\rm X}$ as a function of stellar mass $M_{\star}$. For the symbols of simulated data, they share the same meaning as in Figure~\ref{fig:tx_mstar}. The black and grey crosses are observational data for $L_{\rm X}$ from \citet{Anderson2015} and \citet{Forbes2017}, respectively. Note that all $L_{\rm X}$ points are calculated in the $0.3-8$ keV energy range, except for the observational data from \citet{Anderson2015}, which are in $0.5-2$ keV band. This will result in a small decrease ($<0.3$ dex) of the simulated $L_{\rm X}$ when we calculate $L_{\rm X}$ in the narrower energy band.}\label{fig:lx_mstar}
\end{figure*}

In Figure~\ref{fig:lx_mstar}, we compare the $0.3-8$ keV luminosity for the ISM in the whole galaxy\footnote{We find that the X-ray luminosity inside the virial radius is almost the same as that in the whole galaxy.} with two selected observed data sets from \citet{Anderson2015} and \citet{Forbes2017}. We note that there is slight difference between the two data sets, which could be due to the stellar kinematic and environmental effect \citep{Sarzi2013,Negri2014,Goulding2016}. A similar comparison based on 2D hydrodynamical simulations for a large set of galaxy models with and without AGN feedback has been recently presented \citep{Pellegrini2018}.

For the less massive galaxies, $L_{\rm X}$ is very low when SNe feedback is considered (red symbols). This is because much of the ISM is blown out of the galaxies. When only AGB heating is considered (green ones), a higher $L_{\rm X}$ is obtained, due to the higher $T_{\rm X}$ of the inflowing gas, with the consequent X-ray luminosity peak in the central regions of the galaxies. On average, the $L_{\rm X}$ values in AGB heating models are higher than those of observations, even when the variabilities are considered, while $L_{\rm X}$ in SNe feedback models are slightly smaller than those of observations \citep{Anderson2015,Forbes2017} for galaxy models E190 (circles) and E220 (downward triangles). We will show that this discrepancy can be alleviated with the inclusion of additional gas sources in the initial setup of the galaxies. Note that the flattening of the observational data from  \citet{Anderson2015} below $\sim5\times10^{10}~M_{\sun}$ is largely because the X-ray emission becomes too faint to be distinguishable from the background.

For the massive galaxies, although $L_{\rm X}$  in  AGB (green symbols) and SNe (red ones) feedback models are slightly higher than that in AGN feedback models (blue ones), $L_{\rm X}$ in three feedback models are all roughly consistent with observed values (black and grey crosses) for galaxy models E260 (squares) and E340 (diamonds) after considering their variabilities in the simulated data and the large scatter of observed ones, in agreement with the results in \citet{Pellegrini2018}. However, we cannot conclude that models with only stellar feedback are sufficient to explain the X-ray observations for even massive galaxies. A general trend is that our simulated hot gas X-ray emissions from AGN feedback models are fainter \citep[e.g.,][]{Choi2012} and more close to the the observations from \citet{Forbes2017} (e.g., see also \citealt{Choi2015}). In addition, one contradiction for the no AGN feedback models is the hot gas temperature $T_{\rm X}$ that has been explored in the Section~\ref{sec:tx}. Another one to be checked observationally is the variability of $L_{\rm X}$ (or the scattering of $L_{\rm X}$ for a given $M_{\star}$). Although all of the three feedback models show similar $L_{\rm X}$ values, their time evolutions are apparently different \citep[e.g.,][]{Ostriker2010}. The remarkable oscillations of $L_{\rm X}$ due to the modulation of AGN activities in AGN feedback models are in agreement with the large scatter of $L_{\rm X,obs}$ in observations.

\subsection{Effect of Initial Gas Sources}\label{sec:results:igas}

Up to now, the galaxies we deal with are built up with the cases that all of the gaseous sources come exclusively from late-time stellar mass losses, i.e., secular evolution. Major mergers  could not be important after a cosmological time of 2 Gyr (redshift $z\sim3$; e.g., \citealt{Fan2014}) when we initially setup the galaxies, however, galaxies are constantly accreting gas from the environment, which could be the intergalactic medium (IGM) and/or the accretion of satellite galaxies for field galaxies or the accretion of intracluster medium (ICM) through cluster core's cooling flow for galaxies in clusters. Therefore, it is necessary to consider the external gas sources for galaxy models. Here we include a gaseous component in the entire galaxies in the beginning of the simulations to mimic the external gas sources. This idealized treatment could be appropriate for some situations, e.g., ETGs in clusters where the accretion of satellites is likely to be gas-free. The radial dependence of the gaseous component follows a $\beta$ profile
\begin{equation}\label{eq:gas}
  n_{\rm gas}(r)={n_{0}}{[1+(r/r_{\rm c})^2]^{-3/2\beta}},
\end{equation}
with a slope parameter $\beta=2/3$ and assuming a core radius $r_{\rm c}=r_{\rm eff}$ \citep[e.g.,][]{Jones1984,Eke1998,Mo2010}. For the normalization of the density profile, observations of galaxy groups or clusters infer the baryon or gas fraction within virial radius of the galaxies \citep[e.g.,][]{Dai2010,Boroson2011,Anderson2015,Lim2018}. However, for the elliptical galaxies we setup at 2 Gyr, it would be inappropriate to adopt these results directly. Alternatively, we choose different values of $n_{0}$ for different galaxy models as shown in Table~\ref{tab:parameters} for a comparison with the corresponding secular evolution cases. Our adopted gas density can make sure that the total baryon mass fraction within the virial radius of the galaxies $f_{\rm b}=(M_{\rm gas}+M_{\star})/M_{\rm DM}$ is a fraction of $0.5-0.8$ of the cosmic baryon fractions $f_{\rm b,cosm}=0.1864$ \citep{PlanckXVI2014} for our high gas density models. Although these gas density values are relatively high compared with observations of ETGs mentioned above, they are just for the purposes of studying its effect on star formation activities and ISM properties.

The initial gas sources are incorporated into four galaxy models (i.e., E190, E220, E260, E340). For low mass galaxies, two feedback schemes are considered, one with only AGB heating (E190.agb+gas and E220.agb+gas), another also included SNe feedback (E190.agb+sn+gas and E220.agb+sn+gas). For each case, we choose two different $n_{0}$ values to explore the possible dependence on it. For the more massive galaxies, we consider the cases with only stellar feedback (E260.agb+sn+gas, E340.agb+sn+gas) and with AGN feedback (E260.agb+sn+agn+gas, E340.agb+sn+agn+gas).

\begin{figure*}[htb]
\centering
\includegraphics[width=0.7\textwidth]{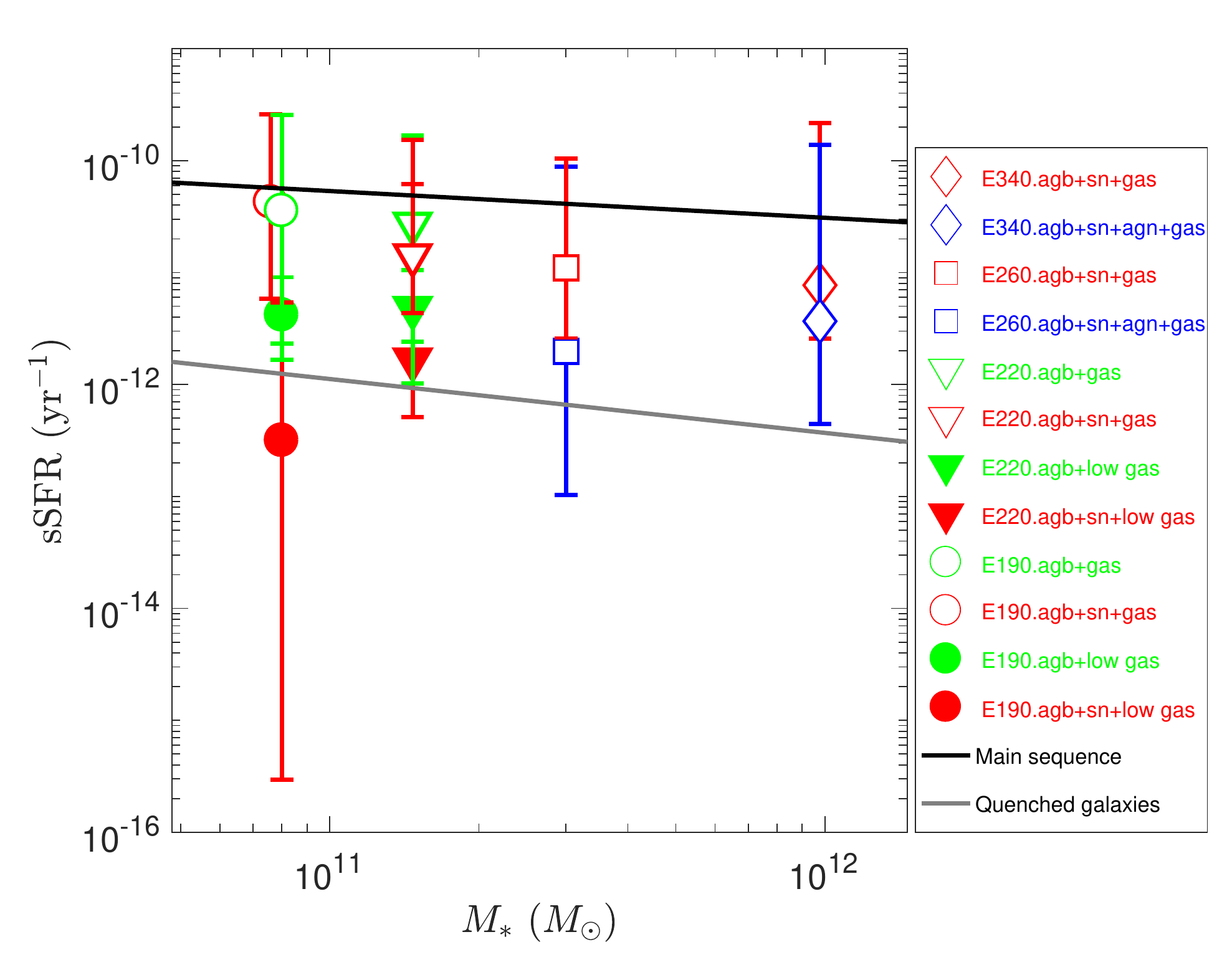}
\caption{Same with Figure~\ref{fig:ssfr_mstar}, but with some fractions of gas included initially (see Table~\ref{tab:parameters}). The green symbols correspond to the AGB heating models with an initial gas source. The red ones represent the models with SNe feedback involved, and blue ones also include AGN feedback. The filled (open) symbols indicate the corresponding low (high) gas density cases when two gas densities are considered for the same galaxy model.} \label{fig:ssfr_mstar_gas}
\end{figure*}

The total sSFRs within the galaxies are shown in Figure~\ref{fig:ssfr_mstar_gas}. The suppression of SFR for low mass systems is contributed by SNe feedback, while this suppression mechanism for high mass systems is provided by AGN feedback. However, the quantitative results for the suppression of SFR compared with secular evolution models are different in two aspects. On the one hand, the separation mass for SNe and AGN feedback dominance slightly decreases to $\lesssim8\times10^{10}~M_{\odot}$, which is actually more consistent with observations \citep[e.g.,][]{Kauffmann2003,Baldry2006,Silk2012} and cosmological simulation results \citep{Taylor2017}. When a massive gaseous component is added into the galaxies initially, this mass scale can become even lower. This can be shown from the feedback models for E190 galaxy with the stellar mass of $M_{\star}=8.0\times10^{10}~M_{\sun}$ (i.e., E190.agb+high gas: open green and E190.agb+sn+high gas: open red), where no significant suppression is seen.  In such a case, AGN feedback is needed for the further suppression of star formation activities. The incorporation of AGN feedback at high mass galaxies can self-regulate sSFR to a very low level, which is consistent with those in the secular evolution cases, even with a large amount of gas sources included initially, as shown with the blue open markers in Figure~\ref{fig:ssfr_mstar_gas}. The exact mass scale for suppression of star formation needs more detailed simulations in the future. On the other hand, the sSFRs at the initial stage reaches the maximum values for all these models, shown as the upper tips of the simulated data points. As galaxies evolved, the sSFRs decline due to different feedback mechanisms. The suppressed sSFRs increase by a factor of a few for SNe feedback models simply because of the inclusion of the initial gas sources, which are now more consistent with the observed data for quenched galaxies \citep{Renzini2015}.

The results for $T_{\rm X}$ are shown in Figure~\ref{fig:tx_mstar}. Here we only show the results of the low gas density cases for low mass galaxies (galaxies E190 and E220).
When SNe feedback is included in the low mass galaxies (black filled markers for E190 and E220 galaxies) or AGN feedback is involved in the high mass galaxies (grey filled ones for E260 and E340 galaxies), $T_{\rm X}$ are close to $T_{\sigma}$ and $T_{\rm X,obs}$. But when a tremendous gaseous component is added initially for SNe feedback models (i.e., E190.agb+sn+high gas, which are not presented in Figure~\ref{fig:tx_mstar}), $T_{\rm X}$ are again much higher than observations due to the compression work of the cooling-induced inflowing, which are similar to the overcooling AGB models. This indicates that a fine tuning of the initial gas density is necessary for SNe feedback models to be compatible with observations. This is because the SNe Ia rate is independent of the current thermal state of the X-ray emitting plasma; therefore, SNe Ia heating cannot act as a self-regulating mechanism.
In addition, AGN feedback can self-regulate $T_{\rm X}$ close to $T_{\sigma}$ and $T_{\rm X,obs}$ even when we adopt a relativity high gas density for high mass galaxies (grey filled markers for E260 and E340 galaxies). Accordingly, we could expect that the inclusion of AGN feedback in the low mass galaxies can also resolve the higher-predicated-$T_{\rm X}$ problem. X-ray observations show that the angular momentum of galaxies could also influence the plasma temperature $T_{\rm X}$.  Fast rotators generally have lower $T_{\rm X}$.

In Figure~\ref{fig:lx_mstar}, we show the $0.3-8$ keV ISM luminosity $L_{\rm X}$ for different feedback models with initial gas sources included. With a tenuous gaseous component for low mass galaxies ($n_{0}=0.01~{\rm cm^{-3}}$), SNe feedback can make $L_{\rm X}$ roughly consistent with X-ray observations \citep{Anderson2015}, especially close to \citet{Forbes2017}. For the high mass galaxies, AGN feedback can still make $L_{\rm X}$ close to the values of the secular evolution models, and be roughly in agreement with observations \citep{Forbes2017}, although they are slightly smaller than the observed values from \citet{Anderson2015}. In addition to the IGM accretion history, it is found that the presence of density structures in 3D simulations could be responsible for such inconsistency \citep{Tang2009,Tang2010}.

The gas masses within $10~r_{\rm eff}$ of the galaxies are tabulated in Table~\ref{tab:parameters}.
This basically shows similar trends as secular evolution models. As expected, the final gas masses retained within the galaxies are higher for the SNe feedback models in the low mass galaxies (e.g., E220.agb+sn+low gas vs. E220.agb+sn) and for the AGN feedback models in the high mass galaxies (E260.agb+sn+agn+gas vs. E260.agb+sn+agn; E340.agb+sn+agn+gas vs. E340.agb+sn+agn). This is straightforward to understand since there exists more gas sources initially throughout the galaxies. It can also explain why $L_{\rm X}$ of the SNe feedback models in low mass galaxies are higher than those of secular evolution cases.

\section{Conclusions and Discussions}\label{sec:summary}

We present 2D hydrodynamical numerical simulations to explore the role of different feedback models in suppressing star formation activities for different isolated ETGs and in regulating ISM properties. Different galaxy models are built with the fundamental plane of elliptical galaxies and \citeauthor{Faber1976} relation. Three different feedback mechanisms (i.e., AGB heating, SNe feedback and AGN feedback) are incorporated for the galaxies with different stellar masses. In most cases, we study the secular evolution cases, where all the gas sources exclusively come from the mass losses of evolved stars.

We find that SNe feedback can suppress star formation for low mass galaxies, while AGN feedback is efficient in regulating star formation activities for high mass galaxies (Figure~\ref{fig:ssfr_mstar} and Figure~\ref{fig:ssfr_mstar_gas}). Especially, AGN feedback can be very effective in suppressing star formation activities in the inner region of the massive galaxies (Figure~\ref{fig:sfrinner_f3}). The mass scale to separate these two feedback mechanisms is around a stellar mass of $M_{\star}\sim10^{11}~M_{\odot}$.  In any case, AGB heating cannot play a dominant role in preventing star formation in all our simulated ETGs. SNe can efficiently thermalize the ISM to very high temperatures without connection to the gravitational potential of the galaxies, which is, however, the limitation of AGB heating by definition. This physical difference results in the different feedback efficiency for these two feedback models.
The inclusion of a tenuous initial gaseous component shifts the SNe feedback dominating stellar mass downward slightly ($M_{\star}\lesssim8.0\times10^{10}~M_{\sun}$), which is more consistent with observational results \citep[e.g.,][]{Kauffmann2003,Baldry2006,Silk2012}.


The consistency between the X-ray plasma temperature $T_{\rm X}$ and the observed values (and $T_{\sigma}$ in some cases) can be built up by SNe feedback in low mass galaxies, and by AGN feedback in high mass galaxies (Figure~\ref{fig:tx_mstar}). This is similar to the energy balance and star formation suppression analysis.  Such a similarity is irrelevant to the question of AGB heating as suggested by some previous studies. This is because $T_{\rm X}$ can be more physically defined as the galactic potential energy in a hydrodynamical equilibrium system state. It is the SNe feedback in low mass systems and AGN feedback in high mass systems that can establish such a balanced state. These conclusions still hold when a certain amount of gas is added into the entire galaxy in the beginning of the simulation.

We further investigate the $0.3-8$ keV ISM luminosity $L_{\rm X}$ to compare with X-ray observations (Figure~\ref{fig:lx_mstar}). Although we find that $L_{\rm X}$ for SNe feedback models in low mass galaxies for secular evolution cases are underluminous, this discrepancy can be resolved when an additional gaseous component is included initially in the galaxy. AGN feedback can self-regulate $L_{\rm X}$ close to the observed values in high mass galaxies even when some gas sources are included. Due to the large uncertainties of the observed data, it seems that most of our models are consistent with observations. The remarkable difference among different models can still make $L_{\rm X}$ a good diagnostic tool to discriminate these models if good observational data are available in the future.

Although some gas sources are included initially when setting up the galaxy models, we do not incorporate the accretion of external gas sources from merger events, the accretion of ICM and/or IGM when galaxies evolve. The accretion process and violent events will allow galaxies to replenish with external gas material even when the galactic gas sources are depleted by star formation activities and blown out by feedback processes. This is the major limitation in the current work. Such effects will be explored in a separate work in the future.

We neglect the effect of the jet in the current work. This can be justified by the well collimated structure of the jet, which may simply pierce through the galaxy and have negligible interaction with the galaxy for the feedback study of a single galaxy, although the jet should be important for the evolution of large-scale structure such as galaxy clusters (e.g., the cooling flow problem discussed by \citealt{Yang2016,Bourne2017,Weinberger2017b}).
However, the assumption is needed to be examined in future work since there are still some debates for this issue \citep{Gaibler2012,Wagner2012}.

We should mention that the stellar component we adopt in this work is simply a \citeauthor{Jaffe1983} profile. However, the existence of a diffuse stellar component (DSC) in groups and clusters of galaxies is now well established \citep[e.g.,][]{Gonzalez2005}.  This component, also known as intracluster stars, mainly dominates the stellar light in the outer region of the galaxy. The dynamical difference between this component and the central dominant one has been found by cosmological simulations due to its velocity distribution \citep{Dolag2010}. It could be important for the central galaxy in cluster via the thermalization of the stellar wind. We will defer the implementation of this component in isolated galaxies to our future work.

\acknowledgments

We thank the referee for careful reading and constructive comments, which have significantly improved our paper. We are very grateful to Q. Daniel Wang and Fulai Guo for very helpful discussions and comments during the preparation of the paper.  This work is supported in part by the National Key Research and Development Program of China (Grant No. 2016YFA0400704), the Natural Science Foundation of China (grants 11573051, 11633006, 11703064, 11650110427, 11661161012), the Key Research Program of Frontier Sciences of CAS (No. QYZDJSSW- SYS008), and Shanghai Sailing Program (grant No. 17YF1422600). HM acknowledges the support from NSF AST-1517528 and  NSFC-11673015. ZG is also supported by the Natural Science Foundation of Shanghai (grant 18ZR1447200). YPL thanks the hospitality of Los Alamos National Laboratory where part of this work is performed. This work made use of the High Performance Computing Resource in the Core Facility for Advanced Research Computing at Shanghai Astronomical Observatory.

\appendix

\section{AGN feedback model}

The readers can refer to \citet{Yuan2018} for the details of AGN physics adopted in the work, which is more updated compared to previous works. Following \citet{Yuan2018}, we classify the AGN feedback modes into two categories, namely cold mode and hot mode feedback, according to the AGN luminosity (or mass accretion rate at Bondi radius $\dot{M}_{\rm Bondi}$) Eddington ratio. $\dot{M}_{\rm Bondi}$ can be estimated based on the inflow rate at the innermost grid. In both modes, the radiative and wind feedback are both incorporated in our AGN feedback models.

When $\dot{M}_{\rm Bondi}$ is larger than a critical value $\dot{M}_{\rm c}$ of $2\% \dot{M}_{\rm Edd}$, the accretion flow stays in the cold mode. We can calculate the wind mass, energy, and momentum flux based on \citet{Gofford2015}. After considering the viscosity timescale of accretion, we can then further obtain the black hole accretion rate $\dot{M}_{\rm BH}$ and AGN luminosity $L_{\rm BH}$ as well by assuming a radiative efficiency of $10\%$. For the Compton heating/cooling term, we simply use the result based on the observed spectrum of quasars \citep{Sazonov2004}, which gives the Compton temperature $T_{\rm C}=2\times10^{7}~{\rm K}$.

When $\dot{M}_{\rm Bondi}<\dot{M}_{\rm c}$, the accretion flow transfers to the hot mode. In this case, a truncation disk geometry, i.e., an inner hot accretion flow plus a truncated standard thin disk in the outer region, is adopted \citep{Yuan2014}. The wind mass, momentum and energy flux and their angular distribution are calculated according to \citet{Yuan2015}. The black hole mass accretion rate can also be obtained self-consistently after considering the disk wind. Using the radiative efficiency of a hot accretion flow \citep{Xie2012}, we can calculate the radiative output from the hot accretion flow $L_{\rm BH}$. Since the spectrum of a hot accretion flow is quite different from that of a cold disk, the Compton temperature is modified accordingly to obtain the Compton heating/cooling term \citep{Xie2017}.


With the updated AGN physics above, the hydrodynamics equations we solve are modified as follows,

\begin{equation} \label{eq:massconsvr_agn}
   \frac{\partial \rho}{\partial t} + \nabla\cdot(\rho{\bf v})
        = \alpha\rho_{\star} + \dot{\rho}_{\rm II} - \dot{\rho}_{\star}^{+},
\end{equation}
\begin{equation} \label{eq:momconsvr_agn}
   \frac{\partial {\bf m}}{\partial t} + \nabla\cdot({\bf m v})
        = - \nabla p_{\rm gas} + \rho {\bf g} -\nabla p_{\rm rad} - \dot{\bf m}^{+}_{\star},
\end{equation}
\begin{equation} \label{eq:engconsvr_agn}
   \frac{\partial E}{\partial t} + \nabla\cdot(E{\bf v})
        =  -p_{\rm gas} \nabla \cdot {\bf v} + H_{\rm AGN} - C  + \dot{E}_{\rm S} +\dot{E}_{\rm I}+\dot{E}_{\rm II} -\dot{E}^{+}_{\star},
\end{equation}
We can see that two additional terms are incorporated in momentum and energy equations, compared with Equations~(\ref{eq:momconsvr}-\ref{eq:engconsvr}), while the mass conservation equation remains the same. All other terms share the same meanings with those of Equations~(\ref{eq:massconsvr}$-$\ref{eq:engconsvr}).

The first obvious modification is that we include an extra radiative heating source due to the central AGN $H_{\rm AGN}$, which is the radiative heating rate per unit volume contributed by AGN radiation. The radiative heating and cooling term $H_{\rm AGN}-C$ in Equation~(\ref{eq:engconsvr_agn}) is computed following \citeauthor{Sazonov2005} (\citeyear{Sazonov2005}, see also Equations~($4.54-4.60$) in \citealt{Ciotti2012}), except that we further update the Compton temperature $T_{\rm C}$ as \citet{Xie2017}, according to the AGN Eddington ratio \citep{Yuan2018} as we have discussed  above.

The wind feedback is introduced by injecting the desired mass, momentum, and energy into the innermost grids of the simulation domain and then self-consistently calculate their radial transport (see also \citet{Ciotti2017} for a similar method in incorporating wind feedback by including these terms in Equations~(\ref{eq:massconsvr_agn}$-$\ref{eq:engconsvr_agn})).


For the radiation force $\nabla p_{\rm rad}$ in Equation~(\ref{eq:momconsvr_agn}), we follow \citet{Novak2011} and include both radiation pressure due to electron scattering and absorption of AGN photos by atomic lines. The radiation pressure contributed by electron scattering can be expressed as
\begin{equation} \label{eq:rad_es}
   (\nabla p_{\rm rad})_{\rm es}=-\frac{\rho\kappa_{\rm es}}{c}\frac{L_{\rm BH}}{4\pi r^2},
\end{equation}
where $\kappa_{\rm es}=0.35~{\rm cm^{2}~g^{-1}}$ is the electron scattering opacity. The photon absorption term can be computed as
\begin{equation} \label{eq:rad_photo}
   (\nabla p_{\rm rad})_{\rm photo}=-\frac{H_{\rm AGN}}{c}.
\end{equation}

\end{document}